\documentclass[11pt,a4paper]{article}
\usepackage{jheppub}
\usepackage[T1]{fontenc}
\usepackage{slashed}
\usepackage{empheq}
\usepackage{feynmp-auto}
\usepackage{shuffle}
\usepackage{float}
\usepackage[compat=1.1.0]{tikz-feynman}

\title{Five-point spinor-helicity Compton amplitudes in gauge theory and gravity}

\author{Raikhik Das}
\author{and Mao Zeng}

\affiliation{The Higgs Centre for Theoretical Physics, James Clerk Maxwell Building,\\
The University of Edinburgh,\\
Peter Guthrie Tait Road, King's Buildings,\\
Edinburgh EH9 3FD, United Kingdom}

\emailAdd{R.Das-4@sms.ed.ac.uk}
\emailAdd{mao.zeng@ed.ac.uk}

\abstract{We calculate the five-point tree-level Compton
  amplitudes for minimally coupled spinning matter in gauge theory and gravity in the massive
  spinor-helicity formalism, using BCFW recursion and KLT
  relations. The results are written in terms of symmetrized spinor products and are
  simultaneously valid for a range of matter spins, which directly generalizes the amplitudes of Arkani-Hamed, Huang, and Huang to the five-point case. Unlike the three- and
  four-point cases, each amplitude can be a sum of two or more
  distinct spin structures. Spurious poles are manifestly canceled through a mechanism that holds for all
  massive spins $s\leq 1$ in gauge theory and $s\leq 2$ in
  gravity. While we keep the amplitudes exact, their classical limits are
  expected to be relevant for binary systems of Kerr
  black holes in the post-Minkowskian expansion.}


\begin{document}

\maketitle

\section{Introduction}\label{sec:introduction}
\nocite{Arkani-Hamed:2017jhn}
Modern scattering-amplitude methods and closely related worldline methods have become a powerful approach to the post-Minkowskian expansion for the dynamics of compact binary systems \cite{Cheung:2018wkq, Kosower:2018adc, Bern:2019nnu, Bern:2019crd, Cristofoli:2019neg, Bjerrum-Bohr:2019kec, Brandhuber:2021eyq, Bern:2021dqo, Bern:2021yeh, Damgaard:2023ttc, Kalin:2020mvi, Kalin:2020fhe, Kalin:2022hph, Dlapa:2023hsl, Dlapa:2021npj, Mogull:2020sak, Jakobsen:2021smu, Jakobsen:2022psy, Jakobsen:2023oow, Driesse:2024xad, Driesse:2024feo, Bini:2017xzy, Bini:2018ywr, Vines:2017hyw, Vines:2018gqi, Guevara:2017csg, Guevara:2018wpp, Chung:2018kqs, Arkani-Hamed:2019ymq, Guevara:2019fsj, Chung:2019duq, Damgaard:2019lfh, Aoude:2020onz, Chung:2020rrz, Guevara:2020xjx, Bern:2020buy, Kosmopoulos:2021zoq, Chen:2021kxt, FebresCordero:2022jts, Bern:2022kto, Bern:2023ity, Menezes:2022tcs, Riva:2022fru, Damgaard:2022jem, Aoude:2022thd, Aoude:2022trd, Bautista:2022wjf, Gonzo:2023goe, Aoude:2023vdk, Lindwasser:2023zwo, Brandhuber:2023hhl, DeAngelis:2023lvf, Aoude:2023dui, Bohnenblust:2023qmy, Gatica:2024mur, Cristofoli:2021jas, Luna:2023uwd, Gatica:2023iws, Liu:2021zxr, Jakobsen:2021lvp, Jakobsen:2021zvh, Jakobsen:2022fcj, Jakobsen:2022zsx, Jakobsen:2023ndj, Jakobsen:2023hig, Heissenberg:2023uvo, Lindwasser:2023dcv, Bautista:2023sdf, Cangemi:2023ysz, Brandhuber:2024bnz, Chen:2024mmm, Correia:2024jgr, Bhattacharyya:2024kxj, Alaverdian:2024spu, Brandhuber:2024qdn, Brandhuber:2024lgl, Akpinar:2024meg, Bohnenblust:2024hkw, Haddad:2024ebn, Bonocore:2024uxk, Akpinar:2025huz, Bohnenblust:2025gir, Bern:2025wyd, Driesse:2026qiz, Dlapa:2026oyq, Haddad:2025cmw, Ben-Shahar:2025tiz}. In these developments, Compton amplitudes play a key role, both as inputs to generalized unitarity calculations of the scattering of massive bodies \cite{Bern:2011qt, Bjerrum-Bohr:2022blt, Kosower:2018adc, Bern:2019crd, Bern:2020buy, Aoude:2022thd, Aoude:2020onz, Chiodaroli:2021eug, Bjerrum-Bohr:2023jau, Bjerrum-Bohr:2023iey} and as probes of finite-size effects for compact bodies in the presence of massless perturbations \cite{Pound:2021qin, Loutrel:2020wbw, Nicasio:2000ge, Georgoudis:2025vkk, Sasaki:2003xr, Masaood:2020uhi, Bjerrum-Bohr:2025bqg, Anastasiou:2002zn, Braun:2024srt, Ji:2023xzk, Braun:2020yib}. Gauge-theory Compton amplitudes serve as simple and useful examples and are intimately connected with graviton Compton amplitudes via the double copy \cite{Kawai:1985xq, Bern:2008qj, Bern:2010ue, Bern:2010yg, Bern:2017yxu, Bern:2019prr, Bern:2022wqg, Adamo:2022dcm, Chiodaroli:2021eug, Bjerrum-Bohr:2020syg, Bjerrum-Bohr:2024fbt}.

In this paper, we focus on calculating tree-level five-point Compton amplitudes in the massive spinor-helicity (MSH) formalism \cite{Arkani-Hamed:2017jhn}. This formalism naturally extends the massless spinor-helicity formalism \cite{Mangano:1990by, Dixon:1996wi, Xu:1986xb} and has opened up several new avenues in amplitude calculations. 
The massless and massive spinor-helicity formalisms, especially when combined with on-shell recursion relations \cite{Britto:2005fq}, significantly reduce the computational burden associated with traditional Feynman-diagram calculations in both gauge and gravity theories, yielding extremely compact results with manifest little-group transformation properties for both massless and massive particles.

Over the years, Compton amplitudes for spinning matter have been studied using many different approaches, including recursion relations \cite{Aoude:2020onz, Ballav:2020ese, Ballav:2021ahg, Haddad:2023ylx, Ema:2025qgd}, covariant bootstrap methods \cite{Vazquez-Holm:2025ztz, Bjerrum-Bohr:2023iey}, a string-theoretic framework \cite{Azevedo:2024rrf}, and massive higher-spin gauge symmetry \cite{Cangemi:2022bew, Cangemi:2023bpe, Cangemi:2023ysz}. The main focus of this paper is the computation of \emph{five-point} Compton amplitudes at tree level. While these amplitudes have been explored in the classical or heavy-mass limit in the literature \cite{Bjerrum-Bohr:2023jau, Vazquez-Holm:2025ztz}, we present exact finite-mass tree-level expressions, written in a form that is simultaneously valid for a range of spins of the massive quantum particles.

We calculate Yang--Mills five-point Compton amplitudes for minimal coupling in the MSH formalism using Britto-Cachazo-Feng-Witten (BCFW) recursion relations. The color ordering in Yang--Mills theory keeps the number of BCFW channels small.  The five-point Yang--Mills Compton amplitudes obtained in this work are valid for massive particles with spin $s\leq 1$ and have the following form as a sum over BCFW channels:
\begin{equation}\label{eq:bcfw-spin-structure-decomposition}
    A^{(s)}=\sum_{j\,\in\ \text{channels}} A_j^{(0)}\: Z_j^{2s} \ .
\end{equation}
Here the coefficients $A^{(0)}_j$ are the massive scalar amplitudes, and all spin dependence comes from the factors $Z_j$, which is the spin structure corresponding to the BCFW channel $j$. 
We will also calculate the same QED and gravity amplitudes with BCFW recursion, and the amplitudes take the same form of Eq.~\eqref{eq:bcfw-spin-structure-decomposition}. In particular, $Z_j$ is a universal expression which depends only on the chosen BCFW shift, the factorization channel, and the helicities of the massless legs, but is unchanged across gauge theory and gravity, and unchanged between different color orderings of the Yang--Mills amplitudes.

However, unlike the Yang--Mills case, QED and gravity amplitudes have a large number of BCFW channels, which makes it harder to analytically demonstrate the cancellation of spurious poles. Therefore, we use permutation sums and Kawai-Lewellen-Tye (KLT) relations \cite{Kawai:1985xq}, respectively, to obtain our main results for QED and gravity amplitudes.

The structure of the same-helicity Compton amplitudes is very simple for both Yang--Mills theory and gravity for minimally coupled massive particles. In the all-plus case, the amplitudes are \cite{Aoude:2020onz, Ballav:2021ahg, Ochirov:2018uyq}:
\begin{equation}
    A^{(s)}=A^{(0)} \left(\frac{\langle \boldsymbol 1 \boldsymbol n \rangle}{m}\right)^{2s} \ , \label{eq:all-plus-spin-exponential}
\end{equation}
where legs 1 and $n$ are massive legs with mass $m$ and spin $s$. In the convention of ref.~\cite{Arkani-Hamed:2017jhn}, an external massive particle $k$ has its wavefunction written in the following basis of symmetrized products of $2s$ two-component spinors
\begin{equation}
  |k^{(I_1} \rangle \dots | k ^{I_{2s})} \rangle \ ,
\end{equation}
where $I_a=+,-$ labels a basis of the fundamental representation of the massive SU(2) little group.
The spin-exponential notation on the RHS of eq.~\eqref{eq:all-plus-spin-exponential} is defined in terms of spinor products symmetrized over little group indices,
\begin{equation}
\left(\langle \boldsymbol{1}\boldsymbol{n}\rangle^{2s}\right)^{I_1\cdots I_{2s};J_1\cdots J_{2s}}
\equiv
\sum_{\sigma\in S_{2s}}
\prod_{a=1}^{2s}
\langle 1^{I_a} n^{J_{\sigma(a)}}\rangle \ .
\end{equation}
It is understood that the LHS of eq.~\eqref{eq:all-plus-spin-exponential} also has multiple components, $A^{(s)\,I_1\cdots I_{2s};J_1\cdots J_{2s}}$.
The all-negative-helicity case follows trivially by charge conjugation. A key insight of ref.~\cite{Arkani-Hamed:2017jhn} is the convenient ``spin exponential'' notation under which the amplitudes in eqs.~\eqref{eq:all-plus-spin-exponential} are valid for all spins of the massive particle under a suitable definition of minimal coupling.

The structure of different-helicity Compton amplitudes, on the contrary, is less simple. As we will see, starting at five points, a different-helicity Compton amplitude contains multiple independent spin structures accompanied by spin-independent factors. Moreover, while these amplitudes again hold for a range of different spins, they are valid only for $s\leq1$ in the Yang--Mills case and $s\leq2$ in the gravity case. In the allowed spin ranges, we elucidate the mechanism through which the spurious poles manifestly cancel between contributions from different spin structures. The mechanism stops working outside the allowed spin ranges, in which case uncanceled spurious poles develop in eqs.~\eqref{eq:bcfw-spin-structure-decomposition}.

It is expected that Compton amplitudes (including five-point ones) of minimally coupled massive matter in $D$ dimensions follow from dimensional reduction of massless pure gluon or graviton amplitudes in $D+2$ dimensions \cite{Bern:2019crd, Johansson:2019dnu, Bautista:2019evw, Chiodaroli:2021eug}. Therefore, the massive amplitudes inherit a large number of relations from massless gluon / graviton amplitudes, including Kleiss--Kuijf (KK) relations \cite{Kleiss:1988ne}, Bern-Carrasco-Johansson (BCJ) relations \cite{Bern:2008qj, Bjerrum-Bohr:2009ulz, Feng:2010my, delaCruz:2015dpa}, and KLT relations \cite{Kawai:1985xq, Bjerrum-Bohr:2004vlu, Bjerrum-Bohr:2010mia, Bjerrum-Bohr:2016axv, Cao:2021dcd}, as reviewed in Appendices \ref{app:dimensional-reduction} and \ref{app:klt-double-copy}.

The main result obtained by this work is the five-point tree-level amplitudes in a form that simultaneously satisfies the following properties:
\begin{itemize}
\item The amplitudes are given as compact expressions in the massive spinor-helicity formalism.
\item They are written in spin-exponentiated forms that hold across a range of different spins.
\item Manifest absence of spurious poles when working below a maximum spin bound.
\item No classical soft / large-mass expansion.
\end{itemize}

We have performed several nontrivial consistency checks on the five-point Compton amplitudes obtained in this work, using analytic parametrizations of spinors presented in app.~\ref{app:analytic-parametrization}. The analytic parametrization writes all massless and massive spinor components explicitly in terms of a set of independent complex variables while preserving kinematic constraints such as momentum conservation. This allows spinor expressions and Lorentz dot products to be evaluated as rational functions in these variables without ambiguities from e.g.\ Schouten identities, so that two amplitudes are only equal if they evaluate to the same rational function.
Known covariant $D$-dimensional results, while being much more complex, provide a valuable cross-check for our compact 4-dimensional spinor-helicity results for individual massive spin representations. Specifically, we compare our Yang--Mills results for massive spin-$0$ and spin-$1$ particles with amplitudes computed using the Mathematica packages \texttt{IncreasingTrees} \cite{Edison:2020ehu} followed by dimensional reduction. The $D$-dimensional results are also used to verify the large-$z$ behavior of amplitudes under various BCFW shifts (see refs.~\cite{Arkani-Hamed:2008bsc, Cheung:2008dn}), with the help of the Mathematica packages \texttt{FeynCalc} and \texttt{SpinorHelicity4D} \cite{AccettulliHuber:2023ldr}. We have also performed internal cross-checks. In particular, resolving spurious poles from BCFW is only performed for Yang--Mills amplitudes, while for QED and gravity, we obtain amplitudes free of spurious poles by relating them to Yang--Mills amplitudes, and the expressions are checked against BCFW calculations of these QED and gravity amplitudes, again using analytic parametrization of spinors.

The outline of the paper is as follows. In sec.~\ref{sec:bcfw-recursion}, we review the BCFW recursion method and the massless--massless and massive--massless shifts used throughout the paper. In sec.~\ref{sec:ym-compton-bcfw}, we derive the Yang--Mills Compton amplitudes using BCFW recursion. In sec.~\ref{sec:qed-compton}, we derive the QED amplitudes by summing gluon partial amplitudes over suitable leg permutations. In sec.~\ref{sec:graviton-compton-klt}, we derive the gravity amplitudes from the gluon amplitudes using KLT relations. Thus, both secs.~\ref{sec:qed-compton} and \ref{sec:graviton-compton-klt} use the gluon amplitudes of sec.~\ref{sec:ym-compton-bcfw} as their starting point. The results are checked in sec.~\ref{sec:bcfw-verification}, where some of the QED, Yang--Mills, and gravity amplitudes are re-derived from scratch using BCFW recursion. We conclude in sec.~\ref{sec:conclusion}; acknowledgments follow. The appendices collect our conventions in app.~\ref{app:conventions}, the three-point Yang--Mills and gravity amplitudes in app.~\ref{app:three-point-ym-gravity-amplitudes}, the four-vector construction from bi-spinors in app.~\ref{app:bispinor-four-vector-construction}, analytic spinor parametrizations in app.~\ref{app:analytic-parametrization}, dimensional reduction checks in app.~\ref{app:dimensional-reduction}, Bern-Carrasco-Johansson (BCJ) relations \cite{Bern:2008qj} for Yang--Mills Compton amplitudes in app.~\ref{app:bcj-relations}, the KLT double copy in app.~\ref{app:klt-double-copy}, and as a pedagogical review, Feynman-rule derivations of three- and four-point QED amplitudes in the massive spinor-helicity notation in app.~\ref{app:qed-feynman-rules}.

%----------------------------------------------------------------------------

\section{A Brief Review of BCFW Recursion Methodology}\label{sec:bcfw-recursion}
The inefficiency of traditional methods for Feynman diagrams and Feynman rules in calculating higher-point gauge and gravity amplitudes is well known. Modern on-shell methods have proven extremely useful for tackling such problems. The BCFW recursion relation \cite{Britto:2005fq} is one such on-shell method, which is very effective in calculating higher-point tree amplitudes.
%Although the BCFW recursion relations proposed shifting two massless legs in the amplitude, there have been other variants of the BCFW recursion relations where more than two legs or massive legs are shifted.
Since our amplitudes involve both massless and massive particles, we use both the massless--massless \cite{Britto:2005fq, Elvang:2013cua} shifts and the massive--massless \cite{Wu:2021nmq, Ballav:2020ese, Ballav:2021ahg} shifts to calculate the Compton amplitudes.
%The same result can obviously be derived by using the other variants of the BCFW recursion.
Here, we present a very brief derivation of the BCFW recursion relations involving the shift of two legs. 

Let us shift the momenta of two legs:
\begin{equation}\label{eq:bcfw-momentum-shift}
    \hat{p}_i^\mu = p_i^\mu + z \, r^\mu \hspace{10mm}\text{and}\hspace{10mm} \hat{p}_j^\mu = p_j^\mu - z \, r^\mu \ ,
\end{equation}
such that the total momentum remains conserved in the shifted amplitude. Here, $z \in \mathbb{C}$ is a deformation parameter. As we wish to keep the on-shell property intact, we demand
\begin{equation}\label{eq:bcfw-shift-vector-constraints}
    p_i \cdot r = p_j \cdot r = 0 \ ,\quad r^2 =0 \ .
\end{equation}

The scattering amplitude $\mathcal{A}_{n}$ with undeformed momenta is related to the deformed amplitude $\widehat{\mathcal{A}}_{n}(z)$ by Cauchy's theorem:
\begin{equation}\label{eq:bcfw-cauchy-relation}
\mathcal{A}_{n}=\widehat{\mathcal{A}}_{n}(0)=\frac{1}{2 \pi i} \oint_{\Gamma} \frac{\widehat{\mathcal{A}}_{n}(z)}{z} d z=-\sum_{z_{I}} \operatorname{Res}_{z=z_{I}}\left(\frac{\widehat{\mathcal{A}}_{n}(z)}{z}\right)+\mathcal{R}_{n} \ .
\end{equation}

The contour $\Gamma$ encloses the pole at the origin. $\mathcal{R}_{n}$ is the boundary term (i.e., the residue at infinity). All other simple poles of the amplitude are denoted by $z_{I}$.

In tree-level scattering amplitudes, the poles in the $z$-plane correspond to the poles observed in the kinematic space. The modified amplitude $\widehat{\mathcal{A}}_{n}(z)$ exhibits simple poles in kinematic space whenever an internal propagator of the type \[\frac{i}{\widehat{P}^{2}-m^{2}}\] becomes on-shell. When this happens, the amplitude $\widehat{\mathcal{A}}_{n}(z)$ decomposes into two lower-point on-shell sub-amplitudes. This is a direct consequence of the unitarity of the theory. Hence, we have
\begin{equation}\label{eq:bcfw-factorization}
\widehat{\mathcal{A}}_{n}(z)=\sum_{I} \widehat{\mathcal{A}}_{l+1} \, \frac{i}{\widehat{P}_{I}^{2}-m^{2}}\, \widehat{\mathcal{A}}_{r+1} \ ,
\end{equation}
where each $I$ represents a factorization channel in which the shifted internal momentum $\widehat P_I$ becomes on shell, and $n=l+r$. It is crucial to recognize that the sub-amplitudes depend on shifted momenta. To achieve the desired pole structure, we can express the shifted propagator using the propagator with unshifted momenta:
\begin{equation}\label{eq:bcfw-shifted-propagator}
\widehat{P}_{I}^{2}|_{z=z_{I}}=m^{2} \Rightarrow\left(P_{I}+z_{I}\, p_{j}\right)^{2}=m^{2} \Longrightarrow \frac{1}{\widehat{P}_{I}^{2}-m^{2}}=-\frac{z_{I}}{z-z_{I}} \frac{1}{P_{I}^{2}-m^{2}} \ .
\end{equation}

We will now concentrate on the boundary term $\mathcal{R}_{n}$ as $z$ approaches infinity. This term cannot be determined using just this recursion relation. Therefore, a BCFW shift like the one described in eq.~\eqref{eq:bcfw-momentum-shift} is considered valid when this boundary term disappears. It is important to note that $\mathcal{R}_{n}$ may become zero under certain conditions.
\begin{equation}\label{eq:bcfw-vanishing-boundary-term}
\widehat{\mathcal{A}}_{n}(z) \rightarrow 0 \ , \quad \text { as } z \rightarrow \infty \ .
\end{equation}
Eq.~\eqref{eq:bcfw-vanishing-boundary-term} is treated as the condition of validity for BCFW recursions.

Finally, we will focus on the scheme to compute the amplitude using the BCFW recursion.
\begin{equation}\label{eq:bcfw-recursion-relation}
\mathcal{A}_{n}=-\operatorname{Res}_{z=z_{I}}\left(\frac{1}{z} \sum_{I} \widehat{\mathcal{A}}_{l+1}(z) \frac{i}{\widehat{P}_{I}^{2}-m^{2}} \widehat{\mathcal{A}}_{r+1}(z)\right)=\sum_{I} \widehat{\mathcal{A}}_{l+1}\left(z_{I}\right)\, \frac{i}{P_{I}^{2}-m^{2}}\, \widehat{\mathcal{A}}_{r+1}\left(z_{I}\right) \ .
\end{equation}

Only figures where the two deformed momenta are positioned on opposite sides of the on-shell propagator contribute, as the momentum shift of the propagator is otherwise zero. This reduces the number of factorization channels and leads to significant computational simplifications. In this paper, we will employ this method to work with spinors in calculating the gluon and graviton Compton amplitudes.

%----------------------------------------------------------------------------

\subsection{Massless--massless Shift}\label{subsec:bcfw-massless-massless-shift}
For massless particles $i$ and $j$, we will look at the two-line massless--massless shift. We will do a holomorphic shift for leg $i$ and an anti-holomorphic shift for leg $j$ \cite{Elvang:2013cua}.
\begin{equation}\label{eq:bcfw-massless-shift-spinors}
    |\hat{i}]=|i]+z|j] \ , \quad |\hat{i}\rangle=|i\rangle \ ;\quad
    |\hat{j}\rangle=|{j}\rangle-z|i\rangle \ , \quad |\hat{j}]=|j] \ .
\end{equation}

The shift in eq.~\eqref{eq:bcfw-massless-shift-spinors} shifts the momentum in the following way:
\begin{equation}\label{eq:bcfw-massless-shift-momenta}
    \hat{p}_i^\mu=p_i^\mu + z\, r^\mu , \qquad \hat{p}_j=p_j - z \,
    r^\mu \ ; \quad \text{where } r^\mu = \frac{\langle i|\gamma^\mu| j]}{2} \ .
\end{equation}
The shift in eq.~\eqref{eq:bcfw-massless-shift-spinors} is called a $[i,j\rangle$ shift.

%----------------------------------------------------------------------------

\subsection{Massive--massless Shift}\label{subsec:bcfw-massive-massless-shift}
For a massless particle $i$ and a massive particle $j$, let us look at two-line shifts. They can be performed in two different ways. The $[i, \mathbf{j}\rangle$ shift \cite{Wu:2021nmq, Ballav:2020ese, Ballav:2021ahg} is an option; it shifts the massive line $j$ anti-holomorphically and the massless line $i$ holomorphically. An unknown $\zeta^{J}$ is introduced; thus, the shifted spinors are
\begin{equation}\label{eq:bcfw-massive-massless-shift-ansatz}
| \hat{i}]=| i]+z | \mathbf{j}^{J}] \zeta_{J} \ , \quad|\hat{\mathbf{j}}^{J}\rangle=|\mathbf{j}^{J}\rangle-z|i\rangle \zeta^{J} \ .
\end{equation}

The shift vector $2 r^{\mu}=\langle i| \gamma^{\mu} | \mathbf{j}^{J}] \zeta_{J}$ is orthogonal to the massless momentum $p_{i}$ as $\langle i| p_{i}=0$. To preserve the on-shell nature of the amplitude, we need to have
\begin{equation}\label{eq:bcfw-massive-massless-shift-constraint}
2 p_{j} \cdot r=\langle i| p_{j} | \mathbf{j}^{J}] \zeta_{J}=m_{j}\langle i \mathbf{j}^{J}\rangle \zeta_{J}=0 \ .
\end{equation}
Hence, $\zeta^{J}=\langle i \mathbf{j}^{J}\rangle$. Substituting this in eq.~\eqref{eq:bcfw-massive-massless-shift-ansatz}, we get the explicit form of the $[i, \mathbf{j}\rangle$ shift.
\begin{equation}\label{eq:bcfw-massive-massless-shift-spinors}
| \hat{i}]=| i]+z | \mathbf{j}^{J}]\langle i \mathbf{j}_{J}\rangle, \quad|\hat{\mathbf{j}}^{J}\rangle=|\mathbf{j}^{J}\rangle-z|i\rangle\langle i \mathbf{j}^{J}\rangle  \ .
\end{equation}

For this deformation, the shifted momenta are 
\begin{equation}\label{eq:bcfw-massive-massless-shift-momenta}
    \hat{p}_{j}^{\mu}=p_{j}^{\mu}-z\,r^{\mu} \ , \quad
    \hat{p}_{i}^{\mu}=p_{i}^{\mu}+z\,r^{\mu} \ , \quad
    2r^{\mu}= -\langle i|\gamma^\mu\mathbf{j}|i\rangle \ .
\end{equation}

We can similarly perform the $[\mathbf{j}, i\rangle$ shift \cite{Wu:2021nmq, Ballav:2020ese, Ballav:2021ahg}:
\begin{equation}\label{eq:bcfw-conjugate-massive-massless-shift-spinors}
| \hat{\mathbf{j}}^{J}]=| \mathbf{j}^{J}]+z | i][i \mathbf{j}^{J}], \quad|\hat{i}\rangle=|i\rangle-z|\mathbf{j}^{J}\rangle[i \mathbf{j}_{J}].
\end{equation}
For this deformation, the shifted momenta are 
\begin{equation}\label{eq:bcfw-conjugate-massive-massless-shift-momenta}
    \hat{p}_{j}^{\mu}=p_{j}^{\mu}+z\,r^{\mu} \ , \quad
    \hat{p}_{i}^{\mu}=p_{i}^{\mu}-z\,r^{\mu} \ , \quad
    2r^{\mu}= -[i|\gamma^\mu\mathbf{j}|i] \ .
\end{equation}
The corresponding bispinor is of the form $r_{\alpha\dot\alpha}=|\mathbf{j}^{J}\rangle_{\alpha}[i\mathbf{j}_{J}][i|_{\dot\alpha}$, so $r^{2}=0$. It is also orthogonal to the massless momentum $p_i$:
\begin{equation}\label{eq:bcfw-conjugate-shift-null-vector}
    2p_i\cdot r=\langle \mathbf{j}^{J}|p_i|i]\,[i\mathbf{j}_{J}]
    =\langle \mathbf{j}^{J} i\rangle [ii]\,[i\mathbf{j}_{J}]=0 \ .
\end{equation}
Finally, using the massive on-shell relation in app.~\ref{app:conventions},
\begin{equation}\label{eq:bcfw-conjugate-shift-massive-onshell-condition}
    2p_j\cdot r=\langle \mathbf{j}^{J}|p_j|i]\,[i\mathbf{j}_{J}]
    =m_j [i\mathbf{j}^{J}][i\mathbf{j}_{J}]=0 \ ,
\end{equation}
where the last equality follows from the antisymmetric $SU(2)$ little-group contraction. Thus the $[\mathbf{j},i\rangle$ shift preserves total momentum conservation and keeps both shifted legs on shell.

%----------------------------------------------------------------------------

\section{Yang--Mills Compton Amplitudes from BCFW}\label{sec:ym-compton-bcfw}
A generic $n$-point Yang--Mills gluon amplitude can be decomposed into color structures in a trace basis:
\begin{equation}\label{eq:YM_Compton_Full_Amplitude}
    \begin{aligned}
    & A(1,2,\hdots,n-1,n)\: \\
    =& \: g^{n-2}\, \sum_{\sigma \in S_n / Z_n} \operatorname{Tr}( \Tilde{T}^{a_{\sigma(1)}} \Tilde{T}^{a_{\sigma(2)}} \hdots \Tilde{T}^{a_{\sigma(n-1)}} \Tilde{T}^{a_{\sigma(n)}} ) \, A[\sigma(1), \sigma(2),\hdots,\sigma(n-1),\sigma(n)] \ ,
    \end{aligned}
\end{equation}
where $\Tilde{T}^a = \sqrt{2} \, T^a$ are the generators of the $SU(N)$ generators and $g$ is the coupling constant for three-point Yang--Mills interaction. The expression $A[\sigma(1), \sigma(2),\hdots,\sigma(n-1),\sigma(n)]$ is called the \emph{color-ordered amplitude} or \emph{partial amplitude}, and it is cyclically invariant.

When leg-$1$ and leg-$n$ have massive spin-$s$ particles charged under the adjoint representation and the other legs $(2,3,\hdots,n-1)$ are massless gluons, the color basis remains identical to pure-gluon amplitudes, so we can naturally write down color-ordered amplitudes like eq.~\eqref{eq:ym-three-point-positive-helicity}. However, cyclic invariance is modified by fermion exchange signs, which give an extra factor of $(-1)^{2s}$ whenever $\boldsymbol n^s$ appears before $\boldsymbol 1^s$ in the list of numerical particle labels in the color-ordered amplitude.

In this section, we focus on color-ordered Compton amplitudes with two massive spin-$s$ legs and one to three gluon legs.
In sec.~\ref{subsec:ym-three-point-compton}, we present the well-known three-point Yang--Mills Compton amplitudes \cite{Arkani-Hamed:2017jhn}. These amplitudes serve as essential building blocks for the four-point Compton amplitudes discussed in sec.~\ref{subsec:ym-four-point-compton}. Finally, by utilizing the three- and four-point results, we construct the five-point Yang--Mills Compton amplitudes in sec.~\ref{subsec:ym-five-point-compton}.

%----------------------------------------------------------------------------

\subsection{Three-Point Compton Amplitudes}\label{subsec:ym-three-point-compton}
The three-point Yang--Mills Compton amplitudes involving a massive particle of mass $m$ and spin $s$ are \cite{Arkani-Hamed:2017jhn}
\begin{empheq}[box=\fbox]{align}
A[\boldsymbol{1}^s,2^+,\boldsymbol{3}^s]
&=  i\,\frac{\langle \eta | \boldsymbol{1} | 2]}{\langle \eta 2 \rangle}
   \frac{\langle \boldsymbol{13} \rangle^{2s}}{m^{2s}} \ ,
   \label{eq:ym-three-point-positive-helicity} \\[4pt]
A[\boldsymbol{1}^s,2^-,\boldsymbol{3}^s]
&=  i \, \frac{\langle 2 | \boldsymbol{1} | \eta ]}{[2 \eta]}
   \frac{[\boldsymbol{13}]^{2s}}{m^{2s}} \ .
   \label{eq:ym-three-point-negative-helicity}
\end{empheq}
Here, $|\eta\rangle$ and $|\eta]$ are reference spinors.
\begin{figure}[H]
    \centering
    \begin{fmffile}{3_0_d01}
        \begin{fmfgraph*}(90,60)
            \fmfleft{i1,i2}
            \fmfright{o1,o2}
            \fmf{plain}{v1,i1}
            \fmfv{label=$\boldsymbol{1}^s$}{i1}
            \fmf{plain}{v1,o1}
            \fmfv{label=$\boldsymbol{3}^s$}{o1}
            \fmf{phantom}{i2,v2,o2}
            \fmf{gluon}{v2,v1}
            \fmfv{label=$2^\pm $}{v2}
            \fmfdot{v1}
        \end{fmfgraph*}
    \end{fmffile}
    \vspace{5mm}
    \caption{Three-point amplitude $A[\boldsymbol{1}^s,2^\pm,\boldsymbol{3}^s]$}
    \label{fig:3pt-compton}
\end{figure}
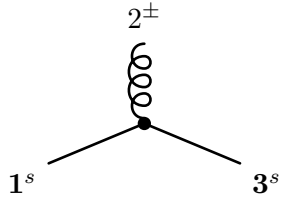

%----------------------------------------------------------------------------

\subsection{Four-Point Compton Amplitudes}\label{subsec:ym-four-point-compton}

As discussed previously, three-point Yang--Mills Compton amplitudes can be used to construct four-point amplitudes via BCFW recursion relations.

Unlike photon Compton scattering, four-point gluon Compton amplitudes depend on color ordering. Consequently, different orderings exhibit distinct pole structures. This feature becomes increasingly important at higher multiplicity.

\vspace{5mm}

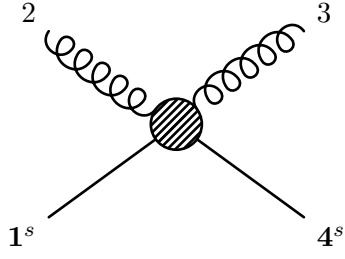
\begin{figure}[H]
    \centering
    \begin{fmffile}{3_0_d02}
        \begin{fmfgraph*}(120,80)
            \fmfleft{i1,i2}
            \fmfright{o1,o2}
            \fmfblob{.16w}{g}
            \fmf{plain}{i1,g}
            \fmfv{label=$\boldsymbol{1}^s$}{i1}
            \fmf{plain}{g,o1}
            \fmfv{label=$\boldsymbol{4}^s$}{o1}
            \fmf{gluon}{i2,g}
            \fmfv{label=$2$}{i2}
            \fmf{gluon}{o2,g}
            \fmfv{label=$3$}{o2}
        \end{fmfgraph*}
    \end{fmffile}
    \vspace{5mm}
    \caption{Four-point amplitude $A[\boldsymbol{1}^s,2,3,\boldsymbol{4}^s]$}
    \label{fig:4pt-1234}
\end{figure}                                                                              

We first consider the ordering: $A[\boldsymbol{1}^s,2,3,\boldsymbol{4}^s]$, where particles 2 and 3 are gluons and particles 1 and 4 are massive spin-$s$ states. 

\vspace{5mm}

\begin{figure}[H]
    \centering
    \begin{fmffile}{3_0_d02_1}
        \begin{fmfgraph*}(200,60)
            \fmfleft{i1,i2}
            \fmfright{o1,o2,o3}
            \fmfblob{.08w}{g1,g2}
            \fmf{plain}{i1,g1}
            \fmfv{label=$\boldsymbol{1}^s$}{i1}
            \fmf{gluon}{i2,g1}
            \fmfv{label=$\hat{2}$}{i2}
            \fmf{plain,label=$-\hat{\boldsymbol{P}}^s$}{g1,v1}
            \fmf{phantom}{v1,v2}
            \fmf{plain, label=$\hat{\boldsymbol{P}}^s$}{v2,g2}
            \fmf{gluon}{o3,g2}
            \fmfv{label=$\hat{3}$}{o3}
            \fmf{plain}{g2,o1}
            \fmfv{label=$\boldsymbol{4}^s$}{o1}
        \end{fmfgraph*}
    \end{fmffile}
    \vspace{5mm}
    \caption{Factorization channel contributing to
    $A[\boldsymbol{1}^s,\hat{2} \,|\, \hat{3},\boldsymbol{4}^s]$}
    \label{fig:ym-four-point-bcfw-channel-12}
\end{figure}
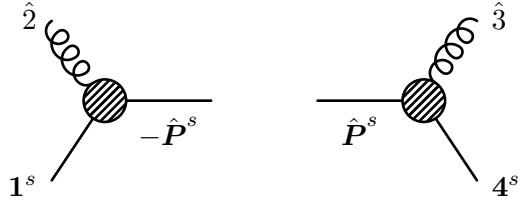

We define the kinematic invariants
\begin{equation}
  s_{ij} = (p_i+p_j)^2 \ , \quad \tau_{ij} = 2\,p_i\!\cdot\!p_j \ .
  \label{eq:Mandef}
\end{equation}
Throughout this paper we use $s_{ij}$ when legs $i$ and $j$ are both massless or both massive, and $\tau_{ij}$ in the mixed massive--massless case. $s_{ij}=0$ and $\tau_{ij}=0$ both indicate factorization channels. When either index is shifted (denoted by a hat), the same rule applies.

To calculate $A[\boldsymbol{1}^s,2^+,3^-,\boldsymbol{4}^s]$, we can do a $\langle 2,3]$ shift:
\begin{equation*}
\begin{aligned}
|\hat{2}\rangle &= |2\rangle - z\,|3\rangle \ ,
&\quad
|\hat{2}] &= |2] \ , \\
|\hat{3}] &= |3] + z\,|2] \ ,
&\quad
|\hat{3}\rangle &= |3\rangle \ .
\end{aligned}
\end{equation*}

The only channel that contributes due to this shift is the $\boldsymbol{1}\hat{2}$-channel as we can see in fig.~\ref{fig:ym-four-point-bcfw-channel-12}. In this cut,
\begin{equation}
  \tau_{1\hat{2}}=0 \implies z=\frac{\tau_{12}}{\langle 3|\boldsymbol{1}|2]} \ .
  \label{eq:deformation_param}
\end{equation}
The four-point amplitude from BCFW is
{\allowdisplaybreaks
\begin{align}
\nonumber A[\boldsymbol{1}^s,2^+,3^-,\boldsymbol{4}^s]
=&\nonumber A[\boldsymbol{1}^s,\hat{2}^+,\hat{\boldsymbol{P}}^s]\frac{i}{\tau_{12}}A[-\hat{\boldsymbol{P}}^s,\hat{3}^-,\boldsymbol{4}^s]\\
=&\nonumber \left(i \, \frac{\langle 3|\boldsymbol{1}|2]}{\langle32\rangle}\frac{\langle \boldsymbol{ 1} \hat{\boldsymbol{P}}^I\rangle^{2s}}{m^{2s}}\right)\frac{i}{\tau_{12}} \left(i \, \frac{\langle 3|- \hat{\boldsymbol{P}}|2]}{[32]}\frac{[-\hat{\boldsymbol{P}}_I \, \boldsymbol{4}]^{2s}}{m^{2s}}\right)\\
=&\nonumber i\frac{\langle 3|\boldsymbol{1}|2]^2}{\tau_{12}s_{23}}\left(\frac{[\boldsymbol{4|\hat{\boldsymbol{P}}|1}\rangle}{m^2}\right)^{2s}\\
=&\nonumber i \, \frac{\langle 3|\boldsymbol{1}|2]^2}{\tau_{12}s_{23}} \left(-\frac{[\boldsymbol{4}|(\boldsymbol{1}+\hat{2})|\boldsymbol{1}\rangle}{m^2}\right)^{2s}\\
=&\nonumber i \, \frac{\langle 3|\boldsymbol{1}|2]^2}{\tau_{12}s_{23}}\left(-\frac{m[\boldsymbol{41}]+[\boldsymbol{4}|2|\boldsymbol{1}\rangle-z[\boldsymbol{4}2]\langle3\boldsymbol{1}\rangle}{m^2}\right)\\
\implies \Aboxed{A[\boldsymbol{1}^s,2^+,3^-,\boldsymbol{4}^s]=& i \, \frac{\langle 3 | \boldsymbol{1} | 2 ]^{2-2s}}{\tau_{12}\,s_{23}}\big(\langle \boldsymbol{1}3 \rangle [\boldsymbol{4}2]+ \langle\boldsymbol{4}3 \rangle [\boldsymbol{1}2] \big)^{2s}}\ .
\label{eq:ym-four-point-positive-negative-helicity}
\end{align}
}

After charge conjugation, we also obtain
\begin{align}
\Aboxed{A[\boldsymbol{1}^s,2^-,3^+,\boldsymbol{4}^s]= i \, \frac{\langle 2 | \boldsymbol{1} | 3 ]^{2-2s}}{\tau_{12}\,s_{23}}\big(\langle \boldsymbol{4}2 \rangle [\boldsymbol{1}3]+ \langle\boldsymbol{1}2 \rangle [\boldsymbol{4}3] \big)^{2s}} \ .
\label{eq:ym-four-point-negative-positive-helicity}
\end{align}

To achieve eq.~\eqref{eq:ym-four-point-positive-negative-helicity} from the previous step, we simplify the parenthetical expression by substituting the deformation parameter in eq.~\eqref{eq:deformation_param}. (See ref.~\cite{Arkani-Hamed:2017jhn} for an alternative indirect derivation based on parameterizing the form of the final result and fixing free coefficients.) Focusing strictly on its numerator, $\mathcal{N}$, we have:
\begin{equation}
    \mathcal{N} = m[\boldsymbol{41}] + [\boldsymbol{4}2]\langle 2\boldsymbol{1}\rangle - \frac{\langle 2|\boldsymbol{1}|2]}{\langle 3|\boldsymbol{1}|2]} [\boldsymbol{4}2]\langle 3\boldsymbol{1}\rangle = m[\boldsymbol{41}] + \frac{[\boldsymbol{4}2]}{\langle 3|\boldsymbol{1}|2]} \Big( \langle 2\boldsymbol{1}\rangle \langle 3|\boldsymbol{1}|2] - \langle 3\boldsymbol{1}\rangle \langle 2|\boldsymbol{1}|2] \Big) \ .
\end{equation}
By Schouten identities, the term in the bracket becomes
\begin{equation}
    \langle 23\rangle \langle \boldsymbol{1} | \boldsymbol 1 | 2] = -m [\boldsymbol{1}2] \langle 23\rangle \ ,
\end{equation}
where we applied the massive Dirac equation $\langle \boldsymbol{1}| \boldsymbol 1 = -m[\boldsymbol{1}|$. Substituting this back into the numerator, we obtain:
\begin{equation}
    \mathcal{N} = m[\boldsymbol{41}] - m \frac{[\boldsymbol{4}2][\boldsymbol{1}2]\langle 23\rangle}{\langle 3|\boldsymbol{1}|2]} = \frac{m}{\langle 3|\boldsymbol{1}|2]} \Big( [\boldsymbol{41}]\langle 3|\boldsymbol{1}|2] - [\boldsymbol{4}2][\boldsymbol{1}2]\langle 23\rangle \Big) \ .
\end{equation}
To apply momentum conservation, we must align the square brackets. We apply the Schouten identity a second time, now on $|\boldsymbol{1}], |2], |\boldsymbol{4}]$:
\begin{equation}
  [41] \langle 3|\boldsymbol{1}|2] = 
  - [2\boldsymbol{4}] \langle 3| \boldsymbol 1 |\boldsymbol{1}] - [\boldsymbol{1}2] \langle 3|\boldsymbol 1 | \boldsymbol{4}] \ .
\end{equation}
The numerator simplifies to
\begin{equation}
    \mathcal{N} = \frac{m}{\langle 3|\boldsymbol{1}|2]} \Big( - [2\boldsymbol{4}] \langle 3| \boldsymbol 1 |\boldsymbol{1}] - [\boldsymbol{1}2] \langle 3| (\boldsymbol 1 + \boldsymbol 2)|\boldsymbol{4}] \Big) \ .
\end{equation}
The first term evaluates to $-m [2\boldsymbol{4}] \langle 3\boldsymbol{1}\rangle$ via the Dirac equation $p_1|\boldsymbol{1}] = m|\boldsymbol{1}\rangle$. For the second term, momentum conservation $p_1 + p_2 = -p_3 - p_4$ and the fact that particle 3 is massless ($\langle 3|p_3 = 0$) leaves $\langle 3|(-p_4)|\boldsymbol{4}] = -m \langle 3\boldsymbol{4}\rangle$. Thus, the numerator fully collapses to:
\begin{equation}
    \mathcal{N} = \frac{-m^2}{\langle 3|\boldsymbol{1}|2]} \Big( [2\boldsymbol{4}]\langle 3\boldsymbol{1}\rangle - [\boldsymbol{1}2]\langle 3\boldsymbol{4}\rangle \Big) = \frac{-m^2}{\langle 3|\boldsymbol{1}|2]} \Big( \langle \boldsymbol{1}3\rangle[\boldsymbol{4}2] + \langle \boldsymbol{4}3\rangle[\boldsymbol{1}2] \Big) \ .
\end{equation}
This allows us to finally arrive at eq.~\eqref{eq:ym-four-point-positive-negative-helicity}.

To calculate $A[\boldsymbol{1}^s,2^+,3^+,\boldsymbol{4}^s]$, we again do  a $\langle 2,3]$ shift. Hence, similar to the previous case, only $1\hat{2}$-channel contributes.
\begin{align}
    &A[\boldsymbol{1}^s,2^+,3^+,\boldsymbol{4}^s]=\nonumber A[\boldsymbol{1}^s,\hat{2}^+,\hat{\boldsymbol{P}}^s]\frac{i}{\tau_{12}}A[-\hat{\boldsymbol{P}}^s,\hat{3}^+,\boldsymbol{4}^s]\\
    =&\nonumber \left( i \, \frac{\langle 3|\boldsymbol{1}|2]}{\langle32\rangle} \frac{\langle \boldsymbol{1} \hat{\boldsymbol{P}}^I \rangle^{2s}}{m^{2s}}\right) \frac{i}{\tau_{12}} \left( i \, \frac{\langle\hat{2}|-\hat{\boldsymbol{P}}|\hat{3}]}{\langle23\rangle} \frac{\langle  - \hat{\boldsymbol{P}}_I \, \boldsymbol{4}\rangle^{2s}}{m^{2s}} \right)\\
    =&\nonumber i \, \frac{\langle 3|\boldsymbol{1}|2](\langle 2|\boldsymbol{1}|{3}]+z\langle 2|\boldsymbol{1}|{2}] -z\langle 3|\boldsymbol{1}|{3}] -z^2 \langle 3|\boldsymbol{1}|{2}])}{\tau_{12}\langle23\rangle^2} \frac{\langle\boldsymbol{14}\rangle^{2s}}{m^{2s}}\\
    \implies\Aboxed{&A[\boldsymbol{1}^s,2^+,3^+,\boldsymbol{4}^s]
= - i \, \frac{m^{2-2s} [23]^2}
        {\tau_{12}\,s_{23}}
   \langle \boldsymbol{14} \rangle^{2s}}\ . \label{eq:ym-four-point-all-plus-helicity}
\end{align}

Similarly one can calculate
\begin{align}
\Aboxed{A[\boldsymbol{1}^s,2^-,3^-,\boldsymbol{4}^s]
&= -i \, \frac{m^{2-2s} \langle 23 \rangle^2}
        {\tau_{12}\,s_{23}}
   [\boldsymbol{14}]^{2s}} \ .
\label{eq:ym-four-point-all-minus-helicity}
\end{align}

\vspace{6pt}

%----------------------------------------------------------------------------

\subsection{Five-Point Compton Amplitudes}\label{subsec:ym-five-point-compton}

We study five-point Yang--Mills Compton amplitudes involving two massive spin-$s$ particles and three gluons. The possible helicity configurations of the three gluons are
\begin{enumerate}
    \item All positive-helicity gluons, $(+,+,+)$.
    \item All negative-helicity gluons, $(-,-,-)$, related to the previous configuration by charge conjugation.
    \item Two positive-helicity gluons and one negative-helicity gluon, $(+,+,-)$.
    \item Two negative-helicity gluons and one positive-helicity gluon, $(-,-,+)$, related to the previous configuration by charge conjugation.
\end{enumerate}

We will calculate the amplitudes for the $(+, +, -)$ configuration first and then the $(+, +, +)$ configuration.

For the $(+,+,-)$ helicity sector, without loss of generality, we will assign positive helicity to gluon legs 2 and 4 and negative helicity to gluon leg 3. Once we have analytically calculated amplitudes with external legs $(\boldsymbol 1^s, 2^+, 3^-, 4^+, \boldsymbol 5^s)$, other amplitudes in the $(+,+,-)$ helicity sector can be obtained by exchanging external leg labels: for example, $2 \leftrightarrow 3$ exchange gives amplitudes with external legs $(\boldsymbol 1^s, 3^+, 2^-, 4^+, \boldsymbol 5^s)$.

Note that we have been specifying only the external states rather than their color ordering so far, and the full amplitude for $(\boldsymbol 1^s, 2^+, 3^-, 4^+, \boldsymbol 5^s)$ decomposes into a set of color-ordered amplitudes. As a consequence Kleiss-Kuijf (KK) relations \cite{Kleiss:1988ne} and Bern-Carrasco-Johansson (BCJ) relations \cite{Bern:2008qj, Bjerrum-Bohr:2009ulz, Feng:2010my, delaCruz:2015dpa}, all these color-ordered amplitudes can be written as linear combinations of a BCJ basis of only $(5-3)! = 2$ independent amplitudes. The BCJ relations allow us to fix the positions of three legs, $\boldsymbol 1^s$, $3^-$, and $\boldsymbol n^s$ in the color ordering. Therefore, the basis of two independent amplitudes is chosen to be
$A[\boldsymbol{1}^s,2^+,3^-,4^+,\boldsymbol{5}^s]$ and $A[\boldsymbol{1}^s,4^+,3^-,2^+,\boldsymbol{5}^s]$.
Both of them are for the helicity ordering $[\boldsymbol{s},+,-,+,\boldsymbol{s}]$ if we suppress external leg labels, so it is sufficient to calculate the former,
\begin{equation}
    \boxed{A[\boldsymbol{1}^s,2^+,3^-,4^+,\boldsymbol{5}^s]} \ ,
\end{equation}
which will be carried out in Section \ref{subsubsec:ym-five-point-12345-single-minus}, and obtain $A[\boldsymbol{1}^s,2^+,3^-,4^+,\boldsymbol{5}^s]$ by $2 \leftrightarrow 4$ exchange. Although other color-ordered amplitudes can be written as linear combinations of these two, we find it advantageous to also calculate
\begin{equation}
    \boxed{A[\boldsymbol{1}^s,3^-,2^+,4^+,\boldsymbol{5}^s]}
\end{equation}
using BCFW recursion directly, which will be carried out in Section \ref{subsubsec:ym-five-point-13245-single-minus}.

For the $(+,+,+)$ helicity sector, KK and BCJ relations imply that all color-ordered amplitudes are linear combinations of a basis of the schematic form $[\boldsymbol{s},+,+,+,\boldsymbol{s}]$, suppressing external leg labels. We will explicitly calculate
\begin{equation}
    \boxed{A[\boldsymbol{1}^s,2^+,3^+,4^+,\boldsymbol{5}^s]}
\end{equation}
to be carried out in Section \ref{subsubsec:ym-five-point-12345-all-plus}. Then other color-ordered amplitudes in the $(+,+,+)$ helicity sector can be obtained by exchanging particle labels in the analytic expression.

Finally, amplitudes in the $(---)$ and $(+--)$ helicity sectors are obtained through charge conjugation, implemented as the following transformation on spinor brackets,
\begin{equation*}
\langle a b \rangle \;\longleftrightarrow\; [b a] \ .
\end{equation*}

The amplitudes under color orderings not directly calculated in this section can be obtained from BCJ relations, whose explicit forms are given in app.~\ref{app:bcj-relations}. 
% --------------------------------------------------

\subsubsection{Single-minus sector: Case I, 
\texorpdfstring{$A[\boldsymbol{1}^s,2^+,3^-,4^+,\boldsymbol{5}^s]$}
               {}
}
\label{subsubsec:ym-five-point-12345-single-minus}

Here, we calculate the $(\boldsymbol{s},+,-,+,\boldsymbol{s})$ helicity configuration using the massless--massless BCFW shift $[3,2\rangle$, which is enough for generating all color orderings with the help of KK and BCJ relations, as discussed above. Later in the paper, in sec.~\ref{subsubsec:ym-five-point-13245-bcfw-check}--\ref{subsubsec:ym-five-point-31425-bcfw-check}, other orderings are calculated directly from the same BCFW shift as a cross-check. The $[3, 2 \rangle$ BCFW shift is
\begin{equation}\label{eq:ym-five-point-12345-bcfw-shift}
\begin{aligned}
|\hat{2}\rangle &= |2\rangle - z\,|3\rangle \ ,
&\qquad
|\hat{2}] &= |2] \ , \\
|\hat{3}] &= |3] + z\,|2] \ ,
&\qquad
|\hat{3}\rangle &= |3\rangle \ .
\end{aligned}
\end{equation}

Under the $[3,2\rangle$ shift in eq.~\eqref{eq:ym-five-point-12345-bcfw-shift}, $A[\boldsymbol{1}^s,2^+,3^-,4^+,\boldsymbol{5}^s]$ decomposes into two different channels.
\begin{equation}\label{eq:ym-five-point-12345-bcfw-decomposition}
\boxed{
\begin{aligned}
A[\boldsymbol{1}^s,2^+,3^-,4^+,\boldsymbol{5}^s]=A[\boldsymbol{1}^s,\hat{2}^+|\hat{3}^-,4^+,\boldsymbol{5}^s]+A[\boldsymbol{5}^s,\boldsymbol{1}^s,\hat{2}^+|\hat{3}^-,4^+]
\end{aligned} 
}\ .
\end{equation}
%---------------------------------------------------------------------
\paragraph{Channel 1: $A[\boldsymbol{1}^s,\hat{2}^+ \,|\, \hat{3}^-,4^+,\boldsymbol{5}^s]$}

We calculate the contribution from $A[\boldsymbol{1}^s,\hat{2}^+ \,|\, \hat{3}^-,4^+,\boldsymbol{5}^s]$. In fig.~\ref{fig:ym-five-point-12345-bcfw-channel-12},
\begin{flalign*}
    \tau_{1\hat{2}}=0 \implies z=\frac{\tau_{12}}{\langle3|\boldsymbol{1}|2]} \ .
\end{flalign*}
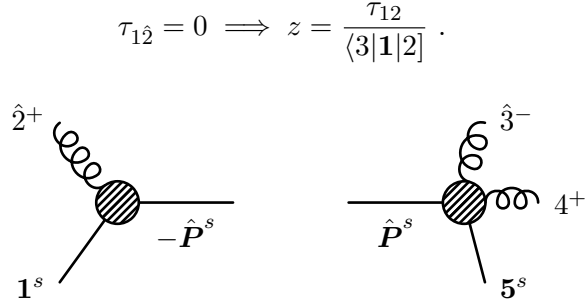
\begin{figure}[H]
    \centering
    \begin{fmffile}{3_1_d01}
        \begin{fmfgraph*}(200,60)
            \fmfleft{i1,i2}
            \fmfright{o1,o2,o3}
            \fmfblob{.08w}{g1,g2}
            \fmf{plain}{i1,g1}
            \fmfv{label=$\boldsymbol{1}^s$}{i1}
            \fmf{gluon}{i2,g1}
            \fmfv{label=$\hat{2}^+$}{i2}
            \fmf{plain,label=$-\hat{\boldsymbol{P}}^s$}{g1,v1}
            \fmf{phantom}{v1,v2}
            \fmf{plain, label=$\hat{\boldsymbol{P}}^s$}{v2,g2}
            \fmf{gluon}{o3,g2}
            \fmfv{label=$\hat{3}^-$}{o3}
            \fmf{gluon}{o2,g2}
            \fmfv{label=$4^+$}{o2}
            \fmf{plain}{g2,o1}
            \fmfv{label=$\boldsymbol{5}^s$}{o1}
        \end{fmfgraph*}
    \end{fmffile}
    \vspace{5mm}
    \caption{Factorization channel contributing to
    $A[\boldsymbol{1}^s,\hat{2}^+ \,|\, \hat{3}^-,4^+,\boldsymbol{5}^s]$.}
    \label{fig:ym-five-point-12345-bcfw-channel-12}
\end{figure}
{\allowdisplaybreaks
\begin{align}
    &A[\boldsymbol{1}^s,\hat{2}^+|\hat{3}^-,4^+,\boldsymbol{5}^s]
    =A[\boldsymbol{1}^s,\hat{2}^+,\hat{\boldsymbol{P}}^s]\frac{i}{\tau_{12}}A[-\hat{\boldsymbol{P}}^s,\hat{3}^-,4^+,\boldsymbol{5}^s]\nonumber\\
    \nonumber=& \left(  i\, \frac{\langle3|\boldsymbol{1}|2]}{\langle32\rangle}\frac{\langle \boldsymbol{1\,\hat{P}}^I \rangle^{2s}}{m^{2s}} \right) \frac{i}{\tau_{12}} \left( i \, \frac{\langle3|\boldsymbol{5}|4]^2}{s_{\hat{3}4}\tau_{54}}\left(\frac{\langle-\hat{\boldsymbol{P}}_I 3\rangle[\boldsymbol{5}4]+\langle\boldsymbol{5}3\rangle[-\boldsymbol{\hat{P}}_I 4]}{\langle3|-\boldsymbol{\hat{P}}|4]}\right)^{2s}\right)\\
    =&\nonumber -i \, \frac{\langle 3 |\boldsymbol{1}|2]^2 \langle 3 |\boldsymbol{5}|4]^2 }{\tau _{12} \tau _{54} \langle 2  3\rangle \langle 3 4\rangle [2|\boldsymbol{1}(2+3)|4]}\left(\frac{-\langle\boldsymbol{1}\hat{\boldsymbol{P}}^I\rangle\langle\boldsymbol{\hat{P}}_I 3\rangle[\boldsymbol{5}4]+\langle\boldsymbol{5}3\rangle\langle\boldsymbol{1}\hat{\boldsymbol{P}}^I\rangle[\boldsymbol{\hat{P}}_I 4]}{\langle3|-\boldsymbol{5}|4]}\right)^{2 s}\\
    =&\nonumber -i \, \frac{\langle 3 |\boldsymbol{1}|2]^2 \langle 3 |\boldsymbol{5}|4]^2 }{\tau _{12} \tau _{54} \langle 2  3\rangle \langle 3 4\rangle [2|\boldsymbol{1}(2+3)|4]}\left(\frac{m\langle\boldsymbol{1} 3\rangle[\boldsymbol{5}4]+\langle\boldsymbol{5}3\rangle\langle\boldsymbol{1}|\boldsymbol{\hat{P}}| 4]}{\langle3|-\boldsymbol{5}|4]}\right)^{2 s}\\
    %=\nonumber&\frac{\langle 3 |\boldsymbol{1}|2]^2 \langle 3 |\boldsymbol{5}|4]^2 }{\tau _{12} \tau _{54} \langle 2  3\rangle  \left(\tau _{12} \langle 3 |4|2]+s_{34} \langle 3 |\boldsymbol{1}|2]\right)}Z_{12}^{2 s}\\
    =& -i \, \frac{\langle 3 |\boldsymbol{1}|2]^2 \langle 3 |\boldsymbol{5}|4]^2 }{\tau _{12} \tau _{54} \langle 2  3\rangle \langle 3 4\rangle [2|\boldsymbol{1}(2+3)|4]}\left(\frac{(\langle \boldsymbol{1} 3\rangle  [\boldsymbol{5} 4]+[\boldsymbol{1} 4] \langle \boldsymbol{5} 3\rangle ) \langle 3|\boldsymbol{1}|2]+\langle 3|2|4] [\boldsymbol{1} 2] \langle \boldsymbol{5} 3\rangle }{\langle 3|-\boldsymbol{5}|4] \langle 3|\boldsymbol{1}|2]}\right)^{2 s} \ .
    \label{eq:ym-five-point-12345-bcfw-channel-12}
\end{align}}
To get to the last step, we use the momentum conservation for the left sub-amplitude and then use the Schouten identity\footnote{Here, we will present the calculation for $\langle\boldsymbol{1}|\boldsymbol{\hat{P}}| 4]$, which is useful for the derivation of eq.~\eqref{eq:ym-five-point-12345-bcfw-channel-12}.
\begin{align*}
    & \langle\boldsymbol{1}|\boldsymbol{\hat{P}}| 4]=-\langle\boldsymbol{1}|(\boldsymbol{1}+\hat{2})| 4]=m[\boldsymbol{1}4] -\langle\boldsymbol{1}|{2}| 4]+\frac{\langle2|\boldsymbol{1}|2]}{\langle3|\boldsymbol{1}|2]}\langle\boldsymbol{1}3\rangle[24]=m[\boldsymbol{1}4]+\frac{-\langle\boldsymbol{1}2\rangle\langle3|\boldsymbol{1}|2]+\langle\boldsymbol{1}3\rangle\langle2|\boldsymbol{1}|2]}{\langle3|\boldsymbol{1}|2]}[24]\\
=&m[\boldsymbol{1}4]+\frac{\langle3\boldsymbol{1}\rangle\langle2|\boldsymbol{1}|2]+\langle23\rangle\langle\boldsymbol{1}|\boldsymbol{1}|2]+\langle\boldsymbol{1}3\rangle\langle2|\boldsymbol{1}|2]}{\langle3|\boldsymbol{1}|2]}[24]=m[\boldsymbol{1}4]+\frac{-m\langle23\rangle[\boldsymbol{1}2]}{\langle3|\boldsymbol{1}|2]}[24]\\
=&m\frac{[\boldsymbol{1}4]\langle3|\boldsymbol{1}|2]+[\boldsymbol{1}2]\langle3|2|4]}{\langle3|\boldsymbol{1}|2]} \ .
\end{align*}
\textbf{-----------------------------------------------------}}.
%--------------------------------------------------------------------- 

\paragraph{Channel 2: $A[\boldsymbol{5}^s,\boldsymbol{1}^s,\hat{2}^+|\hat{3}^-,4^+]$}

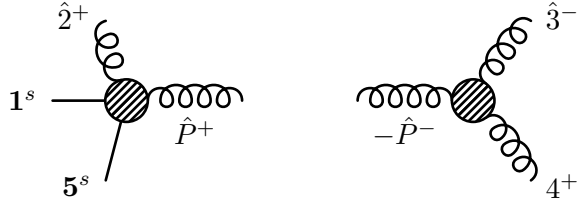
\begin{figure}[H]
    \centering
\begin{fmffile}{3_1_d02}
\begin{fmfgraph*}(200,60)

   % reversed ordering (top ↔ bottom)
   \fmfleft{i2,i1,o1}
   \fmfright{o3,o2}

   \fmfblob{.08w}{g1,g2}

   \fmf{plain}{i1,g1}
   \fmfv{label=$\boldsymbol{1}^s$}{i1}

   \fmf{gluon}{o1,g1}
   \fmfv{label=$\hat{2}^+$}{o1}

   \fmf{plain}{i2,g1}
   \fmfv{label=$\boldsymbol{5}^s$}{i2}

   % flip label.side
   \fmf{gluon, label=$\hat{P}^+$, label.side=left}{v1,g1}
   \fmf{phantom}{v1,v2}
   \fmf{gluon, label=$-\hat{P}^-$, label.side=left}{g2,v2}

   \fmf{gluon}{o2,g2}
   \fmfv{label=$\hat{3}^-$}{o2}

   \fmf{gluon}{o3,g2}
   \fmfv{label=$4^+$}{o3}

\end{fmfgraph*}
\end{fmffile}
\vspace{5mm}
\caption{Factorization channel contributing to $A[\boldsymbol{5}^s,\boldsymbol{1}^s,\hat{2}^+|\hat{3}^-,4^+]$ }
\label{fig:ym-five-point-12345-bcfw-channel-34}
\end{figure}

Now, we calculate the contribution from $A[\boldsymbol{5}^s,\boldsymbol{1}^s,\hat{2}^+|\hat{3}^-,4^+]$. In fig.~\ref{fig:ym-five-point-12345-bcfw-channel-34}, 
\begin{align*}
    s_{\hat{3}4}=0 \implies z=-\frac{[43]}{[42]}\ ,\quad \text{when }\langle34\rangle \neq 0 \ .
\end{align*}Therefore, we will only consider the kinematics that allows for $\langle34\rangle \neq 0$. Hence, the sub-amplitude on the right is MHV.
\vspace{5mm}

{\allowdisplaybreaks
\begin{align}
\nonumber&A[\boldsymbol{5}^s,\boldsymbol{1}^s,\hat{2}^+|\hat{3}^-,4^+] = A[\boldsymbol{5}^s,\boldsymbol{1}^s,\hat{2}^+,\hat{P}^+]\frac{i}{s_{34}}A[-\hat{P}^-,\hat{3}^-,4^+]\\
=\nonumber& -i \, \frac{m^2 [\hat{2}\hat{P}]^2}{\tau_{1\hat{2}} s_{15}} \frac{i}{s_{34}} \frac{i\langle-\hat{P}\,\hat{3}\rangle^4}{\langle-\hat{P}\,\hat{3}\rangle\langle\hat{3}\,4\rangle\langle4\,-\hat{P}\rangle}\left(\frac{\langle \boldsymbol{15}\rangle }{m}\right)^{2 s}\\
=& i \,  \frac{m^2  [24]^4}{s_{15} [2  3][34]   [2|\boldsymbol{1}(2+3)|4]}\left(\frac{\langle \boldsymbol{15}\rangle }{m}\right)^{2 s} \ .
\label{eq:ym-five-point-12345-bcfw-channel-34}
\end{align}}

According to eq.~\eqref{eq:ym-five-point-12345-bcfw-decomposition}, adding Eq.~\eqref{eq:ym-five-point-12345-bcfw-channel-12} from fig.~\ref{fig:ym-five-point-12345-bcfw-channel-12} and Eq.~\eqref{eq:ym-five-point-12345-bcfw-channel-34} from fig.~\ref{fig:ym-five-point-12345-bcfw-channel-34}, we get the full amplitude:
\begin{equation}
   \boxed{A[\boldsymbol{1}^s,2^+,3^-,4^+,\boldsymbol{5}^s]=A_{12}[\boldsymbol{1}^0,2^+,3^-,4^+,\boldsymbol{5}^0]\: Z_{12}^{2s} + A_{34}[\boldsymbol{1}^0,2^+,3^-,4^+,\boldsymbol{5}^0]\: Z_{34}^{2s}} \ ,
    \label{eq:ym-five-point-12345-bcfw-amplitude}
\end{equation}
\begin{flalign*}
\text{where}\quad
&\left\{\quad
\begin{aligned}
A_{12}[\boldsymbol{1}^0,2^+,3^-,4^+,\boldsymbol{5}^0]
&= -i\,\frac{\langle 3 |\boldsymbol{1}|2]^2 \langle 3 |\boldsymbol{5}|4]^2 }{\tau _{12} \tau _{54} \langle 2  3\rangle \langle 3 4\rangle [2|\boldsymbol{1}(2+3)|4]}\ , \\[6pt]
A_{34}[\boldsymbol{1}^0,2^+,3^-,4^+,\boldsymbol{5}^0]
& =i\, \frac{m^2  [24]^4}{s_{15} [2  3][34]   [2|\boldsymbol{1}(2+3)|4]}\ ;
\end{aligned}
\right.&&
\end{flalign*}
and $Z_{12}$ and $Z_{34}$ are given by eq.~\eqref{eq:bcfw-channel-spin-structures}.

We also observe both $A_{12}[\boldsymbol{1}^0,2^+,3^-,4^+,\boldsymbol{5}^0]$ and $A_{34}[\boldsymbol{1}^0,2^+,3^-,4^+,\boldsymbol{5}^0]$ have a spurious singularity when $[2|\boldsymbol{1}(2+3)|4] \to 0$.
To analyze the nature of the spurious singularity, we use analytic parametrization of spinors in app.~\ref{app:analytic-parametrization}. We verify that when $[2|\boldsymbol{1}(2+3)|4] \to 0$, the numerators of $A_{12}[\boldsymbol{1}^0,2^+,3^-,4^+,\boldsymbol{5}^0]$ and $A_{34}[\boldsymbol{1}^0,2^+,3^-,4^+,\boldsymbol{5}^0]$ do not go to zero. Hence, we can conclude that there are nontrivial cancellations of the spurious pole, $[2|\boldsymbol{1}(2+3)|4]$, between the two terms. In the spin-0 case, the spin structures trivially become unity, and it is easy to show that the sum of the two terms can be written in a form free of spurious poles,
\begin{equation}
    A[\boldsymbol{1}^0,2^+,3^-,4^+,\boldsymbol{5}^0]=-\,i \, \left(\frac{\langle3|\boldsymbol{1}|2]\langle3|\boldsymbol{5}|4]\langle 3|\boldsymbol{15}|3\rangle}{\tau_{12}\langle23\rangle\langle34\rangle\tau_{45}s_{15}} + \frac{m^2[24]^2([2|\boldsymbol{1}3|4]-[24]\tau_{45})}{\tau_{12}[23][34]\tau_{45}s_{15}}\right) \ ,
    \label{eq:YM-Compton_pmp_spin0}
\end{equation}
as given in ref.~\cite{Bern:2019crd}.
Having established the cancellation of spurious poles in the spin-0 case, one might still wonder if the cancellation persists in the presence of spin structures in Eq.~\eqref{eq:ym-five-point-12345-bcfw-amplitude}. This is indeed true for any spin $s\leq 1$, due to the fact that
\begin{equation}
    \lim_{[2|\boldsymbol{1}(2+3)|4] \to 0} Z_{12} = \lim_{[2|\boldsymbol{1}(2+3)|4] \to 0}Z_{34}\ ,\label{eq:ym-five-point-12345-spin-structure-coincidence}
\end{equation}
This guarantees that the two spin structures are identical at the location of the spurious pole, and therefore the spurious pole is eliminated for any spin as long as it is eliminated for the spin-0 case. Still, it is necessary to restrict to $s\leq 1$ to prevent the emergence of new spurious poles in eq.~\eqref{eq:ym-five-point-12345-bcfw-channel-12}, the factor $\langle 3|\boldsymbol{1}|2] \langle 3|\boldsymbol{5}|4]$ in the denominator of $Z_{12}$ is canceled by the same expression (squared) in the numerator of $A_{12}[\boldsymbol{1}^0,2^+,3^-,4^+,\boldsymbol{5}^0]$ when $s\leq 1$.
We prove Eq.~\eqref{eq:ym-five-point-12345-spin-structure-coincidence} through a slightly lengthy algebraic manipulation:
\begin{align}
    Z_{12}-Z_{34}
    &=
    \frac{
    m\Big[
      \big(\langle\boldsymbol{1}3\rangle[\boldsymbol{5}4]
      +[\boldsymbol{1}4]\langle\boldsymbol{5}3\rangle\big)\langle 3|\boldsymbol{1}|2]
      +\langle 3|2|4][\boldsymbol{1}2]\langle\boldsymbol{5}3\rangle
    \Big]
    -\langle\boldsymbol{15}\rangle \langle 3|\boldsymbol{1}|2] \langle 3|-\boldsymbol{5}|4]
    }
    {m\,\langle 3|\boldsymbol{1}|2]\,\langle 3|-\boldsymbol{5}|4]} \ .
\end{align}
To simplify the numerator, we use the Schouten identity on the angle spinors $\boldsymbol{1},\boldsymbol{5},3$, contracted with $-\boldsymbol{5}|4]$:
\begin{equation}
    \langle\boldsymbol{15}\rangle\langle 3|-\boldsymbol{5}|4]
    =
    m\,\langle\boldsymbol{1}3\rangle[\boldsymbol{5}4]
    -\langle\boldsymbol{5}3\rangle\langle\boldsymbol{1}|-\boldsymbol{5}|4]\ ,
\end{equation}
where we used $\langle\boldsymbol{5}|p_5=-m[\boldsymbol{5}|$ to evaluate $\langle\boldsymbol{5}|-\boldsymbol{5}|4]=m[\boldsymbol{5}4]$.
Substituting this into the last term of the numerator gives
\begin{align}
    Z_{12}-Z_{34}
    &=
    \frac{
    \langle\boldsymbol{5}3\rangle
    \Big[
       \langle 3|\boldsymbol{1}|2]\big(\langle\boldsymbol{1}|-\boldsymbol{5}|4]
          +m[\boldsymbol{1}4]\big)
       +m\langle 3|2|4][\boldsymbol{1}2]
    \Big]
    }
    {m\,\langle 3|\boldsymbol{1}|2]\,\langle 3|-\boldsymbol{5}|4]} \ .
\end{align}
Next, we use the massive Dirac equation $\langle\boldsymbol{1}| \boldsymbol 1 =-m[\boldsymbol{1}|$. Let
\begin{equation}
    q\equiv k_2+k_3+k_4=-(p_1+p_5)
\end{equation}
be the total outgoing gluon momentum. Then
\begin{equation}
    \langle\boldsymbol{1}|-\boldsymbol{5}|4]
    +m[\boldsymbol{1}4]
    =
    \langle\boldsymbol{1}|-(p_5+p_1)|4]
    =
    \langle\boldsymbol{1}|q|4]\ ,
\end{equation}
so that
\begin{align}
    Z_{12}-Z_{34}
    &=
    \frac{
    \langle\boldsymbol{5}3\rangle
    \Big[
       \langle 3|\boldsymbol{1}|2]\,\langle\boldsymbol{1}|q|4]
       +m\langle 3|2|4][\boldsymbol{1}2]
    \Big]
    }
    {m\,\langle 3|\boldsymbol{1}|2]\,\langle 3|-\boldsymbol{5}|4]} \ .
\end{align}
To simplify the square bracket in the numerator, we apply a Schouten identity on the square spinors $[2|,\,[\boldsymbol{1}|,\,[3|$, contracted with $q|4]$:
\begin{equation}
    \langle 3|\boldsymbol{1}|2]\,\langle\boldsymbol{1}|q|4]
    +m[\boldsymbol{1}2]\,\langle 3|q|4] =
    \langle\boldsymbol{1}3\rangle[2|\boldsymbol{1}q|4] \ ,
\end{equation}
where $\langle 3|q|4]=\langle 3|2|4]$ follows from $q=k_2+k_3+k_4$. Therefore
\begin{equation}
    \boxed{
    Z_{12}
    =
    Z_{34}
    +
    \frac{[2|\boldsymbol{1}q|4]\,
          \langle\boldsymbol{1}3\rangle
          \langle\boldsymbol{5}3\rangle}
         {m\,\langle3|\boldsymbol{1}|2]\,
          \langle3|-\boldsymbol{5}|4]}
    }\ .
    \label{eq:ym-five-point-12345-spin-structure-relation}
\end{equation}
Since $[2|\boldsymbol{1}q|4]=[2|\boldsymbol{1}(2+3)|4]$, we have proved eq.~\eqref{eq:ym-five-point-12345-spin-structure-coincidence}.

Now, we are ready to write down an explicit expression for the amplitude that is free of spurious poles.
From eq.~\eqref{eq:ym-five-point-12345-bcfw-amplitude} and eq.~\eqref{eq:ym-five-point-12345-spin-structure-relation}, we get
{\allowdisplaybreaks
\begin{align}\label{eq:ym-five-point-12345-spurious-pole-free-amplitude}
    \nonumber& A[\boldsymbol{1}^s,2^+,3^-,4^+,\boldsymbol{5}^s]\\
    \nonumber&\\
    =&\nonumber A_{12}[\boldsymbol{1}^0,2^+,3^-,4^+,\boldsymbol{5}^0] \,\left(Z_{34}\,+\,\frac{[2|\boldsymbol{1}q|4]\langle\boldsymbol{1}3\rangle\langle\boldsymbol{5}3\rangle}{m \langle3|\boldsymbol{1}|2]\langle3|-\boldsymbol{5}|4]}\right)^{2s}+A_{34}[\boldsymbol{1}^0,2^+,3^-,4^+,\boldsymbol{5}^0]\,Z_{34}^{2s}\\
    &\nonumber\\
    =&\nonumber A[\boldsymbol{1}^0,2^+,3^-,4^+,\boldsymbol{5}^0]\,Z_{34}^{2s} + \sum_{n=1}^{2s} {}^{2s}C_{n} \:A_{12}[\boldsymbol{1}^0,2^+,3^-,4^+,\boldsymbol{5}^0]\,Z_{34}^{2s-n}\,\left(\frac{[2|\boldsymbol{1}q|4]\langle\boldsymbol{1}3\rangle\langle\boldsymbol{5}3\rangle}{m \langle3|\boldsymbol{1}|2]\langle3|-\boldsymbol{5}|4]}\right)^n\\
    &\nonumber\\
    \implies&\boxed{\begin{aligned}
        & A[\boldsymbol{1}^s,2^+,3^-,4^+,\boldsymbol{5}^s]\\
        =& A[\boldsymbol{1}^0,2^+,3^-,4^+,\boldsymbol{5}^0]\left(\frac{\langle\boldsymbol{15}\rangle}{m}\right)^{2s}\\
        &\qquad -\, i\,\sum_{n=1}^{2s} {}^{2s}C_{n}\frac{m^{-2s}\left(\langle 3 |\boldsymbol{1}|2] \langle 3 |-\boldsymbol{5}|4]\right)^{2-n} [2|\boldsymbol{1}q|4]^{n-1}}{\tau _{12} \tau _{54} \langle 2  3\rangle \langle 3 4\rangle } \langle\boldsymbol{15}\rangle^{2s-n}\,\left(\langle\boldsymbol{1}3\rangle\langle\boldsymbol{5}3\rangle\right)^{n}\ ,
    \end{aligned}}
\end{align}}
where $q=k_2+k_3+k_4$ is the total outgoing gluon momentum, and $A[\boldsymbol{1}^0,2^+,3^-,4^+,\boldsymbol{5}^0]$ is the amplitude with spinless massive legs given in eq.~\eqref{eq:YM-Compton_pmp_spin0}.
%%%%%%%%%%%%%%%%%%%%%%%%%%%%%%%%%%%%%%%%%%%%%%%%%%%%%%%%%%%%%%%%%%%%%%%%%%%%%%%%%%%%%%%%%

\subsubsection{Single-minus sector: Case II,
\texorpdfstring{$A[\boldsymbol{1}^s,3^-,2^+,4^+,\boldsymbol{5}^s]$}
               {}
}
\label{subsubsec:ym-five-point-13245-single-minus}

Now, we use an alternative BCFW shift: $[2,4\rangle$ shift
\begin{equation}\label{eq:ym-five-point-13245-bcfw-shift}
\begin{aligned}
|\hat{4}\rangle &= |4\rangle - z\,|2\rangle \ ,
&\quad
|\hat{4}] &= |4] \  , \\
|\hat{2}] &= |2] + z\,|4] \ ,
&\quad
|\hat{2}\rangle &= |2\rangle \ ,
\end{aligned}
\end{equation}
to obtain a clean result for $A[\boldsymbol{1}^s,3^-,2^+,4^+,\boldsymbol{5}^s]$ as a further demonstration of our methods, even though this amplitude can be obtained from BCJ relations from the $(\boldsymbol{s},+,-,+,\boldsymbol{s})$ case.

Under the $[2,4\rangle$ shift in eq.~\eqref{eq:ym-five-point-13245-bcfw-shift}, $A[\boldsymbol{1}^s,3^-,2^+,4^+,\boldsymbol{5}^s]$ decomposes into two different channels.
\begin{equation}\label{eq:ym-five-point-13245-bcfw-decomposition}
\boxed{A[\boldsymbol{1}^s,3^-,2^+,4^+,\boldsymbol{5}^s]=A[\boldsymbol{1}^s,3^-,\hat{2}^+|\hat{4}^+,\boldsymbol{5}^s]+A[\hat{4}^+,\boldsymbol{5}^s,\boldsymbol{1}^s|3^-,\hat{2}^+]} \ .
\end{equation}

\paragraph{Channel 1: $A[\boldsymbol{1}^s,3^-,\hat{2}^+|\hat{4}^+,\boldsymbol{5}^s]$}
In the BCFW channel shown in fig.~\ref{fig:ym-five-point-13245-bcfw-channel-54},
\begin{flalign*}
    \tau_{5\hat{4}}=0 \implies \langle4|\boldsymbol{5}|4] - z\, \langle2 | \boldsymbol{5}|4] =0 \implies z = \frac{\langle4|\boldsymbol{5}|4]}{\langle2 | \boldsymbol{5}|4]} \ .
\end{flalign*}
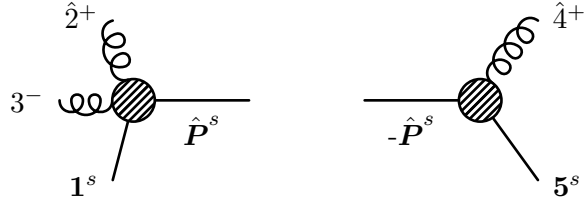
\begin{figure}[H]
    \centering
    \begin{fmffile}{3_2_d01}
 \begin{fmfgraph*}(200,60)% units are now in cm
   \fmfleft{i1,i2,i3}
   \fmfright{o1,o2}
   \fmfblob{.08w}{g1,g2}
   \fmf{plain}{i1,g1}
   \fmfv{label=$\boldsymbol{1}^s$}{i1}
   \fmf{gluon}{i2,g1}
   \fmfv{label=$3^-$}{i2}
   \fmf{gluon}{i3,g1}
   \fmfv{label=$\hat{2}^+$}{i3}
   \fmf{plain, label=$\hat{\boldsymbol{P}}^s$}{g1,v1}
   \fmf{phantom}{v1,v2}
   \fmf{plain, label=-$\hat{\boldsymbol{P}}^s$}{v2,g2}
   \fmf{gluon}{o2,g2}
   \fmfv{label=$\hat{4}^+$}{o2}
   \fmf{plain}{g2,o1}
   \fmfv{label=$\boldsymbol{5}^s$}{o1}
\end{fmfgraph*}
\end{fmffile}
\vspace{5mm}
    \caption{Factorization channel contributing to $A[\boldsymbol{1}^s,3^-,\hat{2}^+|\hat{4}^+,\boldsymbol{5}^s]$}
    \label{fig:ym-five-point-13245-bcfw-channel-54}
\end{figure}

{\allowdisplaybreaks
\begin{align}
    \nonumber & A[\boldsymbol{1}^s,3^-,\hat{2}^+|\hat{4}^+,\boldsymbol{5}^s] = A[\boldsymbol{1}^s,3^-,\hat{2}^+,\hat{\boldsymbol{P}}^s]\frac{i}{\tau_{54}}A[-\hat{\boldsymbol{P}}^s,\hat{4}^+,\boldsymbol{5}^s]\\
    =&\nonumber \left( i\,\frac{\langle 3|\boldsymbol{1}|\hat{2}]^2}{\tau_{13} s_{\hat{2}3}} \left(\frac{\langle \boldsymbol{1} 3 \rangle [\hat{\boldsymbol{P}}^I \hat{2}] + \langle \hat{\boldsymbol{P}}^I 3 \rangle [\boldsymbol{1} \hat{2}]}{\langle 3|\boldsymbol{1}|\hat{2}]}\right)^{2s} \right) \frac{i}{\tau_{54}} \left(i\, \frac{\langle 2| \hat{-\boldsymbol{P}}|\hat{4}]}{\langle2 \hat{4}\rangle} \left(\frac{\langle -\hat{\boldsymbol{P}}_I\, \boldsymbol{5}\rangle}{m}\right)^{2s} \right)\\
    =\nonumber& i \, \frac{\langle 3|-\boldsymbol{1}|\hat{2}]^2 \langle 2| \boldsymbol{5}|4]}{\tau_{13} \tau_{54} \langle 3 2 \rangle [\hat{2}3] \langle2 4\rangle} \left( \frac{-\langle \boldsymbol{1} 3 \rangle \langle\boldsymbol{5}|\hat{\boldsymbol{P}}|\hat{2}] + m \langle5 3\rangle [\boldsymbol{1}\hat{2}]}{m \langle 3|\boldsymbol{1}|\hat{2}]} \right)^{2s}\\
    =\nonumber& i \, \frac{\langle 3|-\boldsymbol{1}|\hat{2}]^2 \langle 2| \boldsymbol{5}|4]}{\tau_{13} \tau_{54} \langle 3 2 \rangle [\hat{2}3] \langle2 4\rangle} \left( \frac{\langle \boldsymbol{51} \rangle \langle3|\hat{\boldsymbol{P}}|\hat{2}] + \langle 3\boldsymbol{5} \rangle \langle1|\hat{\boldsymbol{P}}|\hat{2}] + m \langle5 3\rangle [\boldsymbol{1}\hat{2}]}{m \langle 3|\boldsymbol{1}|\hat{2}]} \right)^{2s}\\
    =\nonumber& i \, \frac{\langle 3|-\boldsymbol{1}|\hat{2}]^2 \langle 2| \boldsymbol{5}|4]}{\tau_{13} \tau_{54} \langle 3 2 \rangle [\hat{2}3] \langle2 4\rangle} \left( \frac{\langle \boldsymbol{51} \rangle \langle3|-\hat{\boldsymbol{1}}|\hat{2}] - \langle 3\boldsymbol{5} \rangle \langle1|(\boldsymbol{1}+3)|\hat{2}] + m \langle5 3\rangle [\boldsymbol{1}\hat{2}]}{m \langle 3|\boldsymbol{1}|\hat{2}]} \right)^{2s}\\
    =\nonumber& i \, \frac{\langle 3|-\boldsymbol{1}|\hat{2}]^2 \langle 2| \boldsymbol{5}|4]}{\tau_{13} \tau_{54} \langle 3 2 \rangle [\hat{2}3] \langle2 4\rangle} \left( \frac{\langle \boldsymbol{15} \rangle }{m}+\frac{ \langle \boldsymbol{5} 3\rangle \langle \boldsymbol{1} 3\rangle [3 \hat{2}]}{m \langle 3|\boldsymbol{1}|\hat{2}]} \right)^{2s}\\
    =& -i \, \frac{\langle 3|\boldsymbol{1}(2 +4)\boldsymbol{5}|4]^2 }{\tau_{13} \tau_{54} \langle 3 2 \rangle [3|q\boldsymbol{5}|4] \langle2 4\rangle} \left( \frac{\langle \boldsymbol{15} \rangle }{m}+\frac{ \langle \boldsymbol{5} 3\rangle \langle \boldsymbol{1} 3\rangle [3|q\boldsymbol{5}|4]}{m \langle 3|\boldsymbol{1}(2 +4)\boldsymbol{5}|4]} \right)^{2s} \ . \label{eq:ym-five-point-13245-bcfw-channel-54}
\end{align}
}

\begin{figure}[H]
    \centering
\begin{fmffile}{3_2_d02}
\begin{fmfgraph*}(200,60)

   % rotated + upside-down mirror
   \fmfleft{i2,i1,o1}
   \fmfright{o3,o2}

   \fmfblob{.08w}{g1,g2}

   \fmf{plain}{i1,g1}
   \fmfv{label=$\boldsymbol{5}^s$}{i1}

   \fmf{plain}{g1,o1}
   \fmfv{label=$\boldsymbol{1}^s$}{o1}

   \fmf{gluon}{i2,g1}
   \fmfv{label=$\hat{4}^+$}{i2}

   % label.side flipped
   \fmf{gluon, label=$\hat{P}^+$, label.side=left}{v1,g1}
   \fmf{phantom}{v1,v2}
   \fmf{gluon, label=-$\hat{P}^-$, label.side=left}{g2,v2}

   \fmf{gluon}{o2,g2}
   \fmfv{label=${3}^-$}{o2}

   \fmf{gluon}{o3,g2}
   \fmfv{label=$\hat{2}^+$}{o3}

\end{fmfgraph*}
\end{fmffile}
\vspace{5mm}
    \caption{Factorization channel contributing to $A[\hat{4}^+,\boldsymbol{5}^s,\boldsymbol{1}^s|3^-,\hat{2}^+]$}
    \label{fig:ym-five-point-13245-bcfw-channel-23}
\end{figure}
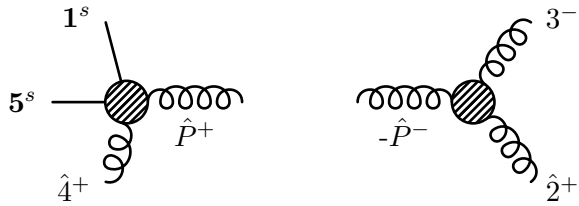

\paragraph{Channel 2: $A[\hat{4}^+,\boldsymbol{5}^s,\boldsymbol{1}^s|3^-,\hat{2}^+]$}
In the BCFW channel shown in fig.~\ref{fig:ym-five-point-13245-bcfw-channel-23},
\begin{flalign*}
    [\hat{2}3]=0 \implies [23] + z\,[43] =0 \implies z=-\frac{[23]}{[43]} \ .
\end{flalign*}

{\allowdisplaybreaks
\begin{align}
    \nonumber& A[\hat{4}^+,\boldsymbol{5}^s,\boldsymbol{1}^s|3^-,\hat{2}^+] = A[\hat{4}^+,\boldsymbol{5}^s,\boldsymbol{1}^s, \hat{P}^+] \frac{i}{s_{23}} A[-\hat{P}^-,3^-,\hat{2}^+]\\
    = \nonumber& \left(-i\,\frac{m^2 [\hat{4} \hat{P}]^2}{\tau_{5\hat{4}} s_{15}} \left(\frac{\langle
    \boldsymbol{15}\rangle}{m}\right)^{2s} \right) \frac{i}{s_{23}} \left(i\frac{\langle-\hat{P} 3\rangle^3}{\langle3 \hat{2}\rangle \langle\hat{2}\, -\hat{P}\rangle}\right)\\
    = \nonumber& i \, \frac{m^2 [4|\hat{P}|3\rangle^3}{\tau_{5\hat{4}} s_{15} s_{23} \langle32\rangle [4|\hat{P}|2\rangle} \left(\frac{\langle
    \boldsymbol{15}\rangle}{m}\right)^{2s}\\
    = \nonumber& i \, \frac{m^2 [4|2|3\rangle^3}{\tau_{5\hat{4}} s_{15} s_{23} \langle32\rangle [4|3|2\rangle} \left(\frac{\langle
    \boldsymbol{15}\rangle}{m}\right)^{2s}\\
    =& i \, \frac{m^2 [24]^3}{s_{15} [32] [3|q\boldsymbol{5}|4]} \left(\frac{\langle
    \boldsymbol{15}\rangle}{m}\right)^{2s} \ . \label{eq:ym-five-point-13245-bcfw-channel-23}
\end{align}}

According to eq.~\eqref{eq:ym-five-point-13245-bcfw-decomposition}, adding the contribution from both fig.~\ref{fig:ym-five-point-13245-bcfw-channel-54} and fig.~\ref{fig:ym-five-point-13245-bcfw-channel-23}, we get the full amplitude:
\begin{align}
    &\boxed{\begin{aligned}
        & A[\boldsymbol{1}^s,3^-,2^+,4^+,\boldsymbol{5}^s]\\
   = & -i \, \frac{\langle 3|\boldsymbol{1}(2 +4)\boldsymbol{5}|4]^2 }{\tau_{13} \tau_{54} \langle 3 2 \rangle [3|q\boldsymbol{5}|4] \langle2 4\rangle} \left( \frac{\langle \boldsymbol{15} \rangle }{m}+\frac{ \langle \boldsymbol{5} 3\rangle \langle \boldsymbol{1} 3\rangle [3|q\boldsymbol{5}|4]}{m \langle 3|\boldsymbol{1}(2 +4)\boldsymbol{5}|4]} \right)^{2s} +i \,\frac{m^2 [24]^3}{s_{15} [32] [3|q\boldsymbol{5}|4]} \left(\frac{\langle
    \boldsymbol{15}\rangle}{m}\right)^{2s}\\
    = & A[\boldsymbol{1}^0,3^-,2^+,4^+,\boldsymbol{5}^0] \left(\frac{\langle
    \boldsymbol{15}\rangle}{m}\right)^{2s} -i\, \sum_{k=1}^{2s} {}^{2s}C_k \frac{m^{-2s}\langle 3|\boldsymbol{1}(2 +4)\boldsymbol{5}|4]^{2-k} \langle\boldsymbol{15}\rangle^{2s-k} (\langle\boldsymbol{5} 3\rangle \langle \boldsymbol{1} 3\rangle)^k [3|q\boldsymbol{5}|4]^{k-1}}{\tau_{13} \tau_{54} \langle 3 2 \rangle \langle2 4\rangle} 
    \end{aligned}} \  ,
    &\label{eq:ym-five-point-13245-bcfw-amplitude}
\end{align}
where $q=k_2+k_3+k_4$ is the total outgoing gluon momentum. $A[\boldsymbol{1}^0,3^-,2^+,4^+,\boldsymbol{5}^0]$ is the gluon Compton amplitude with spinless massive legs and the mentioned ordering of the external legs \cite{Bern:2019crd}.
\[A[\boldsymbol{1}^0,3^-,2^+,4^+,\boldsymbol{5}^0] =- i \, \frac{[2 3] \langle 3|\boldsymbol{15}|3\rangle  \left(\tau _{13} \langle 3|\boldsymbol{5}|4]+m^2 [2 4] \langle 2 3\rangle \right)+m^2 s_{42} [2 4] \langle 2 3\rangle  \langle 3|\boldsymbol{1}|2]}{s_{15} \tau _{13} s_{23} \tau _{54} \langle 2 4\rangle } \ .\]

We can now also calculate $(s,+,+,-,s)$ ordering by using reflection relations\[A[\boldsymbol{1}^s,4,2,3,\boldsymbol{5}^s] = -(-1)^{2s} A[\boldsymbol{5}^s,3,2,4,\boldsymbol{1}^s]\quad ,\] followed by a $1 \leftrightarrow 5$ exchange, we get
\begin{align}
    &\boxed{
    \begin{aligned}
        & A[\boldsymbol{1}^s,4^+,2^+,3^-,\boldsymbol{5}^s]\\
   = & \,i \, \frac{\langle 3|\boldsymbol{5}(2 +4)\boldsymbol{1}|4]^2 }{\tau_{53} \tau_{14} \langle 3 2 \rangle [3|q\boldsymbol{1}|4] \langle2 4\rangle} \left( \frac{\langle \boldsymbol{15} \rangle }{m}+\frac{ \langle \boldsymbol{1} 3\rangle \langle \boldsymbol{5} 3\rangle [3|q\boldsymbol{1}|4]}{m \langle 3|-\boldsymbol{5}(2 +4)\boldsymbol{1}|4]} \right)^{2s} - i\,\frac{m^2 [24]^3}{s_{15} [32] [3|q\boldsymbol{1}|4]} \left(\frac{\langle
    \boldsymbol{15}\rangle}{m}\right)^{2s}\\
    = & A[\boldsymbol{1}^0,4^+,2^+,3^-,\boldsymbol{5}^0] \left(\frac{\langle
    \boldsymbol{15}\rangle}{m}\right)^{2s} +\,i\, \sum_{k=1}^{2s} {}^{2s}C_k \frac{m^{-2s}\langle 3|-\boldsymbol{5}(2 +4)\boldsymbol{1}|4]^{2-k} \langle\boldsymbol{15}\rangle^{2s-k} (\langle\boldsymbol{1} 3\rangle \langle \boldsymbol{5} 3\rangle)^k [3|q\boldsymbol{1}|4]^{k-1}}{\tau_{53} \tau_{14} \langle 3 2 \rangle \langle2 4\rangle} 
    \end{aligned}
    }\ ,
    &\label{eq:ym-five-point-14235-reflection-amplitude}
\end{align}
where $q=k_2+k_3+k_4$ is the total outgoing gluon momentum. $A[\boldsymbol{1}^0,4^+,2^+,3^-,\boldsymbol{5}^0]$ is the gluon Compton amplitude with spinless massive legs and the mentioned ordering of the external legs.
\[A[\boldsymbol{1}^0,4^+,2^+,3^-,\boldsymbol{5}^0] = \,i \, \frac{[2 3] \langle 3|\boldsymbol{51}|3\rangle  \left(\tau _{53} \langle 3|\boldsymbol{1}|4]+m^2 [2 4] \langle 2 3\rangle \right)+m^2 s_{42} [2 4] \langle 2 3\rangle  \langle 3|\boldsymbol{5}|2]}{s_{15} \tau _{53} s_{23} \tau _{14} \langle 2 4\rangle } \ .\]

\subsubsection{All-plus sector:
\texorpdfstring{$A[\boldsymbol{1}^s, 2^+, 3^+, 4^+, \boldsymbol{5}^s]$}
               {}
}
\label{subsubsec:ym-five-point-12345-all-plus}

Finally, we compute the same-helicity five-point Compton amplitude with $[s,+,+,+,s]$ ordering, from which the $[s,+,+,s,+]$ ordering can also be obtained by KK and BCJ relations (this has been mentioned before).

For the same-helicity five-point Compton amplitude, we calculate $A[\boldsymbol{1}^s,2^+,3^+,4^+,\boldsymbol{5}^s]$ using a massive--massless BCFW shift of the type $[\boldsymbol{1},2\rangle$.
\begin{equation}\label{eq:ym-five-point-all-plus-bcfw-shift}
\begin{aligned}
|\hat{2}\rangle &= |2\rangle - z |\boldsymbol{1}^J\rangle[2\boldsymbol{1}_J] \ ,
&\qquad
|\hat{2}] &= |2] \ , \\
|\hat{\boldsymbol{1}}^J] &= |\boldsymbol{1}^J] + z\, |2] [2 \boldsymbol{1}^J] \ ,
&\qquad
|\hat{\boldsymbol{1}}^J\rangle &= |\boldsymbol{1}^J\rangle \ .
\end{aligned}
\end{equation}

To calculate $A[\boldsymbol{1}^s, 2^+, 3^+, 4^+, \boldsymbol{5}^s]$, we will use the $[\boldsymbol{1},2\rangle$ massive--massless BCFW shift in eq.~\eqref{eq:ym-five-point-all-plus-bcfw-shift}.

Since $A[\boldsymbol{5}^s, \boldsymbol{1}^s| 2^+, 3^+, 4^+]$ does not survive as the 4-point gluon amplitude with 3 positive-helicity gluons goes to zero, 
\begin{equation}\label{eq:ym-five-point-all-plus-bcfw-decomposition}
    \boxed{A[\boldsymbol{1}^s, 2^+, 3^+, 4^+, \boldsymbol{5}^s]=A[ 4^+,\boldsymbol{5}^s, \hat{\boldsymbol{1}}^s| \hat{2}^+ 3^+]} \ .
\end{equation}

In the shift in eq.~\eqref{eq:ym-five-point-all-plus-bcfw-shift}, the only surviving figure is
\vspace{5mm}
\begin{figure}[h]
    \centering
    \begin{fmffile}{3_5_d01}
 \begin{fmfgraph*}(200,60)% units are now in cm
   \fmfleft{o1,i1,i2}
   \fmfright{o2,o3}
   \fmfblob{.08w}{g1,g2}
   \fmf{gluon}{o1,g1}
   \fmfv{label=$4^+$}{o1}
   \fmf{plain}{i1,g1}
   \fmfv{label=$\boldsymbol{5}^s$}{i1}
   \fmf{plain}{i2,g1}
   \fmfv{label=$\hat{\boldsymbol{1}}^s$}{i2}
   \fmf{gluon, label=$\hat{P}^+$,label.side=left}{v1,g1}
   \fmf{phantom}{v1,v2}
   \fmf{gluon, label=$-\hat{P}^-$,label.side=left}{g2,v2}
   \fmf{gluon}{o3,g2}
   \fmfv{label=$\hat{2}^+$}{o3}
   \fmf{gluon}{o2,g2}
   \fmfv{label=$3^+$}{o2}
\end{fmfgraph*}
\end{fmffile}
\vspace{5mm}
    \caption{Factorization channel contributing to $A[ 4^+,\boldsymbol{5}^s, \hat{\boldsymbol{1}}^s| \hat{2}^+ 3^+]$}
    \label{fig:ym-five-point-all-plus-bcfw-channel-23}
\end{figure}
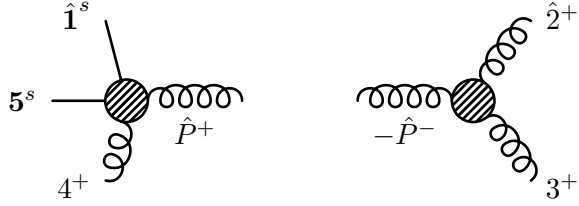
\begin{flalign*}
    &\text{In fig.~\ref{fig:ym-five-point-all-plus-bcfw-channel-23}},\quad s_{\hat{2}3}=0 \implies z = \frac{\langle23\rangle}{\langle3|\boldsymbol{1}|2]}\ .&&
\end{flalign*}
\begin{align}
    & A[ 4^+,\boldsymbol{5}^s, \hat{\boldsymbol{1}}^s| \hat{2}^+ \hat{3}^+]= A[ 4^+,\boldsymbol{5}^s, \hat{\boldsymbol{1}}^s,\hat{P}^+] \frac{i}{s_{23}} A[-\hat{P}^- 2^+ \hat{3}^+]\nonumber\\
    =& i\,\frac{m^{2-2s}[2|\boldsymbol{1}(2+3)|4][23][34]}{\tau_{12}\tau_{54}s_{23}s_{34}} \langle{\boldsymbol{1}\boldsymbol{5}}\rangle^{2s} \ . \label{eq:ym-five-point-all-plus-bcfw-channel}
\end{align}

Following eq.~\eqref{eq:ym-five-point-all-plus-bcfw-decomposition} and eq.~\eqref{eq:ym-five-point-all-plus-bcfw-channel}, we get 
\begin{align}
    & \boxed{A[\boldsymbol{1}^s, 2^+, 3^+, 4^+, \boldsymbol{5}^s] \,=\, i\,\frac{m^{2-2s}[2|\boldsymbol{1}(2+3)|4][23][34]}{\tau_{12}\tau_{54}s_{23}s_{34}} \langle{\boldsymbol{1}\boldsymbol{5}}\rangle^{2s}} \ . \label{eq:ym-five-point-all-plus-amplitude}
\end{align}

\subsubsection{General form of Compton amplitudes}

We present a general structure for five-point single-minus and all-plus Compton amplitudes in Yang--Mills theory. We adopt a convenient notation for this purpose.
\begin{align}
    A_{s, X}^{h_1 h_2 h_3}=A[\boldsymbol{1}^{s},X_1^{h_{1}},X_2^{h_{2}},X_3^{h_{3}},\boldsymbol{5}^{s}] \ , \label{eq_convenient_notation_YM_Comptons}
\end{align}
with
$X = X_1 X_2 X_3 \in \{ 234, 432, 324\}$, with $h_{(1,2,3)}$ representing the helicity for leg $(X_1,X_2,X_3)$.

Now, we rewrite the partial YM-Compton amplitudes, $A^{+-+}_{s, 234}$ in eq.~\eqref{eq:ym-five-point-12345-spurious-pole-free-amplitude}, $A^{+-+}_{s, 432}$ obtained from $2 \leftrightarrow 4$ swap, $A^{-++}_{s, 324}$ in eq.~\eqref{eq:ym-five-point-13245-bcfw-amplitude}, and $A^{+++}_{s, 234}$ in eq.~\eqref{eq:ym-five-point-all-plus-amplitude} in the following way
\begin{align}
    \boxed{ A_{s,X}^{h_1 h_2 h_3} = A_{0,X}^{h_1 h_2 h_3} \left(\frac{\langle\boldsymbol{15}\rangle}{m}\right)^{2s} - i \sum_{n=1}^{2s} \frac{m^{-2s} H_X^{2-n} Q_X^{n-1} \langle\boldsymbol{15}\rangle^{2s-n} (\langle\boldsymbol{1}3\rangle\langle\boldsymbol{5}3\rangle)^n}{D_X} }\ , \label{eq_YM_Compton_single_minus_general_form}
\end{align}
with $H_X$, $Q_X$ and $D_X$ for different $X$-s are presented in the table below:
\[
\begin{array}{|c|c|c|c|c|}
\hline
X & h_1 h_2 h_3 & H_{X} & Q_{X} & D_{X} \\ \hline
234 & +-+ &
\langle3|\boldsymbol{1}|2]\langle3|-\boldsymbol{5}|4] &
[2|\boldsymbol{1}q|4] &
\tau_{12}\tau_{54}\langle23\rangle\langle34\rangle
\\[1mm]
432 & +-+ &
\langle3|\boldsymbol{1}|4]\langle3|-\boldsymbol{5}|2] &
[4|\boldsymbol{1}q|2] &
\tau_{14}\tau_{52}\langle43\rangle\langle32\rangle
\\[1mm]
324 & -++ &
\langle3|\boldsymbol{1}(2+4)\boldsymbol{5}|4] &
[3|q\boldsymbol{5}|4] &
\tau_{13}\tau_{54}\langle32\rangle\langle24\rangle\\
234 & +++ &
0 &
0 &
1 \\
\hline
\end{array}
\ .\]

The second term of eq.~\eqref{eq_YM_Compton_single_minus_general_form} is only present in the single-minus case, and for the all-plus case, we set $H_X=Q_X=0$ to indicate its absence.

%----------------------------------------------------------------------------

\section{QED Amplitudes as Abelian Case of Yang--Mills Amplitudes}\label{sec:qed-compton}
The ordering of external legs is irrelevant for QED Compton amplitudes unlike its non-Abelian counterpart, i.e. gluon Compton amplitudes. Hence, we will compute the full QED Compton amplitudes. 

One can calculate the QED Compton amplitudes by doing the permutation sum\footnote{This relation holds as we are taking a YM-Compton amplitude consisting of gluons which has $SU(N)$-gauge symmetry and going to a theory where the massless gauge bosons have $U(1)$-gauge symmetry. Therefore, taking the permutation sum in eq.~\eqref{eq:Photon_Compton_Perm} gives us the Compton amplitudes in QED.} of the gluon legs in the corresponding YM-partial Compton amplitudes.\footnote{The factor of $2^{\frac{n}{2}-1}$ in eq.~\eqref{eq:Photon_Compton_Perm} appears because of the chosen normalization of the $SU(N)$ generators in the YM theory up to the factor of $\sqrt{2}$ in eq.~\eqref{eq:YM_Compton_Full_Amplitude}. These $\sqrt 2$ factors can be further absorbed in the electro-magnetic charge in QED if one wishes to do so, but it is not done here.}
\begin{align}\label{eq:Photon_Compton_Perm}
        \boxed{\mathcal{A}(\boldsymbol{1}^s,2,\dots,n-1,\boldsymbol{n}^s) = 2^{\frac{n}{2}-1} \sum_{\sigma \in S_{n-2}} A[\boldsymbol{1}^s,\sigma(2,\dots,n-1),\boldsymbol{n}^s]} \ .
\end{align} 

We present the three-point photon Compton amplitudes in sec.~\ref{subsec:qed-three-point-compton}. Then we use the permutation sum relation in eq.~\eqref{eq:Photon_Compton_Perm} to calculate the four-point QED Compton amplitude in sec.~\ref{subsec:qed-four-point-compton}. Later, we verify the four-point QED Compton amplitudes in sec.~\ref{subsec:qed-four-point-compton} with the BCFW calculations in sec.~\ref{subsec:qed-four-point-bcfw-check}. Then we use permutation sum of the gluon legs in YM-Compton amplitudes to calculate the five-point QED Compton amplitudes in sec.~\ref{subsec:qed-five-point-compton} utilizing the BCJ relations and five-point YM-Compton amplitudes. Later, we also match the five-point QED Compton amplitudes in sec.~\ref{subsec:qed-five-point-compton} with the amplitudes calculated using BCFW recursion relations in sec.~\ref{subsec:qed-five-point-bcfw-check} using analytic parametrizations presented in app.~\ref{app:analytic-parametrization}.

\subsection{Three-Point Compton Amplitudes}\label{subsec:qed-three-point-compton}
We present the three-point QED Compton amplitudes involving a massive particle of mass $m$ and spin $s$. The expressions below are valid for arbitrary $m$ and $s$.

The three-point photon Compton amplitudes are
\begin{figure}[H]
    \centering
    \begin{fmffile}{4_0_d01}
        \begin{fmfgraph*}(150,100)
            \fmfleft{i1,i2}
            \fmfright{o1,o2}
            \fmfdot{v1}
            \fmf{plain}{v1,i1}
            \fmfv{label=$\boldsymbol{1}^s$}{i1}
            \fmf{plain}{v1,o1}
            \fmfv{label=$\boldsymbol{3}^s$}{o1}
            \fmf{phantom}{i2,v2,o2}
            \fmf{photon}{v1,v2}
            \fmfv{label=$2\pm $}{v2}
        \end{fmfgraph*}
    \end{fmffile}
    \vspace{5mm}
    \caption{Three-point amplitude $\mathcal{A}(\boldsymbol{1}^s, 2^\pm, \boldsymbol{3}^s)$}
    \label{fig:qed-three-point-compton}
\end{figure}
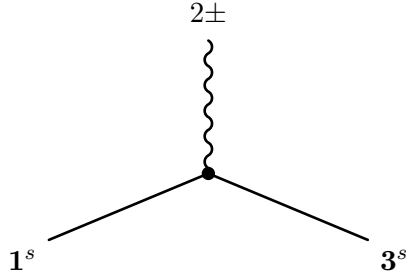

\begin{empheq}[box=\fbox]{align}
\mathcal{A}(\boldsymbol{1}^s,2^+,\boldsymbol{3}^s)
&= i \, \sqrt{2}\; \frac{\langle \eta | \boldsymbol{1} | 2]}{\langle \eta 2 \rangle}
   \frac{\langle \boldsymbol{13} \rangle^{2s}}{m^{2s}} \ ,
   \label{eq:qed-three-point-positive-helicity} \\[4pt]
\mathcal{A}(\boldsymbol{1}^s,2^-,\boldsymbol{3}^s)
&= i \,\sqrt{2}\; \frac{\langle 2 | \boldsymbol{1} | \eta ]}{[2 \eta]}
   \frac{[\boldsymbol{13}]^{2s}}{m^{2s}} \ .
   \label{eq:qed-three-point-negative-helicity}
\end{empheq}

Here, $|\eta\rangle$ and $|\eta]$ are reference spinors.

For the three-point case, eq.~\eqref{eq:Photon_Compton_Perm} trivially reduces to eq.~\eqref{eq:qed-three-point-positive-helicity} -- eq.~\eqref{eq:qed-three-point-negative-helicity} following from eq.~\eqref{eq:ym-three-point-positive-helicity} -- eq.~\eqref{eq:ym-three-point-negative-helicity}.

%----------------------------------------------------------------------------

\subsection{Four-Point Compton Amplitudes}\label{subsec:qed-four-point-compton}

In the case of four-point amplitudes, eq.~\eqref{eq:Photon_Compton_Perm} reduces to
\begin{equation}\label{eq:QED_4pt_Perm_Sum}
    \mathcal{A}(\boldsymbol{1}^s,2,3,\boldsymbol{4}^s) \; = \; 2 \; (A[\boldsymbol{1}^s,2,3,\boldsymbol{4}^s] \, +\, A[\boldsymbol{1}^s,3,2,\boldsymbol{4}^s]) \ .
\end{equation}

The opposite helicity photon Compton amplitude takes the following form:
{\allowdisplaybreaks
\begin{align}\label{eq:QED_4pt_Perm_Sum_Opp_Hel}
    &\nonumber \mathcal{A}(\boldsymbol{1}^s,2^+,3^-,\boldsymbol{4}^s)\;=\;2\,(A[\boldsymbol{1}^s,2^+,3^-,\boldsymbol{4}^s] \, + \, A[\boldsymbol{1}^s,3^-,2^+,\boldsymbol{4}^s])\\
    = &\nonumber\; 2\,i \, \frac{\langle 3 | \boldsymbol{1} | 2 ]^{2-2s}}{\tau_{12}\,s_{23}}\big(\langle \boldsymbol{1}3 \rangle [\boldsymbol{4}2]+ \langle\boldsymbol{4}3 \rangle [\boldsymbol{1}2] \big)^{2s} \, + \, 2\,i \, \frac{\langle 3 | \boldsymbol{1} | 2 ]^{2-2s}}{\tau_{13}\,s_{23}}\big(\langle \boldsymbol{1}3 \rangle [\boldsymbol{4}2]+ \langle\boldsymbol{4}3 \rangle [\boldsymbol{1}2] \big)^{2s}\\
    \implies & \boxed{\mathcal{A}(\boldsymbol{1}^s,2^+,3^-,\boldsymbol{4}^s) \; = \; -2\,i \, \frac{\langle 3 | \boldsymbol{1} | 2 ]^{2-2s}}{\tau_{12}\,\tau_{13}}\big(\langle \boldsymbol{1}3 \rangle [\boldsymbol{4}2]+ \langle\boldsymbol{4}3 \rangle [\boldsymbol{1}2] \big)^{2s}} \ .
\end{align}
}
To get to eq.~\eqref{eq:QED_4pt_Perm_Sum_Opp_Hel}, in the last step we use: $\tau_{12}\,+\,\tau_{13}\,+\,s_{23}\, = \, 0$.

Similarly, we can calculate the all-plus four-point Compton amplitude.
{\allowdisplaybreaks
\begin{align}\label{eq:QED_4pt_Perm_Sum_All_Pos_Hel}
    &\nonumber \mathcal{A}(\boldsymbol{1}^s,2^+,3^+,\boldsymbol{4}^s)\;=\;2\,(A[\boldsymbol{1}^s,2^+,3^+,\boldsymbol{4}^s] \, + \, A[\boldsymbol{1}^s,3^+,2^+,\boldsymbol{4}^s])\\
    = &\nonumber\; - 2 \, i \, \frac{m^{2-2s} [23]^2}
        {\tau_{12}\,s_{23}}
   \langle \boldsymbol{14} \rangle^{2s} \, - 2 \, i \, \frac{m^{2-2s} [23]^2}
        {\tau_{13}\,s_{23}}
   \langle \boldsymbol{14} \rangle^{2s}\\
    \implies & \boxed{\mathcal{A}(\boldsymbol{1}^s,2^+,3^+,\boldsymbol{4}^s) \; = \;  2 \, i \, \frac{m^{2-2s} [23]^2}
        {\tau_{12}\,\tau_{13}}
   \langle \boldsymbol{14} \rangle^{2s}} \ .
\end{align}
}

A similar calculation leads to the all-minus four-point Compton amplitude.
\begin{equation}\label{eq:QED_4pt_Perm_Sum_All_Neg_Hel}
    \boxed{\mathcal{A}(\boldsymbol{1}^s,2^-,3^-,\boldsymbol{4}^s) \; = \;  2 \, i \, \frac{m^{2-2s} \langle 23 \rangle^2}
        {\tau_{12}\,\tau_{13}}
   [\boldsymbol{14} ]^{2s}} \ .
\end{equation}

\subsection{Five-Point Compton Amplitudes}\label{subsec:qed-five-point-compton}
We will use eq.~\eqref{eq:Photon_Compton_Perm} to calculate five-point QED Compton amplitudes with different helicities. We use the BCJ relations to reduce the six terms in the permutation sum to a basis consisting of $A[\boldsymbol{1}^s,2,3,4,\boldsymbol{5}^s]$ and $A[\boldsymbol{1}^s,4,3,2,\boldsymbol{5}^s]$, as in eqs.~\eqref{eq:bcj-relation-ordering-13245} and \eqref{eq:bcj-relation-ordering-12435} and two more similar equations from the $(2 \leftrightarrow 4)$ exchange. The permutation sum then gives
\begin{align}\label{eq:fivePtQEDPerms}
    \boxed{\mathcal{A}(\boldsymbol{1}^s,2,3,4,\boldsymbol{5}^s)=2\sqrt{2}\left(1 + \frac{\tau_{12}}{\tau_{13}}+\frac{\tau_{54}}{\tau_{53}}\right)A[\boldsymbol{1}^s,2,3,4,\boldsymbol{5}^s]+ (2\leftrightarrow4)} \ .
\end{align}

Just like the Yang--Mills case in sec.~\ref{subsec:ym-five-point-bcfw-checks}, we get spurious singularities for $s\leq 1$ upon using BCFW recursion in sec.~\ref{subsec:qed-five-point-bcfw-check}. To resolve these spurious singularities, we use BCJ relations for YM-partial Compton amplitudes. Using the permutation sum for the five-point case in eq.~\eqref{eq:fivePtQEDPerms} and the partially ordered YM-Compton amplitude in eq.~\eqref{eq:ym-five-point-12345-spurious-pole-free-amplitude}, we get: 
{
\allowdisplaybreaks
\begin{align}
    &\boxed{
    \begin{aligned}
        &\nonumber\mathcal{A}(\mathbf{1}^s,{2}^+,{3}^-,4^+,\mathbf{5}^s)\\
    &\nonumber\\
    =&\nonumber\nonumber 2\sqrt{2}\,i \, \left(-\frac{\langle 3 |\boldsymbol{1}|2]^2 \langle 3 |\boldsymbol{5}|4]^2 }{\tau _{12} \tau _{54} \langle 2  3\rangle \langle 3 4\rangle [2|\boldsymbol{1}(2+3)|4]}\left(\frac{\langle\boldsymbol{15}\rangle}{m}\,+\,\frac{[2|\boldsymbol{1}(2+3)|4]\langle\boldsymbol{1}3\rangle\langle\boldsymbol{5}3\rangle}{m \langle3|\boldsymbol{1}|2]\langle3|-\boldsymbol{5}|4]}\right)^{2s}\right.\\
    &\hspace{3cm}\nonumber\left.+\frac{m^2  [24]^4}{s_{15} [2  3][34]   [2|\boldsymbol{1}(2+3)|4]}\left(\frac{\langle\boldsymbol{15}\rangle}{m}\right)^{2s}\right)\left(1 +\frac{\tau_{12}}{\tau_{13}}+\frac{\tau_{54}}{\tau_{53}}\right)+ (2\leftrightarrow4)\\
    &\nonumber\\
    =&\nonumber 2\sqrt{2}\,i \, \left(\left(\frac{\langle3|\boldsymbol{1}|2]\langle3|\boldsymbol{5}|4]\langle 3|\boldsymbol{15}|3\rangle}{\tau_{12}\langle23\rangle\langle34\rangle\tau_{45}s_{15}} + \frac{m^2[24]^2([2|\boldsymbol{1}3|4]-[24]\tau_{45})}{\tau_{12}[23][34]\tau_{45}s_{15}}\right)\left(\frac{\langle\boldsymbol{15}\rangle}{m}\right)^{2s}\right.\\
    &\nonumber\left.\qquad-\sum_{n=1}^{2s} {}^{2s}C_{n}\frac{m^{-2s}\left(\langle 3 |\boldsymbol{1}|2] \langle 3 |-\boldsymbol{5}|4]\right)^{2-n} [2|\boldsymbol{1}q|4]^{n-1}}{\tau _{12} \tau _{54} \langle 2  3\rangle \langle 3 4\rangle } \langle\boldsymbol{15}\rangle^{2s-n}\,\left(\langle\boldsymbol{1}3\rangle\langle\boldsymbol{5}3\rangle\right)^{n}\right)\left(1 +\frac{\tau_{12}}{\tau_{13}}+\frac{\tau_{54}}{\tau_{53}}\right)\\
    &\nonumber\hspace{14cm}+ (2\leftrightarrow4)
    \end{aligned}
    } \ .\\
    & \label{eq_five-point_single_minus_QED_Perm_sum}
\end{align}
}

For the all-plus case, we can directly sum over the 6 terms in eq.~\eqref{eq:Photon_Compton_Perm} because they can all be obtained from simple permutations of one analytic all-plus color-ordered amplitude.
\begin{align}
    \boxed{\mathcal{A}({\mathbf{1}}^s,2^+,3^+,4^+,\mathbf{5}^s)=2\sqrt{2}\,i \, \sum_{\sigma \in S_3}\frac{m^{2-2s}[\sigma(2)|\boldsymbol{1}\,q|\sigma(4)]}{\tau_{1\sigma(2)}\langle\sigma(2)\,\sigma(3)\rangle\langle\sigma(3)\,\sigma(4)\rangle \tau_{5\sigma(4)}} \langle{\boldsymbol{1}\boldsymbol{5}}\rangle^{2s}} \ . \label{eq:qed-five-point-all-plus-permutation-sum}
\end{align}

%----------------------------------------------------------------------------

\section{Graviton Compton Amplitude from KLT Relations}\label{sec:graviton-compton-klt}

There is no color ordering for graviton Compton amplitudes. Consequently, we compute the full Compton amplitudes without decomposing into partial amplitudes.

In sec.~\ref{subsec:graviton-three-point-compton}, we present the well-known three-point graviton Compton amplitudes. These three-point amplitudes serve as building blocks for the four-point Compton amplitudes discussed in sec.~\ref{subsec:graviton-four-point-bcfw-check}. In sec.~\ref{subsec:graviton-three-point-compton}, we also verify that the three-point Compton amplitudes in sec.~\ref{subsec:graviton-three-point-compton} agree with the three-point tree-level KLT relation in eq.~\eqref{eq:KLT_double_copy_relation_three_point}.

We then use the four-point tree-level KLT relation in eq.~\eqref{eq:KLT_double_copy_relation_four_point} to calculate the tree-level four-point graviton Compton amplitude in sec.~\ref{subsec:graviton-four-point-compton}. We later verify these amplitudes in sec.~\ref{subsec:graviton-four-point-compton} with our BCFW calculation in sec.~\ref{subsec:graviton-four-point-bcfw-check}.

In sec.~\ref{subsec:graviton-five-point-compton}, we use the KLT double copy relation and BCJ relations to construct the five-point graviton Compton amplitudes which we later match with the graviton Compton amplitudes calculated by BCFW recursion relations in sec.~\ref{subsec:graviton-five-point-bcfw-check} using analytic parametrizations presented in app.~\ref{app:analytic-parametrization}.

\subsection{Three-Point Compton Amplitude}\label{subsec:graviton-three-point-compton}

The three-point graviton Compton amplitudes involving a graviton and a massive particle of spin $s$ and mass $m$ are given in ref.~\cite{Arkani-Hamed:2017jhn}:
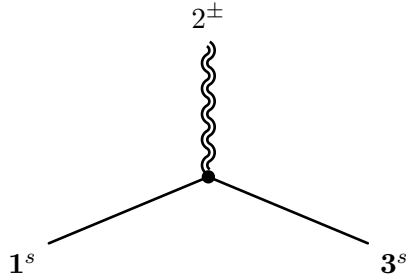
\begin{figure}[H]
\centering
\begin{fmffile}{5_0_d01}
\begin{fmfgraph*}(150,100)
    \fmfleft{i1,i2}
    \fmfright{o1,o2}

    \fmf{plain}{v1,i1}
    \fmfv{label=$\boldsymbol{1}^s$}{i1}
    \fmf{plain}{v1,o1}
    \fmfv{label=$\boldsymbol{3}^s$}{o1}

    \fmf{phantom}{i2,v2,o2}
    \fmf{dbl_wiggly}{v1,v2}
    \fmfv{label=$2^{\pm}$}{v2}
    \fmfdot{v1}

\end{fmfgraph*}
\end{fmffile}
\vspace{5mm}
\caption{Three-point Graviton Compton Amplitude}
\label{fig:graviton-three-point-compton}
\end{figure}

\begin{empheq}[box=\fbox]{align}
M(\boldsymbol{1}^s,2^+,\boldsymbol{3}^s)
&= -i \, \frac{\langle \eta | \boldsymbol{1} | 2 ]^2}{[2\eta]^2}
   \frac{\langle \boldsymbol{13} \rangle^{2s}}{m^{2s}} \ , \label{eq:graviton-three-point-positive-helicity}\\[1ex]
M(\boldsymbol{1}^s,2^-,\boldsymbol{3}^s)
&= -i \, \frac{\langle 2 | \boldsymbol{1} | \eta ]^2}{\langle 2\eta \rangle^2}
   \frac{[ \boldsymbol{13} ]^{2s}}{m^{2s}} \ . \label{eq:graviton-three-point-negative-helicity}
\end{empheq}
Here, $|\eta\rangle$ and $|\eta]$ denote arbitrary reference spinors.

Eq.~\eqref{eq:graviton-three-point-positive-helicity} and eq.~\eqref{eq:graviton-three-point-negative-helicity} can also be verified utilizing the three-point Compton amplitudes in eq.~\eqref{eq:ym-three-point-positive-helicity} and eq.~\eqref{eq:ym-three-point-negative-helicity} through the following double copy relation for three-point amplitudes,
\begin{equation}\label{eq:KLT_double_copy_relation_three_point}
    M(\boldsymbol{1}^s,\, 2,\, \boldsymbol{3}^s) = i\, A[\boldsymbol{1}^{s_L},\, 2,\, \boldsymbol{3}^{s_L}]\: A[\boldsymbol{1}^{s_R},\, 2,\, \boldsymbol{3}^{s_R}]\ ,
\end{equation}
where $s=s_L+s_R$.

\subsection{Four-Point Compton Amplitude}\label{subsec:graviton-four-point-compton}

Now, we calculate the four-point graviton Compton amplitude using the KLT relation for four-point tree-level Compton amplitudes:
\begin{equation}\label{eq:KLT_double_copy_relation_four_point}
    M(\boldsymbol{1}^s,\, 2,\, 3,\, \boldsymbol{4}^s)=  -i\, s_{23} \: A[\boldsymbol{1}^{s_L},\, 2,\, 3,\, \boldsymbol{4}^{s_L}]\: A[\boldsymbol{1}^{s_R},\, 3,\, 2,\, \boldsymbol{4}^{s_R}]\ ,
\end{equation}
where $s=s_L+s_R$.

Following the double copy relation and using eq.~\eqref{eq:ym-four-point-positive-negative-helicity} and \eqref{eq:ym-four-point-negative-positive-helicity} we get the following opposite helicity Compton amplitude:
\begin{align}
    &\nonumber M(\boldsymbol{1}^s,\, 2^+,\, 3^-,\, \boldsymbol{4}^s)=  -i\,s_{23} \: A[\boldsymbol{1}^{s_L},\, 2^+,\, 3^-,\, \boldsymbol{4}^{s_L}]\: A[\boldsymbol{1}^{s_L},\, 3^-,\, 2^+,\, \boldsymbol{4}^{s_L}]\\
    =\;&\nonumber -i \, s_{23} \, \left( i \, 
   \frac{\langle 3 | \boldsymbol{1} | 2 ]^{2-2s_L}}
        {\tau_{12}\,s_{23}}
   \big(
      \langle \boldsymbol{1}3 \rangle [\boldsymbol{4}2]
      + \langle \boldsymbol{4}3 \rangle [\boldsymbol{1}2]
   \big)^{2s_L} \right)\: \left(i \, 
   \frac{\langle 3 | \boldsymbol{1} | 2 ]^{2-2s_R}}
        {\tau_{13}\,s_{23}}
   \big(
      \langle \boldsymbol{1}3 \rangle [\boldsymbol{4}2]
      + \langle \boldsymbol{4}3 \rangle [\boldsymbol{1}2]
   \big)^{2s_R}\right)\\
   \implies \;& \boxed{M(\boldsymbol{1}^s,2^+,3^-,\boldsymbol{4}^s)
= i\,\frac{\langle 3 | \boldsymbol{1} | 2 ]^{4-2s}}
        {\tau_{12}s_{23}\tau_{13}}
   \bigl(
      \langle \boldsymbol{1}3 \rangle [\boldsymbol{4}2]
      + \langle \boldsymbol{4}3 \rangle [\boldsymbol{1}2]
   \bigr)^{2s}} \ . \label{eq_GR_4pt_opp_hel_Compton}
\end{align}

Similarly, using the relation in eq.~\eqref{eq:KLT_double_copy_relation_four_point}, from the YM-Compton amplitudes in eq.~\eqref{eq:ym-four-point-all-plus-helicity}, we get the all-plus graviton Compton amplitude:
\begin{align}
    &\nonumber M(\boldsymbol{1}^s,\, 2^+,\, 3^+,\, \boldsymbol{4}^s)=  -i\,s_{23} \: A[\boldsymbol{1}^{s_L},\, 2^+,\, 3^+,\, \boldsymbol{4}^{s_L}]\: A[\boldsymbol{1}^{s_L},\, 3^+,\, 2^+,\, \boldsymbol{4}^{s_L}]\\
    =\;&\nonumber -i \, s_{23} \, \left( - i \, \frac{m^{2-2s_L} [23]^2}
        {\tau_{12}\,s_{23}}
   \langle \boldsymbol{14} \rangle^{2s_L} \right)\: \left(- i \, \frac{m^{2-2s_R} [23]^2}
        {\tau_{13}\,s_{23}}
   \langle \boldsymbol{14} \rangle^{2s_R}\right)\\
   \implies \;& \boxed{M(\boldsymbol{1}^s,2^+,3^+,\boldsymbol{4}^s)
= i\,\frac{m^{4-2s}\,[23]^4}
        {\tau_{12}s_{23}\tau_{13}}
   \langle \boldsymbol{14} \rangle^{2s}} \ . \label{eq_GR_4pt_all_plus_hel_Compton}
\end{align}

The all-negative-helicity four-point graviton Compton amplitude follows from charge conjugation of the above result (or a direct KLT calculation),
\begin{equation}
    \boxed{M(\boldsymbol{1}^s,2^-,3^-,\boldsymbol{4}^s)= i\,\frac{m^{4-2s}\,\langle 23 \rangle^4}
        {\tau_{12}s_{23}\tau_{13}}
   [\boldsymbol{14}]^{2s}} \ . \label{eq_GR_4pt_all_minus_hel_Compton}
\end{equation}

\subsection{Five-Point Compton Amplitudes}\label{subsec:graviton-five-point-compton}

Since gravity amplitudes are cross-symmetric in the gravitons without color ordering, there exists only one independent helicity configuration for graviton Compton amplitudes with two positive-helicity and one negative-helicity gravitons. The same statement applies to the case where all gravitons have positive helicities. Again, we get the graviton Compton amplitudes with opposite helicities by complex conjugation of spinor brackets:
\begin{equation*}
\langle a b \rangle \;\longleftrightarrow\; [b a] \ .
\end{equation*}

To get to the five-point graviton Compton amplitudes, we can start with the five-point KLT relation in eq.~\eqref{eq:klt-five-point-relation}:
\begin{align*}
    & M(\boldsymbol{1}^{s},2,3,4,\boldsymbol{5}^{s}) \\
      = \; & i \,\tau_{12} \,s_{34} \,A[\boldsymbol{1}^{s_L},2,3,4,\boldsymbol{5}^{s_L}]\: A[2,\boldsymbol{1}^{s_R},4,3,\boldsymbol{5}^{s_R}] + i\,\tau_{13}\, s_{24}\, A[\boldsymbol{1}^{s_L},3,2,4,\boldsymbol{5}^{s_L}]\: A[3,\boldsymbol{1}^{s_R},4,2,\boldsymbol{5}^{s_R}] \ ,
\end{align*}
where  $s = s_L + s_R$.

Now, we can use the gauge-invariant BCJ relations for YM amplitudes in app.~\ref{app:bcj-relations} to rewrite the YM amplitudes in the above equation in terms of our chosen BCJ basis, which consists of $A[\boldsymbol{1}^{s},4,3,2,\boldsymbol{5}^{s}]$ and $ A[\boldsymbol{1}^{s},2,3,4,\boldsymbol{5}^{s}]$. The graviton Compton amplitude is now written as
\begin{align}
&\boxed{
    \begin{aligned}
    M(\boldsymbol{1}^{s},2,3,4,\boldsymbol{5}^{s}) = \; & A_L^T \begin{pmatrix}
K_{234,234} & K_{234,432} \\
K_{432,234} & K_{432,432}
\end{pmatrix}  A_R
    \end{aligned}}\ , \label{eq:Five_Point_KLT_BCJ_Basis}
\end{align}
where
{\allowdisplaybreaks
\begin{align*}
    & s=s_L+s_R \ ,\\
    & K_{234,234}
=
i \ \left(\frac{\tau_{12}^2\tau_{54}(\tau_{13}+\tau_{53})(s_{24}+\tau_{54})}
{\tau_{13}s_{24}\tau_{53}}
-
\frac{\tau_{12}s_{34}\tau_{54}}{\tau_{53}} \right) \ ,\\
    & K_{432,432}
=
i \ \left( \frac{\tau_{14}^2\tau_{52}(\tau_{13}+\tau_{53})(s_{24}+\tau_{52})}
{\tau_{13}s_{24}\tau_{53}}
-
\frac{\tau_{14}s_{23}\tau_{52}}{\tau_{53}} \right) \ ,\\
    & K_{234,432} = K_{432,234}
=
- \ 
i \ \frac{
\tau_{12}\tau_{14}\tau_{52}(\tau_{13}+\tau_{53})\tau_{54}
}{
\tau_{13}s_{24}\tau_{53}
} \ ,\\
    & A_L \,= \, \begin{pmatrix}
         A[\boldsymbol{1}^{s_L},2, 3, 4, \boldsymbol{5}^{s_L}] \\
         A[\boldsymbol{1}^{s_L},4, 3, 2, \boldsymbol{5}^{s_L}]
    \end{pmatrix} \qquad \text{and} \qquad A_R \,= \, \begin{pmatrix}
         A[\boldsymbol{1}^{s_R},2, 3, 4, \boldsymbol{5}^{s_R}] \\
         A[\boldsymbol{1}^{s_R},4, 3, 2, \boldsymbol{5}^{s_R}]
    \end{pmatrix} \ .
\end{align*}
}

The single-minus five-point graviton Compton amplitude takes the following form:
\begin{align}
    &\boxed{\begin{aligned}
    &M(\boldsymbol{1}^{s},2^+,3^-,4^+,\boldsymbol{5}^s)\\
={}&
i\,\mathcal M_0\,m^{-2s}\langle\boldsymbol{15}\rangle^{2s} +
\sum_{x=L,R}\sum_{X}
\sum_{\ell=1}^{2s_x}
\binom{2s_x}{\ell}
\mathcal M^{{}x}_{I,\ell}\,
m^{-2s}\langle\boldsymbol{15}\rangle^{2s-\ell}(\langle\boldsymbol{1}3\rangle\langle\boldsymbol{5}3\rangle)^\ell \\
&-
i
\sum_{X,Y}
\sum_{\ell=1}^{2s_L}
\sum_{r=1}^{2s_R}
\binom{2s_L}{\ell}
\binom{2s_R}{r}
\mathcal M^{{}LR}_{IJ,\ell r}\,
m^{-2s}\langle\boldsymbol{15}\rangle^{2s-\ell-r}(\langle\boldsymbol{1}3\rangle\langle\boldsymbol{5}3\rangle)^{\ell+r} 
\end{aligned}}\ , \label{single_minus_five_point_graviton_Compton}
\end{align}
where $X,Y \in \{ 234, 432\}$ and the $\mathcal M$-s are
\[
\boxed{\begin{aligned}
    \mathcal M_0 &=
\sum_{X,Y}
A^{+-+}_{0,X}\,
K_{XY}\,
A^{+-+}_{0,Y} \ ,\\
\mathcal M^{{}L}_{X,\ell}
&=
\frac{Q_X^{\ell-1} H_X^{2-\ell}}{D_X}
\sum_Y K_{XY}A^{+-+}_{0,Y} \ ,\\
\mathcal M^{{}R}_{Y,r}
&=
\frac{Q_Y^{r-1} H_Y^{2-r}}{D_Y}
\sum_X A^{+-+}_{0,X}K_{XY} \ ,\\
\mathcal M^{{}LR}_{XY,\ell r}
&=
\left( \frac{H_X^{2-\ell}}{D_X} K_{XY}\frac{H_Y^{2-r}}{D_Y} \right) Q_Y^{\ell-1}Q_Y^{r-1} \ , 
\end{aligned}}
\]
and $s=s_L+s_R$. Here, we are using the notation mentioned in eq\eqref{eq_convenient_notation_YM_Comptons}. 

Similarly, the all-plus five-point graviton Compton amplitude can be written as
\begin{align}
    \boxed{M(\boldsymbol{1}^{s},2^+,3^+,4^+,\boldsymbol{5}^s) = \sum_{X,Y \in \{234,432\}} A_{0,X}^{+++} K_{XY} A_{0,Y}^{+++} \left(\frac{\langle \boldsymbol{15} \rangle}{m}\right)^{2s}} \ ,\label{all_plus_five_point_graviton_Compton}
\end{align}
using the all-plus partial YM-Compton amplitude in eq.~\eqref{eq:ym-five-point-all-plus-amplitude} and notations in eq.~\eqref{eq_convenient_notation_YM_Comptons}. 

In both the five-point graviton Compton amplitudes in eq.~\eqref{eq:graviton-five-point-single-minus-bcfw-amplitude} and eq.~\eqref{eq:graviton-five-point-all-plus-bcfw-amplitude}, calculated in the later section (sec.~\ref{subsec:graviton-five-point-bcfw-check}) using BCFW recursion relations, we have unphysical spurious poles for $s\leq 2$. But, the amplitudes in the eq.~\eqref{single_minus_five_point_graviton_Compton} and eq.~\eqref{all_plus_five_point_graviton_Compton} are manifestly free of unphysical spurious poles.

We have verified that eq.~\eqref{single_minus_five_point_graviton_Compton} leads to the same result for the graviton Compton amplitude (with massive spin-$s = s_L + s_R$) for different massive spins ($s_L$ and $s_R$) of the single copy gauge amplitudes as long as $s_{L,R} \leq 1$. For example:
\[M_{\frac{1}{2}  \otimes  \frac{1}{2}} = M_{1 \otimes 0} = M_{0 \otimes 1} \quad \text{and} \quad  M_{1  \otimes  \frac{1}{2}} = M_{\frac{1}{2}  \otimes  1} \ .\]
But
\[M_{1  \otimes  \frac{1}{2}} \neq M_{\frac{3}{2}  \otimes  0} \quad \text{and} \quad M_{1  \otimes  1} \neq M_{2  \otimes  0} \ .\]
This is not surprising, as minimally coupled gauge-theories cannot have massive spin-$s > 1$. But there is no such restriction for the all-plus graviton Compton amplitude in eq.~\eqref{all_plus_five_point_graviton_Compton} as all the spin-dependence is factorized. Therefore,
\[M^{\textrm{all-plus}}_{s_L \otimes s_R} = M^{\textrm{all-plus}}_{s'_L \otimes s'_R} \ , \ \text{if } \ s_L +s_R = s'_L +s'_R \ ,  \]
for all-plus graviton Compton amplitudes, as long as $s_L, s_R, s'_L$ and $s'_R$ are non-zero integers.

%----------------------------------------------------------------------------

\section{Checking Gauge and Gravity Amplitudes using BCFW}\label{sec:bcfw-verification}

In this section we present the BCFW calculations of
\begin{itemize}
\item four-point QED and graviton Compton amplitudes with 
\begin{enumerate}
    \item one positive- and one negative-helicity massless leg; 
    \item two positive-helicity massless legs; and
\end{enumerate}

\item five-point QED and graviton Compton amplitudes with 
\begin{enumerate}
    \item two positive- and one negative-helicity massless leg; 
    \item three positive-helicity massless legs.
\end{enumerate}

\item Yang--Mills color-ordered amplitudes that are not directly calculated in sec.~\ref{sec:ym-compton-bcfw} but are derivable from the ones there via well-known relations.
\end{itemize}

These calculations give alternative forms of the amplitudes which do not exhibit manifest cancellation of spurious poles but serve as checks of the results in sec.~\ref{sec:ym-compton-bcfw}, sec.~\ref{sec:qed-compton} and sec.~\ref{sec:graviton-compton-klt}, as well as various amplitude relations including KK relations, reflection relations, BCJ relations and KLT relations. From now on, all the checks mentioned in this section will be performed via the analytic parametrization of spinors in sec.~\ref{app:analytic-parametrization}. For single-minus amplitudes, we will see the emergence of universal spin structures shared between gauge theory and gravity, in one-to-one correspondence with BCFW channels, if we calculate all amplitudes with the same $[3,2\rangle$ shift; this shift will be used throughout the section unless a different shift is mentioned. The amplitudes will take the form given in eq.~\eqref{eq:bcfw-spin-structure-decomposition}.

\subsection{Five-point YM-Compton Amplitudes Using BCFW}\label{subsec:ym-five-point-bcfw-checks}

Under the $[3,2\rangle$ BCFW shift in eq.~\eqref{eq:ym-five-point-12345-bcfw-shift}, for the amplitudes
\[A[\boldsymbol{1}^s,2^+,3^-,4^+,\boldsymbol{5}^s] \ ,
\quad
A[\boldsymbol{1}^s,3^-,2^+,4^+,\boldsymbol{5}^s] \ ,\]
the shifted legs are adjacent, resulting in two factorization channels corresponding to simple physical poles.

In contrast, for
\[A[2^+,\boldsymbol{1}^s,4^+,3^-,\boldsymbol{5}^s] \ ,
\qquad
A[3^-,\boldsymbol{1}^s,4^+,2^+,\boldsymbol{5}^s] \ ,\]
the shifted legs are non-adjacent, leading to four distinct factorization channels. While an alternative adjacent shift could reduce the computational effort, we deliberately employ the same shift in eq.~\eqref{eq:ym-five-point-12345-bcfw-shift} for all cases in order to expose the spin dependence of the amplitudes in a uniform manner.
% As mentioned in sec.~\ref{sec:introduction} \draftnote{(is it necessary to refer back to Introduction? RD: not rellay. I mentioned it as we had the spin structure and few other comments regarding it in the introduction. We can maybe just delete: "As mentioned in sec.~\ref{sec:introduction}".)}, we find that these different helicity amplitudes with two positive and one negative helicity gluons in sec.~\ref{subsubsec:ym-five-point-12345-single-minus} and sec.~\ref{subsubsec:ym-five-point-13245-bcfw-check}--\ref{subsubsec:ym-five-point-31425-bcfw-check} depend on the spin structure $Z_j$-s in eq.~\eqref{eq:bcfw-channel-spin-structures}. Here, $j$ labels the BCFW channel. Just to avoid cluttering we will avoid the hats on shifted legs: $2$ and $3$. So, for example, $Z_{12}$ will represent the spin structure coming from BCFW-channel-$1\hat{2}$.
The $Z_j$ spin structures for all channels $j$ are given below,
\begin{align}\label{eq:bcfw-channel-spin-structures}
    &\left\{
\begin{aligned}
Z_{12}
&= \frac{(\langle \boldsymbol{1} 3\rangle  [\boldsymbol{5} 4]+[\boldsymbol{1} 4] \langle \boldsymbol{5} 3\rangle ) \langle 3|\boldsymbol{1}|2]+\langle 3|2|4] [\boldsymbol{1} 2] \langle \boldsymbol{5} 3\rangle }{\langle 3|-\boldsymbol{5}|4] \langle 3|\boldsymbol{1}|2]}\ , \\[6pt]
Z_{25}
&= \frac{(\langle 3 \boldsymbol{5}\rangle  [\boldsymbol{1} 4]+[\boldsymbol{5} 4] \langle 3 \boldsymbol{1}\rangle ) \langle 3|\boldsymbol{5}|2]+\langle 3|2|4] [\boldsymbol{5} 2] \langle 3 \boldsymbol{1}\rangle }{\langle 3|\boldsymbol{1}|4] \langle 3|-\boldsymbol{5}|2]}\ , \\[6pt]
Z_{24}
&=- \frac{\langle \boldsymbol{5} 3\rangle  ([\boldsymbol{1}|2|3\rangle +[\boldsymbol{1}|4|3\rangle )+\langle \boldsymbol{1} 3\rangle  ([\boldsymbol{5}|2|3\rangle +[\boldsymbol{5}|4|3\rangle )}{\langle 3|\boldsymbol{1}\boldsymbol{5}|3\rangle }\ , \\[6pt]
Z_{34}
&= \frac{\langle \boldsymbol{1}\boldsymbol{5} \rangle }{m}\ , \\[6pt]
Z_{13}
&= -\frac{\langle \boldsymbol{5}\boldsymbol{1} \rangle  \langle 3|\boldsymbol{1}|2]-\langle 3 \boldsymbol{5}\rangle  \langle 3 \boldsymbol{1}\rangle  [3 2]}{m \langle 3|\boldsymbol{1}|2]}\ ,\\[6pt]
Z_{35}
&= -\frac{\langle \boldsymbol{5}\boldsymbol{1} \rangle  \langle 3|\boldsymbol{5}|2]+\langle 3 \boldsymbol{5}\rangle  \langle 3\boldsymbol{1} \rangle  [3 2]}{m \langle 3|\boldsymbol{5}|2]}\ ,
\end{aligned}
\right.
\end{align}
where the legs are $\boldsymbol 1^s, 2^+, 3^-, 4^+, \boldsymbol 5^s$, under various color orderings. The spin structures for opposite helicity configurations can be obtained utilizing charge conjugation. 

%We also analyze the spurious poles that appear due to the BCFW shifts in $A[\boldsymbol{1}^s,2^+,3^-,4^+,\boldsymbol{5}^s]$ and $A[\boldsymbol{1}^s,3^-,2^+,4^+,\boldsymbol{5}^s]$ in sec.~\ref{subsubsec:ym-five-point-12345-single-minus} and sec.~\ref{subsubsec:ym-five-point-13245-bcfw-check}. \draftnote{Do we really re-analyze the spurious poles of Section 3? RD: We don't need to keep this whole paragraph here anymore.} This is enough to probe the spurious pole structure of YM-Compton amplitudes with all other orderings and graviton Compton amplitudes in sec.~\ref{subsec:graviton-three-point-compton} utilizing double copy relations.

In sec.~\ref{subsubsec:ym-five-point-13245-bcfw-check}, we re-calculate $A\left[\mathbf{1}^{s}, 3^{-} , 2^{+}, 4^{+}, \mathbf{5}^{s}\right]$, already calculated in sec.~\ref{subsubsec:ym-five-point-13245-single-minus}, but with the $[3,2\rangle$ shift this time. Then we go on to calculate two other orderings, $A[2^+,\boldsymbol{1}^s,4^+,3^-,\boldsymbol{5}^s]$ and $A[3^-,\boldsymbol{1}^s,4^+,2^+,\boldsymbol{5}^s]$, in sec.~\ref{subsubsec:ym-five-point-21435-bcfw-check}--\ref{subsubsec:ym-five-point-31425-bcfw-check} with the same shift.

\subsubsection{
\texorpdfstring{$A\left[\mathbf{1}^{s}, 3^{-} , 2^{+}, 4^{+}, \mathbf{5}^{s}\right]$}{}}\label{subsubsec:ym-five-point-13245-bcfw-check}

Under the $[3,2\rangle$ shift, $A\left[\mathbf{1}^{s}, 3^{-}, 2^{+}, 4^{+}, \mathbf{5}^{s}\right]$ decomposes into two different terms/cuts.
\begin{equation}\label{eq:ym-five-point-13245-check-decomposition}
\boxed{A\left[\mathbf{1}^{s}, 3^{-}, 2^{+}, 4^{+},  \mathbf{5}^{s}\right]=A\left[\mathbf{1}^{s}, \hat{3}^{-} | \hat{2}^{+}, 4^{+}, \mathbf{5}^{s}\right]+A\left[\mathbf{5}^{s}, \mathbf{1}^{s}, \hat{3}^{-} | \hat{2}^{+}, 4^{+}\right]} \ .
\end{equation}

\paragraph{Channel 1: $A\left[\mathbf{1}^{s}, \hat{3}^{-} \mid \hat{2}^{+}, 4^{+}, \mathbf{5}^{s}\right]$} We calculate $A\left[\mathbf{1}^{s}, \hat{3}^{-} \mid \hat{2}^{+}, 4^{+}, \mathbf{5}^{s}\right]$. In fig.~\ref{fig:ym-five-point-13245-check-channel-13},
$$
\tau_{1 \hat{3}}=0 \Longrightarrow z=-\frac{\tau_{13}}{\langle 3| \mathbf{1} | 2]} \ .
$$
\begin{figure}[H]
    \centering
    \begin{fmffile}{3_2_Alt_d01}
 \begin{fmfgraph*}(200,60)% units are now in cm
   \fmfleft{i1,i2}
   \fmfright{o1,o2,o3}
   \fmfblob{.08w}{g1,g2}
   \fmf{gluon}{i2,g1}
   \fmfv{label=$\hat{3}^-$}{i2}
   \fmf{plain}{i1,g1}
   \fmfv{label=$\boldsymbol{1}^s$}{i1}
   \fmf{plain, label=$\hat{\boldsymbol{P}}^s$}{g1,v1}
   \fmf{phantom}{v1,v2}
   \fmf{plain, label=-$\hat{\boldsymbol{P}}^s$}{v2,g2}
   \fmf{gluon}{o3,g2}
   \fmfv{label=$\hat{2}^+$}{o3}
   \fmf{gluon}{o2,g2}
   \fmfv{label=$4^+$}{o2}
   \fmf{plain}{g2,o1}
   \fmfv{label=$\boldsymbol{5}^s$}{o1}
\end{fmfgraph*}
\end{fmffile}
\vspace{5mm}
    \caption{Factorization channel contributing to $A\left[\mathbf{1}^{s}, \hat{3}^{-} | \hat{2}^{+}, 4^{+}, \mathbf{5}^{s}\right]$}
    \label{fig:ym-five-point-13245-check-channel-13}
\end{figure}

\begin{align}
\nonumber& A\left[\mathbf{1}^{s}, \hat{3}^{-} \mid \hat{2}^{+}, 4^{+}, \mathbf{5}^{s}\right]=A\left[\mathbf{1}^{s}, \hat{3}^{-}, \hat{\boldsymbol{P}}^{s}\right] \frac{i}{\tau_{35}} A\left[-\hat{\boldsymbol{P}}^{s}, \hat{2}^{+}, 4^{+},, \mathbf{5}^{s}\right] \\
= & i\,\frac{\left.m^{2}[24]\langle 3| \mathbf{1} \mid 2\right]^{2}}{\tau_{13} \tau_{54}[23]\langle 3| \mathbf{1}(2+3)|4\rangle}\left(-\frac{\langle\mathbf{5} \mathbf{1}\rangle\langle 3| \mathbf{1} \mid 2]-\langle 3 \mathbf{5}\rangle\langle 3 \mathbf{1}\rangle[32]}{m\langle 3| \mathbf{1} \mid 2]}\right)^{2 s} \ . \label{eq:ym-five-point-13245-check-channel-13}
\end{align}

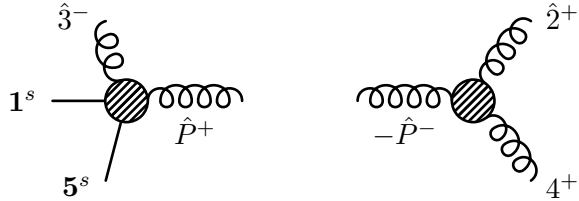
\begin{figure}[H]
\centering
    \begin{fmffile}{3_2_Alt_d02}
 \begin{fmfgraph*}(200,60)% units are now in cm
  \fmfleft{o2,i1,o1}
   \fmfright{i2,o4}
   \fmfblob{.08w}{g1,g2}
   \fmf{gluon}{o1,g2}
   \fmfv{label=$\hat{3}^-$}{o1}
   \fmf{gluon}{o4,g1}
   \fmfv{label=$\hat{2}^+$}{o4}
   \fmf{gluon, label=$-\hat{P}^-$,label.side=left}{g1,v1}
   \fmf{phantom}{v1,v2}
   \fmf{gluon, label=$\hat{P}^+$,label.side=left}{v2,g2}
   \fmf{gluon}{i2,g1}
   \fmfv{label=$4^+$}{i2}
   \fmf{plain}{g2,o2}
   \fmfv{label=$\boldsymbol{5}^s$}{o2}
   \fmf{plain}{g2,i1}
   \fmfv{label=$\boldsymbol{1}^s$}{i1}
\end{fmfgraph*}
\end{fmffile}
\vspace{5mm}
    \caption{Factorization channel contributing to $A\left[\mathbf{5}^{s}, \mathbf{1}^{s}, \hat{3}^{-} |\hat{2}^{+}, 4^{+}\right]$}
    \label{fig:ym-five-point-13245-check-channel-24}
\end{figure}

\paragraph{Channel 2:  $A\left[\mathbf{5}^{s}, \mathbf{1}^{s}, \hat{3}^{-} |\hat{2}^{+}, 4^{+}\right]$} Now, we calculate the contribution of $A\left[\mathbf{5}^{s}, \mathbf{1}^{s}, \hat{3}^{-} \mid \hat{2}^{+}, 4^{+}\right]$. In fig.~\ref{fig:ym-five-point-13245-check-channel-24},
$$
s_{2 \hat{4}}=0 \Longrightarrow z=\frac{\langle 42\rangle}{\langle 43\rangle}, \quad \text {when }[24] \neq 0 \ .
$$
That is why the sub-amplitude on the right in fig.~\ref{fig:ym-five-point-13245-check-channel-24} is anti-MHV.
\begin{align}
\nonumber& A\left[\mathbf{5}^{s}, \mathbf{1}^{s}, \hat{3}^{-} \mid \hat{2}^{+}, 4^{+}\right]=A\left[\mathbf{5}^{s}, \mathbf{1}^{s}, \hat{3}^{-}, \hat{P}^{-}\right] \frac{i}{s_{24}} A\left[-\hat{P}^{+}, \hat{2}^{+}, 4^{+}\right] \\
= & i\,\frac{[24]\langle 3| \mathbf{5 1}|3\rangle^{2}}{s_{15} s_{24}\langle 23\rangle\langle 3| \mathbf{1}(2+3)|4\rangle}\left(-\frac{\langle\mathbf{5} 3\rangle([\mathbf{1}|2| 3\rangle+[\mathbf{1}|4| 3\rangle)+\langle\mathbf{1} 3\rangle([\mathbf{5}|2| 3\rangle+[\mathbf{5}|4| 3\rangle)}{\langle 3| \mathbf{1 5}|3\rangle}\right)^{2 s} \ . \label{eq:ym-five-point-13245-check-channel-24}
\end{align}

According to eq.~\eqref{eq:ym-five-point-13245-check-decomposition}, adding the contribution from eqs.~\eqref{eq:ym-five-point-13245-check-channel-13} and \eqref{eq:ym-five-point-13245-check-channel-24}, we get the full amplitude:
\begin{equation}\label{eq:ym-five-point-13245-check-amplitude}
\boxed{A\left[\mathbf{1}^{s}, 3^{-}, 2^{+}, 4^{+}, \mathbf{5}^{s}\right]=A_{13}\left[\mathbf{1}^{0}, 3^{-}, 2^{+}, 4^{+}, \mathbf{5}^{0}\right] Z_{13}^{2 s}+A_{24}\left[\mathbf{1}^{0}, 3^{-}, 2^{+}, 4^{+}, \mathbf{5}^{0}\right] Z_{24}^{2 s} } \ ,
\end{equation}
\begin{flalign*}
\text{where}\quad
&\left\{\quad
\begin{aligned}
A_{13}\left[\mathbf{1}^{0}, 3^{-}, 2^{+}, 4^{+}, \mathbf{5}^{0}\right] &=\,i\,\frac{\left.m^{2}[24]\langle 3| \mathbf{1} \mid 2\right]^{2}}{\tau_{13} \tau_{54}[23]\langle 3| \mathbf{1}(2+3)|4\rangle} \ , \\[6pt]
A_{24}\left[\mathbf{1}^{0}, 3^{-}, 2^{+}, 4^{+}, \mathbf{5}^{0}\right] &=\,i\,\frac{[24]\langle 3| \mathbf{5 1}|3\rangle^{2}}{s_{15} s_{24}\langle 23\rangle\langle 3| \mathbf{1}(2+3)|4\rangle} \ ;
\end{aligned}
\right.&&\\
& & &
\end{flalign*}
and, $Z_{13}$ and $Z_{24}$ are given in eq.~\eqref{eq:bcfw-channel-spin-structures}. We have checked that eq.~\eqref{eq:ym-five-point-13245-check-amplitude} agrees with eq.~\eqref{eq:ym-five-point-13245-bcfw-amplitude}. As a reminder, in this section we do not derive amplitudes in a form that makes the cancellation of spurious poles manifest, and refer readers to secs.~\ref{sec:ym-compton-bcfw}, \ref{sec:qed-compton} and \ref{sec:graviton-compton-klt} for amplitudes that do exhibit this property.

\subsubsection{
\texorpdfstring{$A[2^+,\boldsymbol{1}^s,4^+,3^-,\boldsymbol{5}^s]$}{}}\label{subsubsec:ym-five-point-21435-bcfw-check}

Under the $[3,2\rangle$ shift, $A[2^+,\boldsymbol{1}^s,4^+,3^-,\boldsymbol{5}^s]$ decomposes into four cuts.
\begin{align}\label{eq:ym-five-point-21435-check-decomposition}
    &\boxed{\begin{aligned}
A[2^+,\boldsymbol{1}^s,4^+,3^-,\boldsymbol{5}^s]=& A[\hat{2}^+,\boldsymbol{1}^s|4^+,\hat{3}^-,\boldsymbol{5}^s]+A[\hat{2}^+,\boldsymbol{1}^s,4^+|\hat{3}^-,\boldsymbol{5}^s]\\
& \qquad\qquad +A[\boldsymbol{1}^s,4^+,\hat{3}^-|\boldsymbol{5}^s,\hat{2}^+]+A[\boldsymbol{5}^s,\hat{2}^+,\boldsymbol{1}^s|4^+,\hat{3}^-]
\end{aligned}} \ .
\end{align}

\paragraph{Channel 1: $A[\hat{2}^+,\boldsymbol{1}^s|4^+,\hat{3}^-,\boldsymbol{5}^s]$}

We calculate the first term in eq.~\eqref{eq:ym-five-point-21435-check-decomposition}. In fig.~\ref{fig:ym-five-point-21435-check-channel-12}, 
\begin{flalign*}
    \tau_{1\hat{2}}=0 \implies z=\frac{\tau_{12}}{\langle3|\boldsymbol{1}|2]}\ .
\end{flalign*}
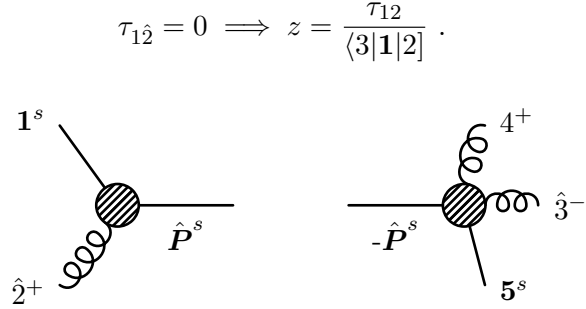
\begin{figure}[H]
    \centering
    \begin{fmffile}{3_3_d01}
 \begin{fmfgraph*}(200,60)% units are now in cm
   \fmfleft{i1,i2}
   \fmfright{o1,o2,o3}
   \fmfblob{.08w}{g1,g2}
   \fmf{gluon}{i1,g1}
   \fmfv{label=$\hat{2}^+$}{i1}
   \fmf{plain}{i2,g1}
   \fmfv{label=$\boldsymbol{1}^s$}{i2}
   \fmf{plain, label=$\hat{\boldsymbol{P}}^s$}{g1,v1}
   \fmf{phantom}{v1,v2}
   \fmf{plain, label=-$\hat{\boldsymbol{P}}^s$}{v2,g2}
   \fmf{gluon}{o3,g2}
   \fmfv{label=$4^+$}{o3}
   \fmf{gluon}{o2,g2}
   \fmfv{label=$\hat{3}^-$}{o2}
   \fmf{plain}{g2,o1}
   \fmfv{label=$\boldsymbol{5}^s$}{o1}
\end{fmfgraph*}
\end{fmffile}
\vspace{5mm}
    \caption{Factorization channel contributing to $A[\hat{2}^+,\boldsymbol{1}^s|4^+,\hat{3}^-,\boldsymbol{5}^s]$}
    \label{fig:ym-five-point-21435-check-channel-12}
\end{figure}
\begin{align}
    &A[\hat{2}^+,\boldsymbol{1}^s|4^+,\hat{3}^-,\boldsymbol{5}^s] = A[\boldsymbol{1}^s,  \hat{\boldsymbol{P}}^s, 2^+] \frac{i}{\tau_{53}} A[-\hat{\boldsymbol{P}}^s,\hat{4}^+, \hat{3}^-, \boldsymbol{5}^s] \nonumber\\
    %=\nonumber&\frac{\langle 3 |\boldsymbol{1}|2]^3 \langle 3 |\boldsymbol{5}|4]^2 }{\tau _{12} \langle 2  3\rangle  \left(\tau _{12} \langle 3 |4|2]+s_{34} \langle 3 |\boldsymbol{1}|2]\right) \left(\tau _{12} \langle 3 |\boldsymbol{5}|2]+\tau _{53} \langle 3 |\boldsymbol{1}|2]\right)} \left(-Z_{12}\right)^{2 s}\\
    =&-i\,\frac{\langle 3 |\boldsymbol{1}|2]^3 \langle 3 |\boldsymbol{5}|4]^2 }{\tau _{12} \langle 2  3\rangle \langle 3 4 \rangle [2|\boldsymbol{1}(2+3)|4] \langle 3|\boldsymbol{5}(2+3)\boldsymbol{1}|2 ]} \left(\frac{(\langle \boldsymbol{1} 3\rangle  [\boldsymbol{5} 4]+[\boldsymbol{1} 4] \langle \boldsymbol{5} 3\rangle ) \langle 3|\boldsymbol{1}|2]+\langle 3|2|4] [\boldsymbol{1} 2] \langle \boldsymbol{5} 3\rangle }{\langle 3|-\boldsymbol{5}|4] \langle 3|\boldsymbol{1}|2]}\right)^{2 s} \ . \label{eq:ym-five-point-21435-check-channel-12}
\end{align}

\paragraph{Channel 2: $A[\hat{2}^+,\boldsymbol{1}^s,4^+|\hat{3}^-,\boldsymbol{5}^s]$}

Now, we calculate the second term in eq.~\eqref{eq:ym-five-point-21435-check-decomposition}. In fig.~\ref{fig:ym-five-point-21435-check-channel-35}, 
\begin{flalign*}
    \tau_{5\hat{3}}=0 \implies z=-\frac{\tau_{53}}{\langle3|\boldsymbol{5}|2]} \ .
\end{flalign*}
\begin{figure}[H]
\centering
    \begin{fmffile}{3_3_d02}
 \begin{fmfgraph*}(200,60)% units are now in cm
   \fmfleft{i1,o1,o2}
   \fmfright{i2,o4}
   \fmfblob{.08w}{g1,g2}
   \fmf{gluon}{i1,g2}
   \fmfv{label=$\hat{2}^+$}{i1}
   \fmf{plain}{i2,g1}
   \fmfv{label=$\boldsymbol{5}^s$}{i2}
   \fmf{plain, label=$-\hat{\boldsymbol{P}}^s$}{g1,v1}
   \fmf{phantom}{v1,v2}
   \fmf{plain, label=$\hat{\boldsymbol{P}}^s$}{v2,g2}
   \fmf{gluon}{o4,g1}
   \fmfv{label=$\hat{3}^-$}{o4}
   \fmf{gluon}{o2,g2}
   \fmfv{label=$4^+$}{o2}
   \fmf{plain}{o1,g2}
   \fmfv{label=$\boldsymbol{1}^s$}{o1}
\end{fmfgraph*}
\end{fmffile}
\vspace{5mm}
    \caption{Factorization channel contributing to $A[\hat{2}^+,\boldsymbol{1}^s,4^+|\hat{3}^-,\boldsymbol{5}^s]$}
    \label{fig:ym-five-point-21435-check-channel-35}
\end{figure}
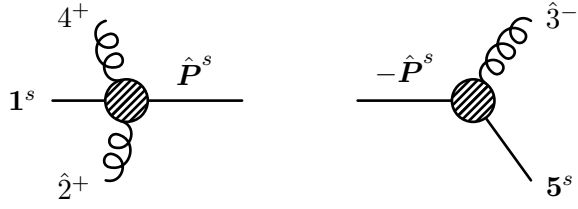
\begin{align}
    &A[\hat{2}^+,\boldsymbol{1}^s,4^+|\hat{3}^-,\boldsymbol{5}^s]= A[\hat{2}^+,\boldsymbol{1}^s,\hat{4}^+,P^s] \frac{i}{\tau_{14}} A[-P^s,\hat{3}^-,\boldsymbol{5}^s] \nonumber\\
    %=\nonumber& -\frac{m^2 [2  4]^2 \langle 3 |\boldsymbol{5}|2]^2 }{\tau _{14} \tau _{53} [2  3] \left(\tau _{12} \langle 3 |\boldsymbol{5}|2]+\tau _{53} \langle 3 |\boldsymbol{1}|2]\right)}\left(-Z_{35}\right)^{2 s}\\
    =& -i\,\frac{m^2 [2  4]^2 \langle 3 |\boldsymbol{5}|2]^2 }{\tau _{14} \tau _{53} [2  3] \langle 3|\boldsymbol{5}(2+3)\boldsymbol{1}|2 ]}\left(-\frac{\langle \boldsymbol{5}\boldsymbol{1} \rangle  \langle 3|\boldsymbol{5}|2]+\langle 3 \boldsymbol{5}\rangle  \langle 3\boldsymbol{1} \rangle  [3 2]}{m \langle 3|\boldsymbol{5}|2]}\right)^{2 s} \ . \label{eq:ym-five-point-21435-check-channel-35}
\end{align}

\paragraph{Channel 3: $A[\boldsymbol{1}^s,4^+,\hat{3}^-|\boldsymbol{5}^s,\hat{2}^+]$}

Now, we calculate the third term in eq.~\eqref{eq:ym-five-point-21435-check-decomposition}. In fig.~\ref{fig:ym-five-point-21435-check-channel-25}, 
\begin{flalign*}
    \tau_{5\hat{2}}=0 \implies z=\frac{\tau_{52}}{\langle3|\boldsymbol{5}|2]}\ .
\end{flalign*}
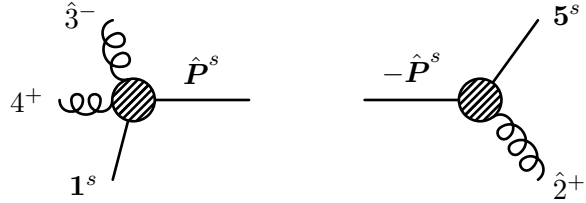
\begin{figure}[H]
    \centering
    \begin{fmffile}{3_3_d03}
 \begin{fmfgraph*}(200,60)% units are now in cm
   \fmfleft{o1,o2,o3}
   \fmfright{i1,i2}
   \fmfblob{.08w}{g1,g2}
   \fmf{gluon}{i1,g1}
   \fmfv{label=$\hat{2}^+$}{i1}
   \fmf{plain}{g1,i2}
   \fmfv{label=$\boldsymbol{5}^s$}{i2}
   \fmf{plain, label=$-\hat{\boldsymbol{P}}^s$}{g1,v1}
   \fmf{phantom}{v1,v2}
   \fmf{plain, label=$\hat{\boldsymbol{P}}^s$}{v2,g2}
   \fmf{gluon}{o3,g2}
   \fmfv{label=$\hat{3}^-$}{o3}
   \fmf{gluon}{o2,g2}
   \fmfv{label=$4^+$}{o2}
   \fmf{plain}{g2,o1}
   \fmfv{label=$\boldsymbol{1}^s$}{o1}
\end{fmfgraph*}
\end{fmffile}
\vspace{5mm}
    \caption{Factorization channel contributing to $A[\boldsymbol{1}^s,4^+,\hat{3}^-|\boldsymbol{5}^s,\hat{2}^+]$}
    \label{fig:ym-five-point-21435-check-channel-25}
\end{figure}
\begin{align}
    &A[\boldsymbol{1}^s,4^+,\hat{3}^-|\boldsymbol{5}^s,\hat{2}^+]= A[\boldsymbol{1}^s,\hat{4}^+,\hat{3}^-,P^s] \frac{i}{\tau_{14}} A[-P^s,\boldsymbol{5}^s,\hat{2}^+] \nonumber\\
    %=\nonumber& -\frac{\langle 3 |\boldsymbol{1}|4]^2 \langle 3 |\boldsymbol{5}|2]^2 }{\tau _{14} \tau _{52} \langle 2  3\rangle  \left(s_{34} \langle 3 |\boldsymbol{5}|2]+\tau _{52} \langle 3 |4|2]\right)}\left(-Z_{25}\right)^{2 s}\\
   =& i\,\frac{\langle 3 |\boldsymbol{1}|4]^2 \langle 3 |\boldsymbol{5}|2]^2 }{\tau _{14} \tau _{52} \langle 2  3\rangle  \langle 3 4 \rangle [2|\boldsymbol{1}(2+3)|4]}\left(\frac{(\langle 3 \boldsymbol{5}\rangle  [\boldsymbol{1} 4]+[\boldsymbol{5} 4] \langle 3 \boldsymbol{1}\rangle ) \langle 3|\boldsymbol{5}|2]+\langle 3|2|4] [\boldsymbol{5} 2] \langle 3 \boldsymbol{1}\rangle }{\langle 3|\boldsymbol{1}|4] \langle 3|-\boldsymbol{5}|2]}\right)^{2 s} \ . \label{eq:ym-five-point-21435-check-channel-25}
\end{align}

\paragraph{Channel 4: $A[\boldsymbol{5}^s,\hat{2}^+,\boldsymbol{1}^s|4^+,\hat{3}^-]$}

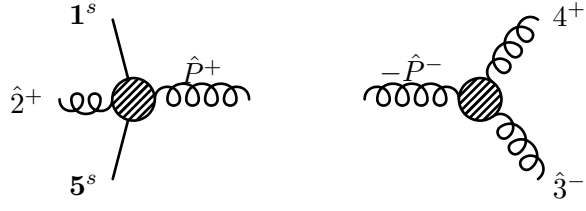
\begin{figure}[H]
    \centering
    \begin{fmffile}{3_3_d04}
 \begin{fmfgraph*}(200,60)% units are now in cm
   \fmfleft{o2,i1,o1}
   \fmfright{i2,o4}
   \fmfblob{.08w}{g1,g2}
   \fmf{gluon}{i1,g2}
   \fmfv{label=$\hat{2}^+$}{i1}
   \fmf{gluon}{o4,g1}
   \fmfv{label=$4^+$}{o4}
   \fmf{gluon, label=$-\hat{P}^-$}{g1,v1}
   \fmf{phantom}{v1,v2}
   \fmf{gluon, label=$\hat{P}^+$}{v2,g2}
   \fmf{gluon}{i2,g1}
   \fmfv{label=$\hat{3}^-$}{i2}
   \fmf{plain}{g2,o2}
   \fmfv{label=$\boldsymbol{5}^s$}{o2}
   \fmf{plain}{g2,o1}
   \fmfv{label=$\boldsymbol{1}^s$}{o1}
\end{fmfgraph*}
\end{fmffile}
\vspace{5mm}
    \caption{Factorization channel contributing to $A[\boldsymbol{5}^s,\hat{2}^+,\boldsymbol{1}^s|4^+,\hat{3}^-]$}
    \label{fig:ym-five-point-21435-check-channel-34}
\end{figure}

Now, we calculate the last term in eq.~\eqref{eq:ym-five-point-21435-check-decomposition}. In fig.~\ref{fig:ym-five-point-21435-check-channel-34}, the kinematics is as follows: $[\hat{3}4]=0$. Hence, the 3-point sub-amplitude is MHV and $z=-\frac{[42]}{[43]}$.
\begin{align}
    &A[\boldsymbol{5}^s,\hat{2}^+,\boldsymbol{1}^s|4^+,\hat{3}^-]= A[\boldsymbol{5}^s,\hat{2}^+,\boldsymbol{1}^s,\hat{P}^+] \frac{i}{\tau_{14}} A[\hat{P}^-,,\hat{4}^+,\hat{3}^-] \nonumber\\
    %=\nonumber& \frac{m^2  \langle 3 |4|2]^5}{s_{34} [2  3] \langle 3  4\rangle ^2 \left(\tau _{12} \langle 3 |4|2]+s_{34} \langle 3 |\boldsymbol{1}|2]\right) \left(s_{34} \langle 3 |\boldsymbol{5}|2]+\tau _{52} \langle 3 |4|2]\right)}\left(\frac{\langle \boldsymbol{ 1 5 }\rangle }{m}\right)^{2 s}\\
    =& i\,\frac{m^2  [24]^5}{[2 3] [3 4] [2|\boldsymbol{1}(2+3)|4]  [2|\boldsymbol{5}(2+3)|4]}\left(\frac{\langle \boldsymbol{1 5}\rangle }{m}\right)^{2 s} \ . \label{eq:ym-five-point-21435-check-channel-34}
\end{align}

According to eq.~\eqref{eq:ym-five-point-21435-check-decomposition}, adding the contribution from fig.~\ref{fig:ym-five-point-21435-check-channel-12}--\ref{fig:ym-five-point-21435-check-channel-34}, we get the full amplitude:
\begin{align}
    &\boxed{\begin{aligned}
    \nonumber A[2^+,\boldsymbol{1}^s,4^+,3^-,\boldsymbol{5}^s]=& A_{12}[2^+,\boldsymbol{1}^0,4^+,3^-,\boldsymbol{5}^0]\:Z_{12}^{2s} + A_{35}[2^+,\boldsymbol{1}^s,4^+,3^-,\boldsymbol{5}^s]\:Z_{35}^{2s} \\
    &\quad+ A_{25}[2^+,\boldsymbol{1}^0,4^+,3^-,\boldsymbol{5}^0]\:Z_{25}^{2s} + A_{34}[2^+,\boldsymbol{1}^0,4^+,3^-,\boldsymbol{5}^0]\:Z_{34}^{2s}
\end{aligned} }\ ,\\
    &\label{eq:ym-five-point-21435-check-amplitude}
\end{align}
\begin{flalign*}
\text{where}\quad
&\left\{\quad
\begin{aligned}
A_{12}[2^+,\boldsymbol{1}^0,4^+,3^-,\boldsymbol{5}^0]
&= %\frac{\langle 3 |\boldsymbol{1}|2]^3\langle 3 |\boldsymbol{5}|4]^2}{\tau_{12}\,\langle 2  3\rangle\left(\tau_{12} \langle 3 |4|2]+s_{43} \langle 3 |\boldsymbol{1}|2]\right)\left(\tau_{12} \langle 3 |\boldsymbol{5}|2]+\tau_{53} \langle 3 |\boldsymbol{1}|2]\right)}
-i\,\frac{\langle 3 |\boldsymbol{1}|2]^3 \langle 3 |\boldsymbol{5}|4]^2 }{\tau _{12} \langle 2  3\rangle \langle 3 4 \rangle [2|\boldsymbol{1}(2+3)|4] \langle 3|\boldsymbol{5}(2+3)\boldsymbol{1}|2 ]} \ , \\[6pt]
A_{35}[2^+,\boldsymbol{1}^0,4^+,3^-,\boldsymbol{5}^0]
&= %-\frac{m^2 [2  4]^2 \langle 3 |\boldsymbol{5}|2]^2 }{ \tau_{14}\,\tau_{53}\,[2  3] \left( \tau_{12} \langle 3 |\boldsymbol{5}|2]+\tau_{53} \langle 3 |\boldsymbol{1}|2] \right)}
-i\,\frac{m^2 [2  4]^2 \langle 3 |\boldsymbol{5}|2]^2 }{\tau _{14} \tau _{53} [2  3] \langle 3|\boldsymbol{5}(2+3)\boldsymbol{1}|2 ]} \ , \\[6pt]
A_{25}[2^+,\boldsymbol{1}^0,4^+,3^-,\boldsymbol{5}^0]
&= %-\frac{\langle 3 |\boldsymbol{1}|4]^2 \langle 3 |\boldsymbol{5}|2]^2 }{ \tau_{14}\,\tau_{52}\, \langle 2  3\rangle \left( s_{43} \langle 3 |\boldsymbol{5}|2] +\tau_{52} \langle 3 |4|2] \right) }
i\,\frac{\langle 3 |\boldsymbol{1}|4]^2 \langle 3 |\boldsymbol{5}|2]^2 }{\tau _{14} \tau _{52} \langle 2  3\rangle  \langle 3 4 \rangle [2|\boldsymbol{5}(2+3)|4]} \ , \\[6pt]
A_{34}[2^+,\boldsymbol{1}^0,4^+,3^-,\boldsymbol{5}^0]
&= %\frac{ m^2 \langle 3 |4|2]^5 }{ s_{34}\,[2  3]\, \langle 3  4\rangle^2 \left( \tau_{12} \langle 3 |4|2] +s_{43} \langle 3 |\boldsymbol{1}|2] \right) \left( s_{43} \langle 3 |\boldsymbol{5}|2] +\tau_{52} \langle 3 |4|2] \right) }
i\,\frac{m^2  [24]^5}{[2 3] [3 4] [2|\boldsymbol{1}(2+3)|4]  [2|\boldsymbol{5}(2+3)|4]} \ ;
\end{aligned}
\right.&&\\
& & &
%\\[10pt]
%\&\quad
%&\left\{
%\begin{aligned}
%Z_{12}
%&= \left(\frac{(\langle \boldsymbol{1} 3\rangle  [\boldsymbol{5} 4]+[\boldsymbol{1} 4] \langle -\boldsymbol{5} 3\rangle ) \langle 3|\boldsymbol{1}|2]+\langle 3|2|4] [\boldsymbol{1} 2] \langle \boldsymbol{5} 3\rangle }{\langle 3|-\boldsymbol{5}|4] \langle 3|\boldsymbol{1}|2]}\right), \\[6pt]
%Z_{35}
%&= \left(\frac{\langle \boldsymbol{1} \boldsymbol{5}\rangle  \langle 3|\boldsymbol{5}|2]+\langle 3 \boldsymbol{5}\rangle  \langle \boldsymbol{1} 3\rangle  [3 2]}{m \langle 3|\boldsymbol{5}|2]}\right), \\[6pt]
%Z_{25}
%&= \left(\frac{(\langle 3 \boldsymbol{5}\rangle  [\boldsymbol{1} 4]+[\boldsymbol{5} 4] \langle 3 \boldsymbol{1}\rangle ) \langle 3|\boldsymbol{5}|2]+\langle 3|2|4] [\boldsymbol{5} 2] \langle 3 \boldsymbol{1}\rangle }{\langle 3|\boldsymbol{1}|4] \langle 3|-\boldsymbol{5}|2]}\right), \\[6pt]
%Z_{34}
%&= \left(\frac{\langle  \boldsymbol{1}\boldsymbol{5} \rangle}{m}\right).
%\end{aligned}
%\right.& &
\end{flalign*}
and, $Z_{12}$, $Z_{35}$, $Z_{25}$ and $Z_{34}$ have the form mentioned in eq.~\eqref{eq:bcfw-channel-spin-structures}.

Spurious singularities are present in eq.~\eqref{eq:ym-five-point-21435-check-amplitude}. We can use KK relations and reflection symmetry to obtain an expression free of spurious singularities,
\begin{align}
    A[2,\boldsymbol{1}^s,4,3,\boldsymbol{5}^s]=A[\boldsymbol{1}^s,4,3,2,\boldsymbol{5}^s] - (-1)^{2s} A[\boldsymbol{5}^s,3,4,2,\boldsymbol{1}^s] - (-1)^{2s} A[\boldsymbol{5}^s,3,2,4,\boldsymbol{1}^s]\ ,
\end{align}
or using the BCJ relations in eq.~\eqref{eq:bcj-relation-ordering-21435}. We have checked that these different forms of the amplitude agree with each other.

\subsubsection{
\texorpdfstring{$A[3^-,\boldsymbol{1}^s,4^+,2^+,\boldsymbol{5}^s]$}
               {}
}
\label{subsubsec:ym-five-point-31425-bcfw-check}

Under the $[3,2\rangle$ shift, $A[\hat{3}^-,\boldsymbol{1}^s,4^+,\hat{2}^+,\boldsymbol{5}^s]$ decomposes into four cuts.
\begin{align}
\boxed{\begin{aligned}
A[3^-,\boldsymbol{1}^s,4^+,2^+,\boldsymbol{5}^s]\:=&\:A[\hat{3}^-,\boldsymbol{1}^s|\,4^+,\hat{2}^+,\boldsymbol{5}^s]\,+\,A[\hat{3}^-,\boldsymbol{1}^s,4^+|\,\hat{2}^+,\boldsymbol{5}^s]\\
&\qquad\qquad+\,A[\boldsymbol{5}^s,\hat{3}^-,\boldsymbol{1}^s|\,4^+,\hat{2}^+]\,+\,A[\boldsymbol{1}^s,4^+,\hat{2}^+|\,\boldsymbol{5}^s,\hat{3}^-]
\end{aligned}} \ . \label{eq:ym-five-point-31425-check-decomposition}
\end{align}

\paragraph{Channel 1: $A[\hat{3}^-,\boldsymbol{1}^s|\,4^+,\hat{2}^+,\boldsymbol{5}^s]$}

First, let us calculate the contribution of the first term in the RHS of eq.~\eqref{eq:ym-five-point-31425-check-decomposition}.
\begin{flalign*}
    & \text{In fig.~\ref{fig:ym-five-point-31425-check-channel-13}}, \quad\tau_{1\hat{3}}=0 \implies z=-\frac{\tau_{13}}{\langle3|\boldsymbol{1}|2]}\ .&&
\end{flalign*}
\begin{figure}[H]
    \centering
    \begin{fmffile}{3_4_d01}
 \begin{fmfgraph*}(200,60)% units are now in cm
   \fmfleft{i1,i2}
   \fmfright{o1,o2,o3}
   \fmfblob{.08w}{g1,g2}
   \fmf{gluon}{i1,g1}
   \fmfv{label=$\hat{3}^-$}{i1}
   \fmf{plain}{i2,g1}
   \fmfv{label=$\boldsymbol{1}^s$}{i2}
   \fmf{plain, label=$\hat{\boldsymbol{P}}^s$}{g1,v1}
   \fmf{phantom}{v1,v2}
   \fmf{plain, label=-$\hat{\boldsymbol{P}}^s$}{v2,g2}
   \fmf{gluon}{o3,g2}
   \fmfv{label=$4^+$}{o3}
   \fmf{gluon}{o2,g2}
   \fmfv{label=$\hat{2}^+$}{o2}
   \fmf{plain}{g2,o1}
   \fmfv{label=$\boldsymbol{5}^s$}{o1}
\end{fmfgraph*}
\end{fmffile}
\vspace{5mm}
    \caption{Factorization channel contributing to $A[\hat{3}^-,\boldsymbol{1}^s|\,4^+,\hat{2}^+,\boldsymbol{5}^s]$}
    \label{fig:ym-five-point-31425-check-channel-13}
\end{figure}
\begin{align}\label{eq:ym-five-point-31425-check-channel-13}
    &A[\hat{3}^-,\boldsymbol{1}^s|\,4^+,\hat{2}^+,\boldsymbol{5}^s] = A[\hat{3}^-, \boldsymbol{1}^s, \hat{\boldsymbol{P}}^s] \frac{i}{\tau_{13}} A[-\hat{\boldsymbol{P}}^s,4^+,\hat{2}^+,\boldsymbol{5}^s] \nonumber\\
    %=\nonumber&-\frac{m^2 [2  4]^2 \langle 3 |\boldsymbol{1}|2]^3 }{\tau _{13} [2  3] \left(\tau _{13} \langle 3 |4|2]+s_{24} \langle 3 |\boldsymbol{1}|2]\right) \left(\tau _{13} \langle 3 |\boldsymbol{5}|2]+\tau _{52} \langle 3 |\boldsymbol{1}|2]\right)} \left(-Z_{13}\right)^{2 s}\\
    =&i\,\frac{m^2 [2  4] \langle 3 |\boldsymbol{1}|2]^3 }{\tau _{13} [2  3] \langle 3|\boldsymbol{1}(2+3)|4\rangle \langle 3| \boldsymbol{1}(2+3)\boldsymbol{5}|2]} \left(-\frac{\langle \boldsymbol{5}\boldsymbol{1} \rangle  \langle 3|\boldsymbol{1}|2]-\langle 3 \boldsymbol{5}\rangle  \langle 3 \boldsymbol{1}\rangle  [3 2]}{m \langle 3|\boldsymbol{1}|2]}\right)^{2 s} \ .
\end{align}

\paragraph{Channel 2: $A[\hat{3}^-,\boldsymbol{1}^s,4^+|\,\hat{2}^+,\boldsymbol{5}^s]$}

Now, we calculate the contribution of the second term in the RHS of eq.~\eqref{eq:ym-five-point-31425-check-decomposition}.
\begin{flalign*}
    \text{In fig.~\ref{fig:ym-five-point-31425-check-channel-25}}, \quad\tau_{5\hat{2}}=0 \implies z=\frac{\tau_{52}}{\langle3|\boldsymbol{5}|2]}\ .&&
\end{flalign*}
\begin{figure}[H]
    \centering
    \begin{fmffile}{3_4_d02}
 \begin{fmfgraph*}(200,60)% units are now in cm
   \fmfleft{i1,i2,i3}
   \fmfright{o1,o2}
   \fmfblob{.08w}{g1,g2}
   \fmf{gluon}{i1,g1}
   \fmfv{label=$\hat{3}^-$}{i1}
   \fmf{plain}{i2,g1}
   \fmfv{label=$\boldsymbol{1}^s$}{i2}
   \fmf{gluon}{i3,g1}
   \fmfv{label=$4^+$}{i3}
   \fmf{plain, label=$\hat{\boldsymbol{P}}^s$}{g1,v1}
   \fmf{phantom}{v1,v2}
   \fmf{plain, label=-$\hat{\boldsymbol{P}}^s$}{v2,g2}
   \fmf{gluon}{o2,g2}
   \fmfv{label=$\hat{2}^+$}{o2}
   \fmf{plain}{g2,o1}
   \fmfv{label=$\boldsymbol{5}^s$}{o1}
\end{fmfgraph*}
\end{fmffile}
\vspace{5mm}
    \caption{Factorization channel contributing to $A[\hat{3}^-,\boldsymbol{1}^s,4^+|\,\hat{2}^+,\boldsymbol{5}^s]$}
    \label{fig:ym-five-point-31425-check-channel-25}
\end{figure}
\begin{align}
    & A[\hat{3}^-,\boldsymbol{1}^s,4^+|\,\hat{2}^+,\boldsymbol{5}^s] = A[\boldsymbol{1}^s,4^+,\hat{\boldsymbol{P}}^s,\hat{3}^-] \frac{i}{\tau_{52}} A[-\hat{\boldsymbol{P}}^s,\hat{2}^+,\boldsymbol{5}^s] \nonumber\\
    %=\nonumber&\frac{\langle 3 |\boldsymbol{1}|4]^2 \langle 3 |\boldsymbol{5}|2]^2 }{\tau _{14} \tau _{52} \langle 2  3\rangle  \left(\tau _{13} \langle 3 |\boldsymbol{5}|2]+\tau _{52} \langle 3 |\boldsymbol{1}|2]\right)} \left(-Z_{25}\right)^{2 s}\\
    =&i\,\frac{\langle 3 |\boldsymbol{1}|4]^2 \langle 3 |\boldsymbol{5}|2]^2 }{\tau _{14} \tau _{52} \langle 2  3\rangle  \langle 3| \boldsymbol{1}(2+3)\boldsymbol{5}|2]} \left(\frac{(\langle 3 \boldsymbol{5}\rangle  [\boldsymbol{1} 4]+[\boldsymbol{5} 4] \langle 3 \boldsymbol{1}\rangle ) \langle 3|\boldsymbol{5}|2]+\langle 3|2|4] [\boldsymbol{5} 2] \langle 3 \boldsymbol{1}\rangle }{\langle 3|\boldsymbol{1}|4] \langle 3|-\boldsymbol{5}|2]}\right)^{2 s} \ . \label{eq:ym-five-point-31425-check-channel-25}
\end{align}

\paragraph{Channel 3: $A[\boldsymbol{5}^s,\hat{3}^-,\boldsymbol{1}^s|\,4^+,\hat{2}^+]$}

Now, we calculate the contribution of the third term in the RHS of eq.~\eqref{eq:ym-five-point-31425-check-decomposition}.
\begin{flalign*}
    &\text{In fig.~\ref{fig:ym-five-point-31425-check-channel-24}},\quad s_{\hat{2}4}=0 \implies z=\frac{s_{24}}{\langle3|\boldsymbol{4}|2]}\ .&&
\end{flalign*}
\begin{figure}[H]
    \centering
    \begin{fmffile}{3_4_d03}
 \begin{fmfgraph*}(200,60)% units are now in cm
   \fmfleft{o1,i1,i2}
   \fmfright{o2,o3}
   \fmfblob{.08w}{g1,g2}
   \fmf{plain}{o1,g1}
   \fmfv{label=$\boldsymbol{5}^s$}{o1}
   \fmf{gluon}{i1,g1}
   \fmfv{label=$\hat{3}^-$}{i1}
   \fmf{plain}{i2,g1}
   \fmfv{label=$\boldsymbol{1}^s$}{i2}
   \fmf{gluon,label=$\hat{P}^+$,label.side=left}{v1,g1}
   \fmf{phantom}{v1,v2}
   \fmf{gluon, label=-$\hat{P}^-$,label.side=left}{g2,v2}
   \fmf{gluon}{o3,g2}
   \fmfv{label=$4^+$}{o3}
   \fmf{gluon}{o2,g2}
   \fmfv{label=$\hat{2}^+$}{o2}
\end{fmfgraph*}
\end{fmffile}
\vspace{5mm}
    \caption{Factorization channel contributing to $A[\boldsymbol{5}^s,\hat{3}^-,\boldsymbol{1}^s|\,4^+,\hat{2}^+]$}
    \label{fig:ym-five-point-31425-check-channel-24}
\end{figure}
\begin{align}
    \nonumber&A[\boldsymbol{5}^s,\hat{3}^-,\boldsymbol{1}^s|\,4^+,\hat{2}^+] =A[\boldsymbol{1}^s,\hat{P},\boldsymbol{5}^s,\hat3^-] \frac{i}{s_{24}} A[-\hat{P},4^+,\hat{2}^+] \nonumber\\
    %=\nonumber&\frac{[2  4]^3 \langle 3 |\boldsymbol{51}|3\rangle ^2 \langle 3 |4|2] }{s_{24} \langle 3 |2|4] \left(\tau _{13} \langle 3 |4|2]+s_{24} \langle 3 |\boldsymbol{1}|2]\right) \left(s_{24} \langle 3 |\boldsymbol{5}|2]+\tau _{53} \langle 3 |4|2]\right)}\left(-Z_{24}\right)^{2 s} \\
   =&i\,\frac{\langle 3 |\boldsymbol{51}|3\rangle ^2 \langle 3 4\rangle }{\langle 24 \rangle \langle 3 2\rangle\langle 3|\boldsymbol{1}(2+3)|4\rangle \langle 3|\boldsymbol{5}(2+3)|4\rangle}\left(-\frac{\langle \boldsymbol{5} 3\rangle  ([\boldsymbol{1}|2|3\rangle +[\boldsymbol{1}|4|3\rangle )+\langle \boldsymbol{1} 3\rangle  ([\boldsymbol{5}|2|3\rangle +[\boldsymbol{5}|4|3\rangle )}{\langle 3|\boldsymbol{1}\boldsymbol{5}|3\rangle }\right)^{2 s} \ . \label{eq:ym-five-point-31425-check-channel-24}
\end{align}

\paragraph{Channel 4: $A[\boldsymbol{1}^s,4^+,\hat{2}^+|\,\boldsymbol{5}^s,\hat{3}^-]$}

Now, we calculate the contribution of the last term on the right-hand side of eq.~\eqref{eq:ym-five-point-31425-check-decomposition}.
\begin{flalign*}
    &\text{In fig.~\ref{fig:ym-five-point-31425-check-channel-35}}, \quad\tau_{5\hat{3}}=0 \implies z=-\frac{\tau_{53}}{\langle3|\boldsymbol{5}|2]}\ . &&
\end{flalign*}
\begin{figure}[H]
    \centering
    \begin{fmffile}{3_4_d04}
 \begin{fmfgraph*}(200,60)% units are now in cm
   \fmfleft{o1,i1,i2}
   \fmfright{o2,o3}
   \fmfblob{.08w}{g1,g2}
   \fmf{plain}{o1,g1}
   \fmfv{label=$\boldsymbol{1}^s$}{o1}
   \fmf{gluon}{i1,g1}
   \fmfv{label=$2^+$}{i1}
   \fmf{gluon}{i2,g1}
   \fmfv{label=$\hat{4}^+$}{i2}
   \fmf{plain, label=$\hat{\boldsymbol{P}}^s$}{g1,v1}
   \fmf{phantom}{v1,v2}
   \fmf{plain, label=-$\hat{\boldsymbol{P}}^s$}{v2,g2}
   \fmf{plain}{g2,o3}
   \fmfv{label=$\boldsymbol{5}^s$}{o3}
   \fmf{gluon}{o2,g2}
   \fmfv{label=$\hat{3}^-$}{o2}
\end{fmfgraph*}
\end{fmffile}
\vspace{5mm}
    \caption{Factorization channel contributing to $A[\boldsymbol{1}^s,4^+,\hat{2}^+|\,\boldsymbol{5}^s,\hat{3}^-]$}
    \label{fig:ym-five-point-31425-check-channel-35}
\end{figure}
\begin{align}
    & A[\boldsymbol{1}^s,4^+,2^+|\,\boldsymbol{5}^s,3^-] = A[\boldsymbol{1}^s,4^+,2^+,\hat{\boldsymbol{P}}^s] \frac{i}{\tau_{53}} A[-\hat{\boldsymbol{P}}^s,\boldsymbol{5}^s,3^-] \nonumber\\
    %=\nonumber& \frac{m^2 [2  4]^2 \langle 3 |\boldsymbol{5}|2]^2 }{\tau _{14} \tau _{53} [2  3] \left(s_{24} \langle 3 |\boldsymbol{5}|2]+\tau _{53} \langle 3 |4|2]\right)}\left(-Z_{35}\right)^{2 s} \\
    =& -i\,\frac{m^2 [2  4] \langle 3 |\boldsymbol{5}|2]^2 }{\tau _{14} \tau _{53} [2  3] \langle 3|\boldsymbol{5}(2+3)|4\rangle}\left(-\frac{\langle \boldsymbol{5}\boldsymbol{1} \rangle  \langle 3|\boldsymbol{5}|2]+\langle 3 \boldsymbol{5}\rangle  \langle 3\boldsymbol{1} \rangle  [3 2]}{m \langle 3|\boldsymbol{5}|2]}\right)^{2 s} \ . \label{eq:ym-five-point-31425-check-channel-35}
\end{align}

According to eq.~\eqref{eq:ym-five-point-31425-check-decomposition}, adding the contribution from fig.~\ref{fig:ym-five-point-31425-check-channel-13}--\ref{fig:ym-five-point-31425-check-channel-35}, we get the full amplitude:
\begin{align}
    \nonumber&\boxed{
    \begin{aligned}
        A[3^-,\boldsymbol{1}^s,4^+,2^+,\boldsymbol{5}^s]=& A_{13}[3^-,\boldsymbol{1}^0,4^+,2^+,\boldsymbol{5}^0]\:Z_{13}^{2s} + A_{25}[3^-,\boldsymbol{1}^0,4^+,2^+,\boldsymbol{5}^0]\:Z_{25}^{2s} \nonumber\\
    &\qquad + A_{24}[3^-,\boldsymbol{1}^0,4^+,2^+,\boldsymbol{5}^0]\:Z_{24}^{2s} + A_{35}[3^-,\boldsymbol{1}^0,4^+,2^+,\boldsymbol{5}^0]\:Z_{35}^{2s}
    \end{aligned}
    }\ ,\\
    &\label{eq:ym-five-point-31425-check-amplitude}
\end{align}
\begin{flalign*}
\text{where }\quad &\left\{\quad
\begin{aligned}
    & A_{13}[3^-,\boldsymbol{1}^0,4^+,2^+,\boldsymbol{5}^0] = %-\frac{m^2 [2  4]^2 \langle 3 |\boldsymbol{1}|2]^3 }{\tau _{13} [2  3] \left(\tau _{13} \langle 3 |4|2]+s_{24} \langle 3 |\boldsymbol{1}|2]\right) \left(\tau _{13} \langle 3 |\boldsymbol{5}|2]+\tau _{52} \langle 3 |\boldsymbol{1}|2]\right)} 
    i\,\frac{m^2 [2  4] \langle 3 |\boldsymbol{1}|2]^3 }{\tau _{13} [2  3] \langle 3|\boldsymbol{1}(2+3)|4\rangle \langle 3| \boldsymbol{1}(2+3)\boldsymbol{5}|2]}\ , \\
    &  A_{25}[3^-,\boldsymbol{1}^0,4^+,2^+,\boldsymbol{5}^0]=%\frac{\langle 3 |\boldsymbol{1}|4]^2 \langle 3 |\boldsymbol{5}|2]^2 }{\tau _{14} \tau _{52} \langle 2  3\rangle  \left(\tau _{13} \langle 3 |\boldsymbol{5}|2]+\tau _{52} \langle 3 |\boldsymbol{1}|2]\right)} 
    i\,\frac{\langle 3 |\boldsymbol{1}|4]^2 \langle 3 |\boldsymbol{5}|2]^2 }{\tau _{14} \tau _{52} \langle 2  3\rangle  \langle 3| \boldsymbol{1}(2+3)\boldsymbol{5}|2]} \ ,\\
    &  A_{24}[3^-,\boldsymbol{1}^0,4^+,2^+,\boldsymbol{5}^0]=%\frac{[2  4]^3 \langle 3 |\boldsymbol{15}|3\rangle ^2 \langle 3 |4|2] }{s_{24} \langle 3 |2|4] \left(\tau _{13} \langle 3 |4|2]+s_{24} \langle 3 |\boldsymbol{1}|2]\right) \left(s_{24} \langle 3 |\boldsymbol{5}|2]+\tau _{53} \langle 3 |4|2]\right)}
    i\,\frac{\langle 3 |\boldsymbol{51}|3\rangle ^2 \langle 3 4\rangle }{\langle 24 \rangle \langle 3 2\rangle\langle 3|\boldsymbol{1}(2+3)|4\rangle \langle 3|\boldsymbol{5}(2+3)|4\rangle} \ , \\
   &  A_{35}[3^-,\boldsymbol{1}^0,4^+,2^+,\boldsymbol{5}^0]=%\frac{m^2 [2  4]^2 \langle 3 |\boldsymbol{5}|2]^2 }{\tau _{14} \tau _{53} [2  3] \left(s_{24} \langle 3 |\boldsymbol{5}|2]+\tau _{53} \langle 3 |4|2]\right)}
   -i\,\frac{m^2 [2  4] \langle 3 |\boldsymbol{5}|2]^2 }{\tau _{14} \tau _{53} [2  3] \langle 3|\boldsymbol{5}(2+3)|4\rangle} \ ; 
\end{aligned}\right.&&%\\
%& & &\\
%\&\quad &\left\{\begin{aligned}
     %&  Z_{13}= \left(\frac{\langle \boldsymbol{1} \boldsymbol{5}\rangle  \langle 3|\boldsymbol{1}|2]+\langle 3 \boldsymbol{5}\rangle  \langle 3 \boldsymbol{1}\rangle  [3 2]}{m \langle 3|\boldsymbol{1}|2]}\right), \quad &&\\
     %&  Z_{25}= \left(\frac{(\langle 3 \boldsymbol{5}\rangle  [\boldsymbol{1} 4]+[\boldsymbol{5} 4] \langle 3 \boldsymbol{1}\rangle ) \langle 3|\boldsymbol{5}|2]+\langle 3|2|4] [\boldsymbol{5} 2] \langle 3 \boldsymbol{1}\rangle }{\langle 3|\boldsymbol{1}|4] \langle 3|-\boldsymbol{5}|2]}\right), \quad &&\\
     %&  Z_{24}=\left(\frac{\langle \boldsymbol{5} 3\rangle  ([\boldsymbol{1}|2|3\rangle +[\boldsymbol{1}|4|3\rangle )+\langle \boldsymbol{1} 3\rangle  ([\boldsymbol{5}|2|3\rangle +[\boldsymbol{5}|4|3\rangle )}{\langle 3|\boldsymbol{1}\boldsymbol{5}|3\rangle }\right), \quad &&\\
     %&  Z_{35}=\left(\frac{\langle \boldsymbol{1} \boldsymbol{5}\rangle  \langle 3|\boldsymbol{5}|2]+\langle 3 \boldsymbol{5}\rangle  \langle \boldsymbol{1} 3\rangle  [3 2]}{m \langle 3|\boldsymbol{5}|2]}\right)\, .&&
%\end{aligned}\right.&&
\end{flalign*}
and, $Z_{13}$, $Z_{25}$, $Z_{24}$, and $Z_{35}$ have the form mentioned in eq.~\eqref{eq:bcfw-channel-spin-structures}.

Spurious singularities are present in eq.~\eqref{eq:ym-five-point-31425-check-amplitude}. We can use KK relations and reflection symmetry to obtain an expression free of spurious singularities,
\begin{align}
A[3,\boldsymbol{1}^s,4,2,\boldsymbol{5}^s] = -(-1)^{2s} A[\boldsymbol{5}^s,3,2,4,\boldsymbol{1}^s] +  A[\boldsymbol{1}^s,3,4,2,\boldsymbol{5}^s] + A[\boldsymbol{1}^s,4,3,2,\boldsymbol{5}^s]\ ,
\end{align}
or using the BCJ relations in eq.~\eqref{eq:bcj-relation-ordering-21435}. We have checked that these different forms of the amplitude agree with each other.

\subsection{Four-point QED Compton Using BCFW}\label{subsec:qed-four-point-bcfw-check}

Just like Yang--Mills Compton amplitudes, three-point QED Compton amplitudes can be used to construct four-point amplitudes via BCFW recursion relations. Here, we use the BCFW recursion relations to verify the results we previously arrived at using eq.~\eqref{eq:Photon_Compton_Perm} in sec.~\ref{subsec:qed-four-point-compton}.

\vspace{5mm}

\begin{figure}[H]
    \centering
    \begin{fmffile}{4_0_d02}
        \begin{fmfgraph*}(120,70)
            \fmfleft{i1,i2}
            \fmfright{o1,o2}
            \fmfblob{.16w}{g}
            \fmf{plain}{i1,g}
            \fmfv{label=$\boldsymbol{1}^s$}{i1}
            \fmf{plain}{g,o1}
            \fmfv{label=$\boldsymbol{4}^s$}{o1}
            \fmf{photon}{i2,g}
            \fmfv{label=$2$}{i2}
            \fmf{photon}{g,o2}
            \fmfv{label=$3$}{o2}
        \end{fmfgraph*}
    \end{fmffile}
    \vspace{5mm}
    \caption{Four-point amplitude $\mathcal{A}[\boldsymbol{1}^s,2,3,\boldsymbol{4}^s]$.}
    \label{fig:qed-four-point-bcfw-channel}
\end{figure}

We first consider the different-helicity configuration: $\mathcal{A}(\boldsymbol{1}^s,2^+,3^-,\boldsymbol{4}^s)$, where particles 2 and 3 are photons and particles 1 and 4 are massive spin-$s$ states. 

To calculate $\mathcal{A}(\boldsymbol{1}^s,2^+,3^-,\boldsymbol{4}^s)$, we use the same $\langle 2,3]$ shift. The two channels that survive in this shift are the $\boldsymbol{1}\hat{2}$-channel and $\boldsymbol{1}\hat{3}$-channel.
{\allowdisplaybreaks
\begin{align}
\nonumber& \mathcal{A}(\boldsymbol{1}^s,2^+,3^-,\boldsymbol{4}^s)\\
=&\nonumber \frac{i\,\left.\left(\mathcal{A}(\boldsymbol{1}^s,\hat{2}^+,\hat{\boldsymbol{P}}^s)\mathcal{A}(-\hat{\boldsymbol{P}}^s,\hat{3}^-,\boldsymbol{4}^s)\right)\right|_{\tau_{\boldsymbol{1}\hat{2}}\to 0}}{\tau_{12}} + \frac{i\,\left.\left(\mathcal{A}(\boldsymbol{1}^s,\hat{3}^-,\hat{\boldsymbol{P}}^s)\mathcal{A}(-\hat{\boldsymbol{P}}^s,\hat{2}^+,\boldsymbol{4}^s)\right)\right|_{\tau_{\boldsymbol{1}\hat{3}}\to 0}}{\tau_{13}}\\
=&\nonumber 2\,i \, \frac{\langle 3 | \boldsymbol{1} | 2 ]^{2-2s}}{\tau_{12}\,s_{23}}\big(\langle \boldsymbol{1}3 \rangle [\boldsymbol{4}2]+ \langle\boldsymbol{4}3 \rangle [\boldsymbol{1}2] \big)^{2s}  + 2\,i\, \frac{\langle 3 | \boldsymbol{1} | 2 ]^{2-2s}}{\tau_{13}\,s_{23}}\big(\langle \boldsymbol{1}3 \rangle [\boldsymbol{4}2]+ \langle\boldsymbol{4}3 \rangle [\boldsymbol{1}2] \big)^{2s}\\
\implies \Aboxed{&\mathcal{A}(\boldsymbol{1}^s,2^+,3^-,\boldsymbol{4}^s)= -2\,i \, \frac{\langle 3 | \boldsymbol{1} | 2 ]^{2-2s}}{\tau_{12}\,\tau_{13}}\big(\langle \boldsymbol{1}3 \rangle [\boldsymbol{4}2]+ \langle\boldsymbol{4}3 \rangle [\boldsymbol{1}2] \big)^{2s}}\ .
\label{eq:qed-four-point-bcfw-positive-negative-helicity}
\end{align}}
The amplitude in eq.~\eqref{eq:qed-four-point-bcfw-positive-negative-helicity}, calculated using BCFW recursion, agrees with the opposite helicity QED Compton amplitude in eq.~\eqref{eq:QED_4pt_Perm_Sum_Opp_Hel}.

To calculate $\mathcal{A}(\boldsymbol{1}^s,2^+,3^+,\boldsymbol{4}^s)$, we again use a $\langle 2,3]$ shift.
{\allowdisplaybreaks
\begin{align}
    &\nonumber \mathcal{A}(\boldsymbol{1}^s,2^+,3^+,\boldsymbol{4}^s)\\
    =&\nonumber \frac{i\, \left.\left(\mathcal{A}(\boldsymbol{1}^s,\hat{2}^+,\hat{\boldsymbol{P}}^s)\mathcal{A}(-\hat{\boldsymbol{P}}^s,\hat{3}^+,\boldsymbol{4}^s)\right)\right|_{\tau_{\boldsymbol{1}\hat{2}}\to 0}}{\tau_{12}} + \frac{i\, \left.\left(\mathcal{A}(\boldsymbol{1}^s,\hat{3}^+,\hat{\boldsymbol{P}}^s)\mathcal{A}(-\hat{\boldsymbol{P}}^s,\hat{2}^+,\boldsymbol{4}^s)\right)\right|_{\tau_{\boldsymbol{1}\hat{3}}\to 0}}{\tau_{13}}\\
    =&\nonumber -2\,i \, \frac{m^{2-2s} [23]^2}
        {\tau_{12}\,s_{23}}
   \langle \boldsymbol{14} \rangle^{2s} -2\,i \, \frac{m^{2-2s} [23]^2}{\tau_{13}\,s_{23}}
   \langle \boldsymbol{14} \rangle^{2s}\\
   \implies\Aboxed{&\mathcal{A}(\boldsymbol{1}^s,2^+,3^+,\boldsymbol{4}^s)
= 2\,i\,\frac{m^{2-2s} [23]^2}
        {\tau_{12}\,\tau_{13}}
   \langle \boldsymbol{14} \rangle^{2s}}\ . \label{eq:qed-four-point-bcfw-all-plus-helicity}
\end{align}
}
The all-plus QED Compton amplitude in eq.~\eqref{eq:qed-four-point-bcfw-all-plus-helicity}, calculated using BCFW recursion, agrees with the amplitude in eq.~\eqref{eq:QED_4pt_Perm_Sum_All_Pos_Hel}.

%----------------------------------------------------------------------------

\subsection{Five-point QED Compton Using BCFW}\label{subsec:qed-five-point-bcfw-check}

In sec.~\ref{subsubsec:qed-five-point-single-minus-bcfw} and sec.~\ref{subsubsec:qed-five-point-all-plus-bcfw}, we use BCFW recursion relations to calculate the photon Compton amplitude with different-helicity photons and same-helicity photons, respectively. Using analytic parametrization, one can verify that these amplitudes agree with the amplitudes calculated in sec.~\ref{subsec:qed-five-point-compton}.

\subsubsection{Case I: \texorpdfstring{$\mathcal{A}(\boldsymbol{1}^s, 2^+, 3^-, 4^+, \boldsymbol{5}^s)$}{}}\label{subsubsec:qed-five-point-single-minus-bcfw}
To calculate $\mathcal{A}(\boldsymbol{1}^s, 2^+, 3^-, 4^+, \boldsymbol{5}^s)$, we again use the  $[3,2\rangle$ BCFW shift. We have to compute contributions from the following four BCFW channels.
\begin{align}  
\boxed{\begin{aligned}
    \mathcal{A}(\mathbf{1}^s,\hat{2}^+,\hat{3}^-,4^+,\mathbf{5}^s)=&\mathcal{A}(\mathbf{1}^s,\hat{2}^+|\hat{3}^-,4^+,\mathbf{5}^s)+\mathcal{A}(\mathbf{1}^s,\hat{3}^-,4^+|\hat{2}^+,\mathbf{5}^s)\\
&\qquad+\mathcal{A}(\mathbf{1}^s,\hat{3}^-|\hat{2}^+,4^+,\mathbf{5}^s)+\mathcal{A}(\mathbf{1}^s,4^+,\hat{2}^+|\hat{3}^-,\mathbf{5}^s)
\end{aligned}} \ . \label{eq:qed-five-point-single-minus-bcfw-decomposition}
\end{align}

\paragraph{Channel 1: $\mathcal{A}(\mathbf{1}^s,\hat{2}^+|\hat{3}^-,4^+,\mathbf{5}^s)$} In fig.~\ref{eq:qed-five-point-single-minus-bcfw-decomposition},
\[\tau_{1\hat{2}}=0 \implies \tau_{12}-z\langle3|\mathbf{1}|2]=0 \nonumber \implies z=\frac{\tau_{12}}{\langle3|\mathbf{1}|2]}\ .\]
\begin{figure}[H]
    \centering
       \begin{fmffile}{fdp03}
 \begin{fmfgraph*}(200,60)% units are now in cm
   \fmfleft{i1,i2}
   \fmfright{o1,o2,o3}
   \fmfblob{.08w}{g1,g2}
   \fmf{plain}{i1,g1}
   \fmfv{label=$\boldsymbol{1}^s$}{i1}
   \fmf{photon}{i2,g1}
   \fmfv{label=$\hat{2}^+$}{i2}
   \fmf{plain, label=$\hat{P}^s$}{g1,v1}
   \fmf{phantom}{v1,v2}
   \fmf{plain, label=-$\hat{P}^s$}{v2,g2}
   \fmf{photon}{g2,o3}
   \fmfv{label=$\hat{3}^-$}{o3}
   \fmf{photon}{g2,o2}
   \fmfv{label=$4^+$}{o2}
   \fmf{plain}{g2,o1}
   \fmfv{label=$\boldsymbol{5}^s$}{o1}
\end{fmfgraph*}
\end{fmffile}
\vspace{5mm}
    \caption{Factorization channel contributing to $\mathcal{A}(\mathbf{1}^s,\hat{2}^+|\hat{3}^-,4^+,\mathbf{5}^s)$}
    \label{fig:qed-five-point-single-minus-bcfw-channel-12}
\end{figure}
\begin{align}
    \mathcal{A}(\mathbf{1}^s\hat{2}^+|\hat{3}^-4^+\mathbf{5}^s)\nonumber& =\mathcal{A}(\mathbf{1}^s,\hat{2}^+, \hat{\mathbf{P}}^s)\frac{i}{\tau_{12}}\mathcal{A}(-\hat{\mathbf{P}}^s,\hat{3}^-,4^+,\mathbf{5}^s)\\
    %&&=\frac{\langle\hat{3}|\mathbf{1}|\hat{2}]}{\langle\hat{3}\hat{2}\rangle}\frac{\langle\boldsymbol{1\hat{P}}\rangle^{2s}}{m^{2s}}\frac{1}{\tau_{12}}\frac{\langle\hat{3}|\mathbf{5}|4]^2}{\tau_{45}\tau_{5\hat{3}}}\left(\frac{\langle-\boldsymbol{\hat{P}}3\rangle[\boldsymbol{5}4]+\langle\boldsymbol{5}3\rangle[-\boldsymbol{\hat{P}}4]}{\langle3|-\boldsymbol{\hat{P}}|4]}\right)^{2s}\nonumber\\
    & =  2\sqrt{2} \, i \, \frac{\left(\langle3|\mathbf{1}|2]\langle3|\mathbf{5}|4]\right)^{2}]}{\langle32\rangle\tau_{12}\tau_{45}\left(\langle3|\mathbf{1}|2]\tau_{53}+\langle3|\mathbf{5}|2]\tau_{12}\right)}\,Z_{12}^{2s} \ . \label{eq:qed-five-point-single-minus-bcfw-channel-12}
\end{align}

\paragraph{Channel 2: $\mathcal{A}(\mathbf{1}^s,\hat{3}^-,4^+|\hat{2}^+,\mathbf{5}^s)$} In fig.~\ref{eq:qed-five-point-single-minus-bcfw-channel-12},
\[\tau_{5\hat{2}}=0 \implies z=\frac{\tau_{52}}{\langle3|\mathbf{5}|2]} \ .\]
\begin{figure}[h]
    \centering
    \begin{fmffile}{fdp04}
 \begin{fmfgraph*}(200,60)% units are now in cm
   \fmfright{i1,i2}
   \fmfleft{o1,o2,o3}
   \fmfblob{.08w}{g1,g2}
   \fmf{plain}{i1,g1}
   \fmfv{label=$\boldsymbol{5}^s$}{i1}
   \fmf{photon}{i2,g1}
   \fmfv{label=$\hat{2}^+$}{i2}
   \fmf{plain, label=$-\hat{P}^s$}{g1,v1}
   \fmf{phantom}{v1,v2}
   \fmf{plain, label=$\hat{P}^s$}{v2,g2}
   \fmf{photon}{g2,o3}
   \fmfv{label=$\hat{3}^-$}{o3}
   \fmf{photon}{g2,o2}
   \fmfv{label=$4^+$}{o2}
   \fmf{plain}{g2,o1}
   \fmfv{label=$\boldsymbol{1}^s$}{o1}
\end{fmfgraph*}
\end{fmffile}
\vspace{5mm}
    \caption{Factorization channel contributing to $\mathcal{A}(\mathbf{1}^s,\hat{3}^-,4^+|\hat{2}^+,\mathbf{5}^s)$}
    \label{fig:qed-five-point-single-minus-bcfw-channel-25}
\end{figure}
\begin{align}
    \mathcal{A}(\mathbf{1}^s,\hat{3}^-,4^+|\hat{2}^+,\mathbf{5}^s)& = \mathcal{A}(\mathbf{1}^s,\hat{3}^-,4^+, \hat{\mathbf{P}}^s) \frac{i}{\tau_{52}}\mathcal{A}(-\hat{\mathbf{P}},\hat{2}^+,\mathbf{5}^s)\nonumber\\
    &=-2\sqrt{2} \,i\, \frac{\left(\langle3|\mathbf{5}|2]\langle3|\mathbf{1}|4]\right)^{2}}{\langle32\rangle\tau_{52}\tau_{14}\left(\langle3|\mathbf{5}|2]\tau_{13}+\langle3|\mathbf{1}|2]\tau_{52}\right)} \, Z_{25}^{2s} \ . \label{eq:qed-five-point-single-minus-bcfw-channel-25}
\end{align}

\paragraph{Channel 3: $\mathcal{A}(\mathbf{1}^s,\hat{3}^-|\hat{2}^+,4^+,\mathbf{5}^s)$} In fig.~\ref{fig:qed-five-point-single-minus-bcfw-channel-13},
\[\langle\hat{3}|\mathbf{1}|\hat{3}]=0\implies \tau_{13}+z\langle3|1|2]=0 \nonumber \implies z=-\frac{\tau_{13}}{\langle3|\mathbf{1}|2]} \ .\]
\begin{figure}[h]
    \centering
    \begin{fmffile}{fdp07}
 \begin{fmfgraph*}(200,60)% units are now in cm
   \fmfleft{i1,i2}
   \fmfright{o1,o2,o3}
   \fmfblob{.08w}{g1,g2}
   \fmf{plain}{i1,g1}
   \fmfv{label=$\boldsymbol{1}^s$}{i1}
   \fmf{photon}{i2,g1}
   \fmfv{label=$\hat{3}^-$}{i2}
   \fmf{plain, label=$\hat{P}^s$}{g1,v1}
   \fmf{phantom}{v1,v2}
   \fmf{plain, label=-$\hat{P}^s$}{v2,g2}
   \fmf{photon}{g2,o3}
   \fmfv{label=$\hat{2}^+$}{o3}
   \fmf{photon}{g2,o2}
   \fmfv{label=$4^+$}{o2}
   \fmf{plain}{g2,o1}
   \fmfv{label=$\boldsymbol{5}^s$}{o1}
\end{fmfgraph*}
\end{fmffile}
    \caption{Factorization channel contributing to $\mathcal{A}(\mathbf{1}^s,\hat{3}^-|\hat{2}^+,4^+,\mathbf{5}^s)$}
    \label{fig:qed-five-point-single-minus-bcfw-channel-13}
\end{figure}
{\allowdisplaybreaks
\begin{align}
    \mathcal{A}(\mathbf{1}^s,\hat{3}^-|\hat{2}^+,4^+,\mathbf{5}^s)&=\mathcal{A}(\mathbf{1}^s,\hat{3}^-,\hat{\boldsymbol{P}}^s) \frac{i}{\tau_{13}} \mathcal{A}(-\hat{\boldsymbol{P}}^s,\hat{2}^+,4^+,\mathbf{5}^s)\nonumber\\
    %&&= \frac{\langle\hat{3}|\mathbf{1}|\hat{2}]}{[\hat{3}\hat{2}]}\frac{[\mathbf{1\hat{P}}]^{2s}}{m^{2s}} \frac{1}{\tau_{13}} \frac{m^2[\hat{2}4]^2}{\tau_{45}\tau_{5\hat{2}}} \frac{\langle-\hat{\mathbf{P}}\mathbf{5}\rangle^{2s}}{m^{2s}} \nonumber \\
    %&&=\frac{m^2[24]^2\langle3|\mathbf{1}|2]^2}{\tau_{45}\left(\tau_{52}+\frac{\tau_{13}}{\langle3|\mathbf{1}|2]}\langle3|\mathbf{5}|2]\right)\tau_{13}[32]}\left(-\frac{\langle\mathbf{5}|\hat{P}|\mathbf{1}]}{m^2}\right)^{2s} \nonumber \\
    &=-2\sqrt{2} \,i\,\frac{m^{2}[24]^2\langle3|\mathbf{1}|2]^{2}}{[32]\tau_{45}\tau_{13}\left(\langle3|\mathbf{1}|2]\tau_{52}+\langle3|\mathbf{5}|2]\tau_{13}\right)}\,Z_{13}^{2s} \ . \label{eq:qed-five-point-single-minus-bcfw-channel-13}
\end{align}
}

\paragraph{Channel 4: $\mathcal{A}(\mathbf{1}^s,\hat{2}^+,4^+|\hat{3}^-,\mathbf{5}^s)$} In fig.~\ref{fig:qed-five-point-single-minus-bcfw-channel-35},
\[\langle\hat{3}|\mathbf{5}|\hat{3}]=0\implies \tau_{53}+z\langle3|5|2]=0 \implies z=-\frac{\tau_{53}}{\langle3|\mathbf{5}|2]} \ .\]
\begin{figure}[h]
    \centering
    \begin{fmffile}{fdp08}
 \begin{fmfgraph*}(200,60)% units are now in cm
   \fmfright{i1,i2}
   \fmfleft{o1,o2,o3}
   \fmfblob{.08w}{g1,g2}
   \fmf{plain}{i1,g1}
   \fmfv{label=$\boldsymbol{5}^s$}{i1}
   \fmf{photon}{i2,g1}
   \fmfv{label=$\hat{3}^-$}{i2}
   \fmf{plain, label=$-\hat{P}^s$}{g1,v1}
   \fmf{phantom}{v1,v2}
   \fmf{plain, label=$\hat{P}^s$}{v2,g2}
   \fmf{photon}{g2,o3}
   \fmfv{label=$\hat{2}^+$}{o3}
   \fmf{photon}{g2,o2}
   \fmfv{label=$4^+$}{o2}
   \fmf{plain}{g2,o1}
   \fmfv{label=$\boldsymbol{1}^s$}{o1}
\end{fmfgraph*}
\end{fmffile}
    \caption{Factorization channel contributing to $\mathcal{A}(\mathbf{1}^s,\hat{2}^+,4^+|\hat{3}^-,\mathbf{5}^s)$}
    \label{fig:qed-five-point-single-minus-bcfw-channel-35}
\end{figure}
\begin{align}
    \mathcal{A}(\mathbf{1}^s,\hat{2}^+,4^+|\hat{3}^-,\mathbf{5}^s)&=\mathcal{A}(\mathbf{1}^s,\hat{2}^+,4^+,\hat{\boldsymbol{P}}^s)\frac{i}{\tau_{53}} \mathcal{A}(-\hat{\boldsymbol{P}}^s,\hat{3}^-,\mathbf{5}^s)\nonumber\\
    %&&=\frac{m^2[\hat{2}4]^2}{\tau_{14}\tau_{1\hat{2}}} \frac{\langle\mathbf{1\hat{P}}\rangle^{2s}}{m^{2s}} \frac{1}{\tau_{13}} \frac{\langle\hat{3}|-\mathbf{5}|\hat{2}]}{[\hat{3}\hat{2}]} \frac{[-\hat{\mathbf{P}}\mathbf{5}]^{2s}}{m^{2s}} \nonumber \\
    &=2\sqrt{2} \,i\,\frac{m^{2}[24]^2\langle3|\mathbf{5}|2]^{2}}{[32]\tau_{14}\tau_{53}\left(\langle3|\mathbf{5}|2]\tau_{12}+\langle3|\mathbf{1}|2]\tau_{53}\right)}\,Z_{35}^{2s} \ . \label{eq:qed-five-point-single-minus-bcfw-channel-35}
\end{align}\\

Now that we have calculated the contributions from all four channels, we get the full amplitude following eq.~\eqref{eq:qed-five-point-single-minus-bcfw-decomposition}.

\begin{align}
    &\boxed{
    \begin{aligned}
        &\nonumber \mathcal{A}(\mathbf{1}^s,{2}^+,{3}^-,4^+,\mathbf{5}^s)\\
    &\nonumber\\
    =& 2\sqrt{2} \,i\,\frac{\left(\langle3|\mathbf{1}|2]\langle3|\mathbf{5}|4]\right)^{2}}{\langle32\rangle\tau_{12}\tau_{45}\left(\langle3|\mathbf{1}|2]\tau_{53}+\langle3|\mathbf{5}|2]\tau_{12}\right)}\,Z_{12}^{2s} - 2\sqrt{2} \,i\,\frac{\left(\langle3|\mathbf{5}|2]\langle3|\mathbf{1}|4]\right)^{2}}{\langle32\rangle\tau_{52}\tau_{14}\left(\langle3|\mathbf{5}|2]\tau_{13}+\langle3|\mathbf{1}|2]\tau_{52}\right)} \, Z_{25}^{2s}\nonumber\\
    &\nonumber\\
    &\quad -2\sqrt{2} \,i\, \frac{m^{2}[24]^2\langle3|\mathbf{1}|2]^{2}}{[32]\tau_{45}\tau_{13}\left(\langle3|\mathbf{1}|2]\tau_{52}+\langle3|\mathbf{5}|2]\tau_{13}\right)}\,Z_{13}^{2s} + 2\sqrt{2} i \,\,\frac{m^{2}[24]^2\langle3|\mathbf{5}|2]^{2}}{[32]\tau_{14}\tau_{53}\left(\langle3|\mathbf{5}|2]\tau_{12}+\langle3|\mathbf{1}|2]\tau_{53}\right)}\,Z_{35}^{2s}\nonumber\\
    &\nonumber\\
    =& 2\sqrt{2} \,i\,\left(\frac{\left(\langle3|\mathbf{1}|2]\langle3|\mathbf{5}|4]\right)^{2}}{\langle32\rangle\tau_{12}\tau_{45}\langle3|\mathbf{5}(2+3)\mathbf{1}|2]}\,Z_{12}^{2s} - \frac{m^{2}[24]^2\langle3|\mathbf{1}|2]^{2}}{[32]\tau_{45}\tau_{13}\langle3|\mathbf{1}(2+3)\mathbf{5}|2]}\,Z_{13}^{2s}\right) - (1\leftrightarrow5)
    \end{aligned}
    } \ . \\
    &\label{eq:qed-five-point-single-minus-bcfw-amplitude}
\end{align}

% \begin{align*}
%     2 i \sqrt{2} \left(\frac{\langle 3|1|4]^2 \langle 3|5|2]^2}{\tau _{1 4} \tau _{2 5} \langle 2 3\rangle  (\langle 3|1,2,5|2]+\langle 3|1,3,5|2])}-\frac{m^2
%    [2 4]^2 \langle 3|5|2]^2}{\tau _{1 4} \tau _{3 5} [2 3] (\langle 3|5,2,1|2]+\langle 3|5,3,1|2])}\right)\\
%    -2 i \sqrt{2} \left(\frac{\langle 3|1|2]^2
%    \langle 3|5|4]^2}{\tau _{1 2} \tau _{4 5} \langle 2 3\rangle  (\langle 3|5,2,1|2]+\langle 3|5,3,1|2])}-\frac{m^2 [2 4]^2 \langle 3|1|2]^2}{\tau _{1 3}
%    \tau _{4 5} [2 3] (\langle 3|1,2,5|2]+\langle 3|1,3,5|2])}\right)
% \end{align*}
We have verified that the single-minus five-point QED Compton amplitude in eq.~\eqref{eq:qed-five-point-single-minus-bcfw-amplitude} agrees with the amplitude in eq.~\eqref{eq_five-point_single_minus_QED_Perm_sum} obtained from the Abelian limit of Yang--Mills amplitudes.

\subsubsection{Case II: \texorpdfstring{$\mathcal{A}(\boldsymbol{1}^s, 2^+, 3^+, 4^+, \boldsymbol{5}^s)$}{}}\label{subsubsec:qed-five-point-all-plus-bcfw}
To calculate the same-helicity five-point Compton amplitude, we use a massive--massless BCFW shift of the type $[\boldsymbol{1},2\rangle$.
\begin{equation*}
\begin{aligned}
& |\hat{2}\rangle = |2\rangle - z |\boldsymbol{1}^J\rangle[2\boldsymbol{1}_J]\ , &&
|\hat{2}] = |2]\ , \\
& |\hat{\boldsymbol{1}}^J] = |\boldsymbol{1}^J] + z\, |2] [2 \boldsymbol{1}^J]\ ,  && |\hat{\boldsymbol{1}}^J\rangle = |\boldsymbol{1}^J\rangle \ .
\end{aligned}
\end{equation*}

In this shift, there are three BCFW channels which contribute to the total amplitude.
\begin{equation}\label{eq:qed-five-point-all-plus-bcfw-decomposition}
    \boxed{\mathcal{A}(\boldsymbol{1}^s, 2^+, 3^+, 4^+, \boldsymbol{5}^s)=\mathcal{A}(\hat{\boldsymbol{1}}^s, 3^+| \hat{2}^+, 4^+, \boldsymbol{5}^s) + \mathcal{A}(\hat{\boldsymbol{1}}^s, 4^+| \hat{2}^+, 3^+, \boldsymbol{5}^s) + \mathcal{A}(\hat{\boldsymbol{1}}^s, 3^+, 4^+| \hat{2}^+, \boldsymbol{5}^s)} \ .
\end{equation}

\paragraph{Channel 1: $\mathcal{A}(\hat{\mathbf{1}}^s,3^+|\hat{2}^+,\hat{4}^+,\mathbf{5}^s)$}

\begin{figure}[h]
    \centering
       \begin{fmffile}{QED_Compton_Same_Helicity_Diag_1}
 \begin{fmfgraph*}(200,60)% units are now in cm
   \fmfleft{i1,i2}
   \fmfright{o1,o2,o3}
   \fmfblob{.08w}{g1,g2}
   \fmf{plain}{i1,g1}
   \fmfv{label=$\hat{\boldsymbol{1}}^s$}{i1}
   \fmf{photon}{i2,g1}
   \fmfv{label=$3^+$}{i2}
   \fmf{plain, label=$\hat{P}^s$}{g1,v1}
   \fmf{phantom}{v1,v2}
   \fmf{plain, label=-$\hat{P}^s$}{v2,g2}
   \fmf{photon}{g2,o3}
   \fmfv{label=$\hat{2}^+$}{o3}
   \fmf{photon}{g2,o2}
   \fmfv{label=$4^+$}{o2}
   \fmf{plain}{g2,o1}
   \fmfv{label=$\boldsymbol{5}^s$}{o1}
\end{fmfgraph*}
\end{fmffile}
\vspace{5mm}
    \caption{Factorization channel contributing to $\mathcal{A}(\hat{\mathbf{1}}^s,3^+|\hat{2}^+,4^+,\mathbf{5}^s)$}
    \label{fig:qed-five-point-all-plus-bcfw-channel-13}
\end{figure}
\begin{align}
    &\nonumber \mathcal{A}(\hat{\mathbf{1}}^s,3^+|\hat{2}^+,4^+,\mathbf{5}^s)=\mathcal{A}(\hat{\mathbf{1}}^s,3^+,\hat{\mathbf{P}^s}) \frac{i}{\tau_{13}} \mathcal{A}(-\hat{\mathbf{P}^s},\hat{2}^+,4^+,\mathbf{5}^s)\\
    =& -2\sqrt{2} \,i\,\frac{m^4 [23]^2 [24]^2}{\tau_{13} \tau_{54} (\tau_{52}[2|3\boldsymbol{1}|2]+\tau_{13}[2|\boldsymbol{51}|2])}\left(\frac{\langle\boldsymbol{15}\rangle}{m}\right)^{2s} \ . \label{eq:qed-five-point-all-plus-bcfw-channel-13}
\end{align}

\paragraph{Channel 2: $\mathcal{A}(\hat{\mathbf{1}}^s,4^+|\hat{2}^+,3^+,\mathbf{5}^s)$}
\begin{figure}[h]
    \centering
       \begin{fmffile}{QED_Compton_Same_Helicity_Diag_2}
 \begin{fmfgraph*}(200,60)% units are now in cm
   \fmfleft{i1,i2}
   \fmfright{o1,o2,o3}
   \fmfblob{.08w}{g1,g2}
   \fmf{plain}{i1,g1}
   \fmfv{label=$\hat{\boldsymbol{1}}^s$}{i1}
   \fmf{photon}{i2,g1}
   \fmfv{label=$4^+$}{i2}
   \fmf{plain, label=$\hat{P}^s$}{g1,v1}
   \fmf{phantom}{v1,v2}
   \fmf{plain, label=-$\hat{P}^s$}{v2,g2}
   \fmf{photon}{g2,o3}
   \fmfv{label=$\hat{2}^+$}{o3}
   \fmf{photon}{g2,o2}
   \fmfv{label=$3^+$}{o2}
   \fmf{plain}{g2,o1}
   \fmfv{label=$\boldsymbol{5}^s$}{o1}
\end{fmfgraph*}
\end{fmffile}
    \caption{Factorization channel contributing to $\mathcal{A}(\hat{\mathbf{1}}^s,4^+|\hat{2}^+,3^+,\mathbf{5}^s)$}
    \label{fig:qed-five-point-all-plus-bcfw-channel-14}
\end{figure}
\begin{align}
    &\nonumber \mathcal{A}(\hat{\mathbf{1}}^s,4^+|\hat{2}^+,3^+,\mathbf{5}^s)=\mathcal{A}(\hat{\mathbf{1}}^s,4^+,\hat{\mathbf{P}}^s) \frac{i}{\tau_{14}} \mathcal{A}(-\hat{\mathbf{P}}^s,\hat{2}^+,3^+,\mathbf{5}^s)\\
    =& -2\sqrt{2} \,i\,\frac{m^4 [24]^2 [23]^2}{\tau_{14} \tau_{53} (\tau_{52}[2|4\boldsymbol{1}|2]+\tau_{14}[2|\boldsymbol{51}|2])}\left(\frac{\langle\boldsymbol{15}\rangle}{m}\right)^{2s} \ . \label{eq:qed-five-point-all-plus-bcfw-channel-14}
\end{align}

\paragraph{Channel 3: $\mathcal{A}(\hat{\mathbf{1}}^s,3^+,4^+|\hat{2}^+,\mathbf{5}^s)$}
\begin{figure}[h]
    \centering
    \begin{fmffile}{QED_Compton_Same_Helicity_Diag_3}
 \begin{fmfgraph*}(200,60)% units are now in cm
   \fmfright{i1,i2}
   \fmfleft{o1,o2,o3}
   \fmfblob{.08w}{g1,g2}
   \fmf{plain}{i1,g1}
   \fmfv{label=$\boldsymbol{5}^s$}{i1}
   \fmf{photon}{i2,g1}
   \fmfv{label=$\hat{2}^+$}{i2}
   \fmf{plain, label=$-\hat{P}^s$}{g1,v1}
   \fmf{phantom}{v1,v2}
   \fmf{plain, label=$\hat{P}^s$}{v2,g2}
   \fmf{photon}{g2,o3}
   \fmfv{label=$4^+$}{o3}
   \fmf{photon}{g2,o2}
   \fmfv{label=$3^+$}{o2}
   \fmf{plain}{g2,o1}
   \fmfv{label=$\hat{\boldsymbol{1}}^s$}{o1}
\end{fmfgraph*}
\end{fmffile}
\vspace{5mm}
    \caption{Factorization channel contributing to $\mathcal{A}(\hat{\mathbf{1}}^s,3^+,4^+|\hat{2}^+,\mathbf{5}^s)$}
    \label{fig:qed-five-point-all-plus-bcfw-channel-12}
\end{figure}
\begin{align}
    &\nonumber\mathcal{A}(\hat{\mathbf{1}}^s,3^+,4^+|\hat{2}^+,\mathbf{5}^s)=\mathcal{A}(\hat{\mathbf{1}}^s,3^+,4^+,\hat{\mathbf{P}^s}) \frac{i}{\tau_{52}} \mathcal{A}(-\hat{\mathbf{P}^s},\hat{2}^+,\mathbf{5}^s)\\
    =& 2\sqrt{2} \,i\,\frac{m^2 [34]^2 [2|\boldsymbol{15}|2]^3}{(\tau_{13}[2|\boldsymbol{15}|2] + \tau_{52} [2|\boldsymbol{1}3|2]) (\tau_{14}[2|\boldsymbol{15}|2] + \tau_{52} [2|\boldsymbol{1}4|2]) \tau_{12} \tau_{52}} \left(\frac{\langle\boldsymbol{15}\rangle}{m}\right)^{2s} \ . \label{eq:qed-five-point-all-plus-bcfw-channel-12}
\end{align}

Now that we have the contribution from the three BCFW channels in eq.~\eqref{eq:qed-five-point-all-plus-bcfw-channel-13}, eq.~\eqref{eq:qed-five-point-all-plus-bcfw-channel-14} and eq.~\eqref{eq:qed-five-point-all-plus-bcfw-channel-12}, we can follow eq.~\eqref{eq:qed-five-point-all-plus-bcfw-decomposition} to get $\mathcal{A}({\mathbf{1}}^s,{2}^+,3^+,4^+,\mathbf{5}^s)$.
\begin{align}
    &\boxed{
    \begin{aligned}
         &\nonumber\mathcal{A}({\mathbf{1}}^s,{2}^+,3^+,4^+,\mathbf{5}^s)\\
    =&\nonumber -2\sqrt{2} \,i\,\left( \frac{m^4 [23]^2 [24]^2}{\tau_{13} \tau_{54} (\tau_{52}[2|3\boldsymbol{1}|2]+\tau_{13}[2|\boldsymbol{51}|2])} +\frac{m^4 [24]^2 [23]^2}{\tau_{14} \tau_{53} (\tau_{52}[2|4\boldsymbol{1}|2]+\tau_{14}[2|\boldsymbol{51}|2])}\right.\\
    &\qquad\qquad\qquad\left.- \frac{m^2 [34]^2 [2|\boldsymbol{15}|2]^3}{(\tau_{13}[2|\boldsymbol{15}|2] + \tau_{52} [2|\boldsymbol{1}3|2]) (\tau_{14}[2|\boldsymbol{15}|2] + \tau_{52} [2|\boldsymbol{1}4|2]) \tau_{12} \tau_{52}}  \right) \left(\frac{\langle\boldsymbol{15}\rangle}{m}\right)^{2s}
    \end{aligned}
    }\quad .\\
    &\label{eq:qed-five-point-all-plus-bcfw-amplitude}
\end{align}
We have verified that the all-plus five-point QED Compton amplitude in eq.~\eqref{eq:qed-five-point-all-plus-bcfw-amplitude} agrees with the manifestly spurious singularity-free amplitude in eq.~\eqref{eq:qed-five-point-all-plus-permutation-sum}.

\subsection{Four-point Graviton Compton Using BCFW}\label{subsec:graviton-four-point-bcfw-check}

Here, we use BCFW recursion to calculate the 4-point graviton Compton amplitudes and verify that they agree with the amplitudes calculated using KLT relations in sec.~\ref{subsec:graviton-four-point-compton}.
\begin{figure}[H]
\centering
\begin{fmffile}{5_1_d02}
\begin{fmfgraph*}(120,70)
    \fmfleft{i1,i2}
    \fmfright{o1,o2}
    \fmfblob{.16w}{g}
    \fmf{plain}{g,i1}
    \fmfv{label=$\boldsymbol{1}^s$}{i1}
    \fmf{plain}{g,o1}
    \fmfv{label=$\boldsymbol{4}^s$}{o1}
    \fmf{dbl_wiggly}{i2,g}
    \fmfv{label=$2$}{i2}
    \fmf{dbl_wiggly}{o2,g}
    \fmfv{label=$3$}{o2}
\end{fmfgraph*}
\end{fmffile}
\vspace{5mm}
\caption{Four-point graviton Compton amplitude with massive legs $\boldsymbol{1}$ and $\boldsymbol{4}$ of spin $s$ and mass $m$.}
\label{fig:graviton-four-point-bcfw-channel}
\end{figure}
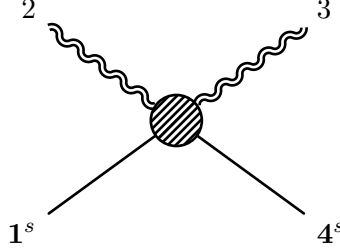

To calculate $M(\boldsymbol{1}^s,2^+,3^-,\boldsymbol{4}^s)$, we again use a $\langle2,3]$ shift. Two BCFW channels contribute:
{\allowdisplaybreaks
\begin{align}
    &\nonumber M(\boldsymbol{1}^s,2^+,3^-,\boldsymbol{4}^s) \\
    =&\nonumber M(\boldsymbol{1},\hat{2}^+,\hat{\boldsymbol{P}}_{12}^s) \frac{i}{\tau_{12}} M(-\hat{\boldsymbol{P}}_{12}^s,\hat{3}^-,\boldsymbol{4}^s)\: + \:  M(\boldsymbol{1},\hat{3}^-,\hat{\boldsymbol{P}}_{13}^s) \frac{i}{\tau_{12}} M(-\hat{\boldsymbol{P}}_{13}^s,\hat{2}^+,\boldsymbol{4}^s)\\
    =&\nonumber i\,(-i)^2\,\frac{\langle3|\boldsymbol{1}|2]^2}{\langle32\rangle^2} \left(\frac{\langle \boldsymbol{1} \, \hat{\boldsymbol{P}}^I_{12} \rangle}{m}\right)^{2s} \frac{1}{\tau_{12}} \frac{\langle3|\boldsymbol{1}|2]^2}{[32]^2} \left(\frac{[ -\hat{\boldsymbol{P}}_{12\,I} \: \boldsymbol{4}]}{m}\right)^{2s} \\
    &\nonumber \hspace{5cm}+ i\,(-i)^2\,\frac{\langle3|\boldsymbol{1}|2]^2}{[32]^2} \left(\frac{[\boldsymbol{1} \, \hat{\boldsymbol{P}}^J_{13} ]}{m}\right)^{2s} \frac{1}{\tau_{13}} \frac{\langle3|\boldsymbol{1}|2]^2}{\langle32\rangle^2} \left(\frac{\langle -\hat{\boldsymbol{P}}_{13\,J} \: \boldsymbol{4}\rangle}{m}\right)^{2s}\\
    =&\nonumber -i\, \frac{\langle3|\boldsymbol{1}|2]^4 (\tau_{12}+ \tau_{13})}{\tau_{12}\tau_{13}s_{23}^2} \left(\frac{\langle \boldsymbol{1}3 \rangle [\boldsymbol{4}2]
      + \langle \boldsymbol{4}3 \rangle [\boldsymbol{1}2]}{\langle3|\boldsymbol{1}|2]}\right)^{2s}\\
    =& i\,\frac{\langle3|\boldsymbol{1}|2]^4}{\tau_{12}\tau_{13}s_{23}} \left(\frac{\langle \boldsymbol{1}3 \rangle [\boldsymbol{4}2]
      + \langle \boldsymbol{4}3 \rangle [\boldsymbol{1}2]}{\langle3|\boldsymbol{1}|2]}\right)^{2s} \ . \label{eq:graviton-four-point-bcfw-positive-negative-helicity}
\end{align}
}
This agrees with eq.~\eqref{eq_GR_4pt_opp_hel_Compton}.
A similar calculation can be performed for the same-helicity graviton Compton amplitudes, and the result agrees with eq.~\eqref{eq_GR_4pt_all_plus_hel_Compton} and eq.~\eqref{eq_GR_4pt_all_minus_hel_Compton}.

\subsection{Five-point Graviton Compton Using BCFW}\label{subsec:graviton-five-point-bcfw-check}

In sec.~\ref{subsubsec:graviton-five-point-single-minus-bcfw} and sec.~\ref{subsubsec:graviton-five-point-all-plus-bcfw}, we use BCFW recursion relations to calculate the Compton amplitudes with different-helicity gravitons and same-helicity gravitons, respectively. Using analytic parametrization, one can verify that these amplitudes agree with the amplitudes calculated in sec.~\ref{subsec:graviton-five-point-compton}.

\subsubsection{
Case I:
\texorpdfstring{$M(\boldsymbol{1}^s, 2^+, 3^-, 4^+, \boldsymbol{5}^s)$}
               {}
}\label{subsubsec:graviton-five-point-single-minus-bcfw}
To calculate $M(\boldsymbol{1}^s, 2^+, 3^-, 4^+, \boldsymbol{5}^s)$, we use the $[3,2\rangle$ BCFW shift with the following six channels.
\begin{align}
&\boxed{
\begin{aligned}
    M(\mathbf{1}^s,\hat{2}^+,\hat{3}^-,4^+,\mathbf{5}^s)=&\: M(\mathbf{1}^s,\hat{2}^+|\hat{3}^-,4^+,\mathbf{5}^s)+M(\mathbf{5}^s,\hat{2}^+|\hat{3}^-,4^+,\mathbf{1}^s)+M(4^+,\hat{2}^+|\hat{3}^-,\mathbf{1}^s,\mathbf{5}^s) \nonumber \\&\hspace{4mm}+M(\mathbf{5}^s,\mathbf{1}^s,\hat{2}^+|\hat{3}^-,4^+)+M(\mathbf{5}^s,4^+,\hat{2}^+|\hat{3}^-,\mathbf{1}^s)+M(\mathbf{1}^s,4^+,\hat{2}^+|\hat{3}^-,\mathbf{5}^s)
\end{aligned}
}\ .\\
&\label{eq:graviton-five-point-single-minus-bcfw-decomposition} 
\end{align}
As we mentioned in sec.~\ref{sec:introduction}, we see that the amplitude in eq.~\eqref{eq:graviton-five-point-single-minus-bcfw-decomposition} depends on the spin structures $Z_j$ in eq.~\eqref{eq:bcfw-channel-spin-structures}.

\paragraph{Channel 1: $M(\mathbf{1}^s,\hat{2}^+|\hat{3}^-,4^+,\mathbf{5}^s)$}

\begin{flalign*}
    \text{In fig.~\ref{fig:graviton-five-point-single-minus-bcfw-channel-12}, } & \tau_{1\hat{2}}=0 \implies \tau_{12}-z\langle3|\mathbf{1}|2]=0 \nonumber \implies z=\frac{\tau_{12}}{\langle3|\mathbf{1}|2]} \ . 
\end{flalign*}
\begin{figure}[H]
    \centering
    \begin{fmffile}{fig_4_1}
 \begin{fmfgraph*}(200,60)% units are now in cm
   \fmfleft{i1,i2}
   \fmfright{o1,o2,o3}
   \fmfblob{.08w}{g1,g2}
   \fmf{plain}{i1,g1}
   \fmfv{label=$\boldsymbol{1}^s$}{i1}
   \fmf{dbl_wiggly}{i2,g1}
   \fmfv{label=$\hat{2}^+$}{i2}
   \fmf{plain, label=$\hat{\boldsymbol{P}}^s$}{g1,v1}
   \fmf{phantom}{v1,v2}
   \fmf{plain, label=-$\hat{\boldsymbol{P}}^s$}{v2,g2}
   \fmf{dbl_wiggly}{g2,o3}
   \fmfv{label=$\hat{3}^-$}{o3}
   \fmf{dbl_wiggly}{g2,o2}
   \fmfv{label=$4^+$}{o2}
   \fmf{plain}{g2,o1}
   \fmfv{label=$\boldsymbol{5}^s$}{o1}
\end{fmfgraph*}
\end{fmffile}
\vspace{5mm}
    \caption{Factorization channel contributing to $M(\mathbf{1}^s,\hat{2}^+|\hat{3}^-,4^+,\mathbf{5}^s)$}
    \label{fig:graviton-five-point-single-minus-bcfw-channel-12}
\end{figure}
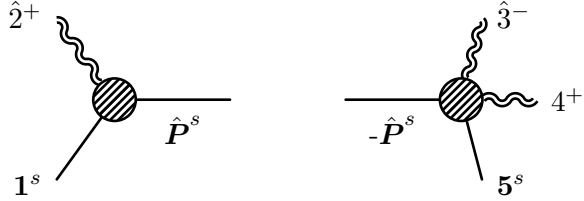
\begin{align}
    & M(\mathbf{1}^s,\hat{2}^+|\hat{3}^-,4^+,\mathbf{5}^s)=\hspace{2mm}M[\mathbf{1}^s,\hat{2}^+, \hat{\mathbf{P}}^s]\frac{i}{\tau_{12}}M[-\hat{\mathbf{P}}^s,\hat{3}^-,4^+,\mathbf{5}^s]\nonumber\\
    =&\nonumber i\,(-1)^{2s}\frac{\langle3|\mathbf{1}|2]^{4-2s}\langle3|\mathbf{5}|4]^{4-2s}((\langle \boldsymbol{1} 3\rangle  [\boldsymbol{5} 4]+[\boldsymbol{1} 4] \langle \boldsymbol{5} 3\rangle ) \langle 3|\boldsymbol{1}|2]+\langle 3|2|4] [\boldsymbol{1} 2] \langle \boldsymbol{5} 3\rangle)^{2s}}{\langle23\rangle^2\tau_{12}\tau_{45}\left(\langle3|\mathbf{1}|2]s_{34}+\langle3|4|2]\tau_{12}\right)\left(\langle3|\mathbf{1}|2]\tau_{53}+\langle3|\mathbf{5}|2]\tau_{12}\right)} \\
    =& M_{12}(\mathbf{1}^0,\hat{2}^+,\hat{3}^-,4^+,\mathbf{5}^0)\: Z_{12}^{2s}\ ,\label{eq:graviton-five-point-single-minus-bcfw-channel-12}
\end{align}
\begin{flalign*}
    &\text{where } \hspace{2mm} M_{12}(\mathbf{1}^0,\hat{2}^+,\hat{3}^-,4^+,\mathbf{5}^0) = -i\,\frac{
\langle3|\mathbf{1}|2]^4
\langle3|\mathbf{5}|4]^4
}{
\langle23\rangle^2\,\langle34\rangle\,\tau_{12}\,\tau_{45}
[2|\boldsymbol{1}(2+3)|4]
\langle 3|\boldsymbol{5}(2+3)\boldsymbol{1}|2 ]
} \ .&& %\\
     %&\text{and } \hspace{2mm} Z_{12} = \left(\frac{(\langle \boldsymbol{1} 3\rangle  [\boldsymbol{5} 4]+[\boldsymbol{1} 4] \langle \boldsymbol{5} 3\rangle ) \langle 3|\boldsymbol{1}|2]+\langle 3|2|4] [\boldsymbol{1} 2] \langle \boldsymbol{5} 3\rangle }{\langle 3|-\boldsymbol{5}|4] \langle 3|\boldsymbol{1}|2]}\right) \quad .&&
\end{flalign*}

\paragraph{Channel 2: $M(\mathbf{1}^s,\hat{3}^-,4^+|\hat{2}^+,\mathbf{5}^s)$}

\begin{flalign*}
    \text{In fig.~\ref{fig:graviton-five-point-single-minus-bcfw-channel-25}, } & \tau_{5\hat{2}}=0 \implies z=\frac{\tau_{52}}{\langle3|\mathbf{5}|2]} \ .&&
\end{flalign*}
\begin{figure}[H]
    \centering
    \begin{fmffile}{fig_4_2}
 \begin{fmfgraph*}(200,60)% units are now in cm
   \fmfright{i1,i2}
   \fmfleft{o1,o2,o3}
   \fmfblob{.08w}{g1,g2}
   \fmf{plain, label=$\boldsymbol{5}^s$}{i1,g1}
   \fmf{dbl_wiggly, label=$\hat{2}^+$}{i2,g1}
   \fmf{plain, label=$-\hat{\boldsymbol{P}}^s$}{g1,v1}
   \fmf{phantom}{v1,v2}
   \fmf{plain, label=$\hat{\boldsymbol{P}}^s$}{v2,g2}
   \fmf{dbl_wiggly, label=$\hat{3}^-$}{g2,o3}
   \fmf{dbl_wiggly, label=$4^+$}{g2,o2}
   \fmf{plain, label=$\boldsymbol{1}^s$}{g2,o1}
\end{fmfgraph*}
\end{fmffile}
    \caption{Factorization channel contributing to $M(\mathbf{1}^s,\hat{3}^-,4^+|\hat{2}^+,\mathbf{5}^s)$}
    \label{fig:graviton-five-point-single-minus-bcfw-channel-25}
\end{figure}
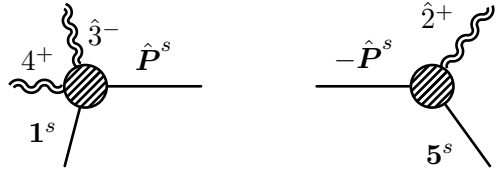
\begin{align}
    &M(\mathbf{1}^s,\hat{3}^-,4^+|\hat{2}^+,\mathbf{5}^s)= \hspace{2mm} M(\mathbf{1}^s,\hat{3}^-,4^+, \hat{\mathbf{P}}^s) \frac{i}{\tau_{52}}M(-\hat{\mathbf{P}},\hat{2}^+,\mathbf{5}^s)\nonumber\\
    =& \, i\,(-1)^{2s}\frac{\langle3|\mathbf{5}|2]^{4-2s}\langle3|\mathbf{1}|4]^{4-2s}\left((\langle 3 \boldsymbol{5}\rangle  [\boldsymbol{1} 4]+[\boldsymbol{5} 4] \langle 3 \boldsymbol{1}\rangle ) \langle 3|\boldsymbol{5}|2]+\langle 3|2|4] [\boldsymbol{5} 2] \langle 3 \boldsymbol{1}\rangle\right)^{2s}}{\langle23\rangle^2\tau_{52}\tau_{14}\left(\langle3|\mathbf{5}|2]s_{34}+\langle3|4|2]\tau_{52}\right)\left(\langle3|\mathbf{5}|2]\tau_{13}+\langle3|\mathbf{1}|2]\tau_{52}\right)}\nonumber\\
    =& M_{25}(\mathbf{1}^0,\hat{2}^+,\hat{3}^-,4^+,\mathbf{5}^0)\: Z_{25}^{2s}\ , \label{eq:graviton-five-point-single-minus-bcfw-channel-25}
\end{align}
\begin{flalign*}
    &\text{where } \hspace{2mm} M_{25}(\mathbf{1}^0,\hat{2}^+,\hat{3}^-,4^+,\mathbf{5}^0)=\, -i\, \frac{
\langle3|\mathbf{5}|2]^4
\langle3|\mathbf{1}|4]^4
}{
\langle23\rangle^2\,\langle34\rangle\,\tau_{52}\,\tau_{14}
[2|\boldsymbol{5}(2+3)|4]
\langle 3|\boldsymbol{1}(2+3)\boldsymbol{5}|2 ]} \ . &&%\\
    %&\text{and } \hspace{2mm} Z_{25}=\left(\frac{(\langle 3 \boldsymbol{5}\rangle  [\boldsymbol{1} 4]+[\boldsymbol{5} 4] \langle 3 \boldsymbol{1}\rangle ) \langle 3|\boldsymbol{5}|2]+\langle 3|2|4] [\boldsymbol{5} 2] \langle 3 \boldsymbol{1}\rangle }{\langle 3|\boldsymbol{1}|4] \langle 3|-\boldsymbol{5}|2]}\right). &&
\end{flalign*}

\paragraph{Channel 3: $M(\mathbf{1}^s,\mathbf{5}^s,\hat{3}^-|\hat{2}^+,4^+)$}

\begin{flalign*}
    &\text{In fig.~\ref{fig:graviton-five-point-single-minus-bcfw-channel-24}, } \hspace{2mm}s_{\hat{2}4}=0\implies \langle2|4|2]-z\langle3|4|2]=0 \implies z=\frac{\langle24\rangle}{\langle34\rangle} \ \text{, when } [24]\neq0. &&
\end{flalign*}
\begin{figure}[H]
    \centering
    \begin{fmffile}{fig_4_3}
 \begin{fmfgraph*}(200,60)% units are now in cm
   \fmfright{i1,i2}
   \fmfleft{o1,o2,o3}
   \fmfblob{.08w}{g1,g2}
   \fmf{dbl_wiggly}{i1,g1}
   \fmfv{label=$4^+$}{i1}
   \fmf{dbl_wiggly}{i2,g1}
   \fmfv{label=$\hat{2}^+$}{i2}
   \fmf{dbl_wiggly, label=$-\hat{P}^-$}{g1,v1}
   \fmf{phantom}{v1,v2}
   \fmf{dbl_wiggly, label=$\hat{P}^+$}{v2,g2}
   \fmf{dbl_wiggly}{g2,o3}
   \fmfv{label=$\hat{3}^-$}{o3}
   \fmf{plain}{g2,o2}
   \fmfv{label=$5^s$}{o2}
   \fmf{plain}{g2,o1}
   \fmfv{label=$\boldsymbol{1}^s$}{o1}
\end{fmfgraph*}
\end{fmffile}
\vspace{5mm}
    \caption{Factorization channel contributing to $M(\mathbf{1}^s,\mathbf{5}^s,\hat{3}^-|\hat{2}^+,4^+)$}
    \label{fig:graviton-five-point-single-minus-bcfw-channel-24}
\end{figure}
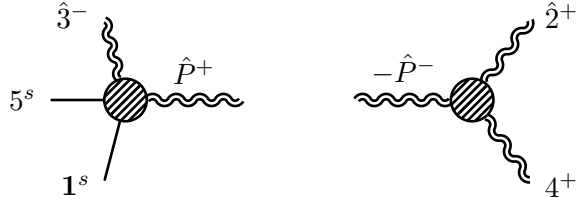
\begin{align}
    &M(\mathbf{1}^s,\mathbf{5}^s,\hat{3}^-|\hat{2}^+,4^+)=M(\mathbf{1}^s,\mathbf{5}^s,\hat{3}^-,\mathbf{P}^+) \frac{i}{s_{24}} M(-\mathbf{P}^-,\hat{2}^+,4^+)\nonumber\\
    =\hspace{2mm}&i\,(-1)^{2s}\frac{[24]^{4}\langle3|\mathbf{15}|3\rangle^{4-2s}(\langle\boldsymbol{5}3\rangle([\boldsymbol{1}|2|3\rangle + [\boldsymbol{1}|4|3\rangle) + \langle\boldsymbol{1}3\rangle([\boldsymbol{5}|2|3\rangle + [\boldsymbol{5}|4|3\rangle))^{2s}}{\langle23\rangle^2s_{24}s_{15}\left(\langle3|4|2]\tau_{35}+s_{24}\langle3|\mathbf{5}|2]\right)\left(\langle3|4|2]\tau_{13}+s_{24}\langle3|\mathbf{1}|2]\right)}\nonumber\\
    =\hspace{2mm}&M_{24}(\mathbf{1}^0,\mathbf{5}^0,\hat{3}^-|\hat{2}^+,4^+)\:Z_{24}^{2s}\ ,\label{eq:graviton-five-point-single-minus-bcfw-channel-24}
\end{align}
\begin{flalign*}
    &\text{where }\hspace{2mm} M_{24}(\mathbf{1}^0,\hat{2}^+,\hat{3}^-,4^+,\mathbf{5}^0)=\,i\,\frac{
[24]^2
\langle3|\mathbf{15}|3\rangle^4
}{
\langle23\rangle^2\,s_{24}\,s_{15}
\langle3|\boldsymbol{5}(2+3)|4\rangle \langle3|\boldsymbol{1}(2+3)|4\rangle
} \ .&&%\\
    %&\text{and } \hspace{2mm} Z_{24}= \frac{\langle\boldsymbol{5}3\rangle([\boldsymbol{1}|2|3\rangle + [\boldsymbol{1}|4|3\rangle) + \langle\boldsymbol{1}3\rangle([\boldsymbol{5}|2|3\rangle + [\boldsymbol{5}|4|3\rangle)}{\langle3|\mathbf{15}|3\rangle}\quad .&&
\end{flalign*}

\paragraph{Channel 4: $M(\mathbf{1}^s,\mathbf{5}^s,\hat{2}^+|\hat{3}^-,4^+)$}

\begin{flalign*}
    &\text{In fig.~\ref{fig:graviton-five-point-single-minus-bcfw-channel-34}, } \hspace{2mm}\langle\hat{3}|4|\hat{3}\rangle=0\implies \langle3|4|3]+z\langle3|4|2]=0 \implies z=-\frac{[43]}{[42]} \ \text{, when } \langle34\rangle\neq0.&&
\end{flalign*}
\begin{figure}[H]
    \centering
     \begin{fmffile}{fig_4_4}
 \begin{fmfgraph*}(200,60)% units are now in cm
   \fmfright{i1,i2}
   \fmfleft{o1,o2,o3}
   \fmfblob{.08w}{g1,g2}
   \fmf{dbl_wiggly}{i1,g1}
   \fmfv{label=$4^+$}{i1}
   \fmf{dbl_wiggly}{i2,g1}
   \fmfv{label=$\hat{3}^-$}{i2}
   \fmf{dbl_wiggly, label=$-\hat{P}^-$}{g1,v1}
   \fmf{phantom}{v1,v2}
   \fmf{dbl_wiggly, label=$\hat{P}^+$}{v2,g2}
   \fmf{dbl_wiggly}{g2,o3}
   \fmfv{label=$\hat{2}^+$}{o3}
   \fmf{plain}{g2,o2}
   \fmfv{label=$5^s$}{o2}
   \fmf{plain}{g2,o1}
   \fmfv{label=$\boldsymbol{1}^s$}{o1}
\end{fmfgraph*}
\end{fmffile}
\vspace{5mm}
    \caption{Factorization channel contributing to $M(\mathbf{1}^s,\mathbf{5}^s,\hat{2}^+|\hat{3}^-,4^+)$}
    \label{fig:graviton-five-point-single-minus-bcfw-channel-34}
\end{figure}
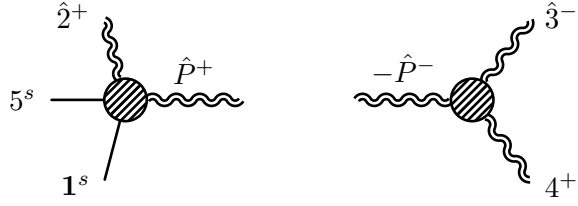
\begin{align}
    &M(\mathbf{5}^s,\mathbf{1}^s,\hat{2}^+|\hat{3}^-,4^+)= M(\mathbf{5}^s,\mathbf{1}^s,\hat{2}^+,\hat{P}^+) \frac{i}{s_{34}}M(-\hat{P}^-,\hat{3}^-,4^+)\nonumber\\
    =&\,i\,(-1)^{2s}\frac{m^{4-2s}[24]^4\langle3|4|2]^4}{[23]^2 s_{15}s_{34}\left(\langle3|4|2]\tau_{12}+\langle3|\mathbf{1}|2]s_{34}\right)\left(\langle3|4|2]\tau_{52}+\langle3|\mathbf{5}|2]s_{34}\right)}\langle\mathbf{51}\rangle^{2s} \nonumber\\
    =&M_{34}(\mathbf{1}^0,\hat{2}^+,\hat{3}^-,4^+,\mathbf{5}^0)\: Z_{34}^{2s}\ , \label{eq:graviton-five-point-single-minus-bcfw-channel-34}
\end{align}
\begin{flalign*}
    &\text{where } \hspace{2mm} M_{34}(\mathbf{1}^0,\hat{2}^+,\hat{3}^-,4^+,\mathbf{5}^0)=\,i\,\frac{
m^4[24]^6\langle3|4|2]^2
}{
[23]^2\,s_{15}\,s_{34}
[2|\boldsymbol{1}(2+3)|4]
[2|\boldsymbol{5}(2+3)|4]
} \ .&&%\\
    %&\text{and } \hspace{2mm} Z_{34} = \frac{\langle\mathbf{15}\rangle}{m} \quad .&&
\end{flalign*}

\paragraph{Channel 5: $M(\mathbf{1}^s,\hat{3}^-|\hat{2}^+,4^+,\mathbf{5}^s)$}

\begin{flalign*}
     &\text{In fig.~\ref{fig:graviton-five-point-single-minus-bcfw-channel-13}, } \langle\hat{3}|\mathbf{1}|\hat{3}]=0\implies \tau_{13}+z\langle3|1|2]=0 \implies z=-\frac{\tau_{13}}{\langle3|\mathbf{1}|2]} \ .&&
\end{flalign*}
\begin{figure}[H]
    \centering
    \begin{fmffile}{fig_4_5}
 \begin{fmfgraph*}(200,60)% units are now in cm
   \fmfleft{i1,i2}
   \fmfright{o1,o2,o3}
   \fmfblob{.08w}{g1,g2}
   \fmf{plain}{i1,g1}
   \fmfv{label=$\boldsymbol{1}^s$}{i1}
   \fmf{dbl_wiggly}{i2,g1}
   \fmfv{label=$\hat{3}^-$}{i2}
   \fmf{plain, label=$\hat{\boldsymbol{P}}^s$}{g1,v1}
   \fmf{phantom}{v1,v2}
   \fmf{plain, label=-$\hat{\boldsymbol{P}}^s$}{v2,g2}
   \fmf{dbl_wiggly}{g2,o3}
   \fmfv{label=$\hat{2}^+$}{o3}
   \fmf{dbl_wiggly}{g2,o2}
   \fmfv{label=$4^+$}{o2}
   \fmf{plain}{g2,o1}
   \fmfv{label=$\boldsymbol{5}^s$}{o1}
\end{fmfgraph*}
\end{fmffile}
\vspace{5mm}
    \caption{Factorization channel contributing to $M(\mathbf{1}^s,\hat{3}^-|\hat{2}^+,4^+,\mathbf{5}^s)$}
    \label{fig:graviton-five-point-single-minus-bcfw-channel-13}
\end{figure}
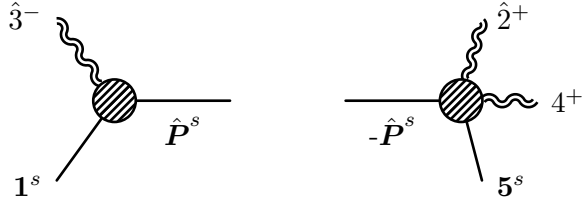
\begin{align}
    &M(\mathbf{1}^s,\hat{3}^-|\hat{2}^+,4^+,\mathbf{5}^s)= M(\mathbf{1}^s,\hat{3}^-,\hat{\boldsymbol{P}}^s)\frac{i}{\tau_{13}}M(-\hat{\boldsymbol{P}}^s,\hat{2}^+,4^+,\mathbf{5}^s)\nonumber\\
    =&\,i\,(-1)^{2s}\frac{m^{4-2s}[24]^4\langle3|\mathbf{1}|2]^{4-2s}\left(\langle \boldsymbol{5}\boldsymbol{1} \rangle  \langle 3|\boldsymbol{1}|2]-\langle 3 \boldsymbol{5}\rangle  \langle 3 \boldsymbol{1}\rangle  [3 2]\right)^{2s}}{[23]^2\tau_{45}\tau_{13}\left(\langle3|\mathbf{1}|2]s_{24}+\langle3|4|2]\tau_{13}\right)\left(\langle3|\mathbf{1}|2]\tau_{52}+\langle3|\mathbf{5}|2]\tau_{13}\right)}\nonumber\\
    =& M_{13}(\mathbf{1}^0,\hat{2}^+,\hat{3}^-,4^+,\mathbf{5}^0)\:Z_{13}^{2s}\ ,\label{eq:graviton-five-point-single-minus-bcfw-channel-13}
\end{align}
\begin{flalign*}
    &\text{where }\hspace{2mm} M_{13}(\mathbf{1}^0,\hat{2}^+,\hat{3}^-,4^+,\mathbf{5}^0)= \,-i\,\frac{
m^4[24]^3\langle3|\mathbf{1}|2]^4
}{
[23]^2\,\tau_{45}\,\tau_{13}
\langle3|\boldsymbol{1}(2+3)|4\rangle
\langle 3|\boldsymbol{1}(2+3)\boldsymbol{5}|2 ]
}\ .&& %\\
    %&\text{and } \hspace{2mm} Z_{13}=\frac{\langle\mathbf{15}\rangle\langle3|\mathbf{1}|2]+\langle\mathbf{1}3\rangle\langle\mathbf{5}3\rangle[32]}{m\langle3|\mathbf{1}|2]} \quad .&&
\end{flalign*}

\paragraph{Channel 6: $M(\mathbf{1}^s,\hat{2}^+,4^+|\hat{3}^-,\mathbf{5}^s)$}

\begin{flalign*}
    &\text{In fig.~\ref{fig:graviton-five-point-single-minus-bcfw-channel-35}, }\langle\hat{3}|\mathbf{5}|\hat{3}]=0\implies \tau_{53}+z\langle3|5|2]=0 \implies z=-\frac{\tau_{53}}{\langle3|\mathbf{5}|2]} \ . &&
\end{flalign*}
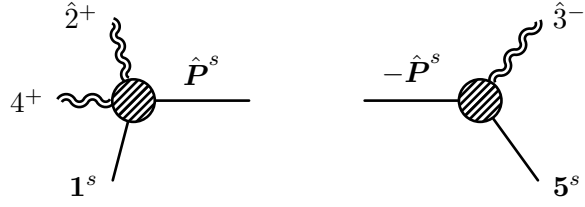
\begin{figure}[H]
    \centering
    \begin{fmffile}{fig_4_6}
 \begin{fmfgraph*}(200,60)% units are now in cm
   \fmfright{i1,i2}
   \fmfleft{o1,o2,o3}
   \fmfblob{.08w}{g1,g2}
   \fmf{plain}{i1,g1}
   \fmfv{label=$\boldsymbol{5}^s$}{i1}
   \fmf{dbl_wiggly}{i2,g1}
   \fmfv{label=$\hat{3}^-$}{i2}
   \fmf{plain, label=$-\hat{\boldsymbol{P}}^s$}{g1,v1}
   \fmf{phantom}{v1,v2}
   \fmf{plain, label=$\hat{\boldsymbol{P}}^s$}{v2,g2}
   \fmf{dbl_wiggly}{g2,o3}
   \fmfv{label=$\hat{2}^+$}{o3}
   \fmf{dbl_wiggly}{g2,o2}
   \fmfv{label=$4^+$}{o2}
   \fmf{plain}{g2,o1}
   \fmfv{label=$\boldsymbol{1}^s$}{o1}
\end{fmfgraph*}
\end{fmffile}
\vspace{5mm}
    \caption{Factorization channel contributing to $M(\mathbf{1}^s,\hat{2}^+,4^+|\hat{3}^-,\mathbf{5}^s)$}
    \label{fig:graviton-five-point-single-minus-bcfw-channel-35}
\end{figure}
\begin{align}
    &M(\mathbf{1}^s,\hat{2}^+,4^+|\hat{3}^-,\mathbf{5}^s)= M(\mathbf{1}^s,\hat{2}^+,4^+,\hat{\boldsymbol{P}}^s) \frac{i}{\tau_{53}}M(-\hat{\boldsymbol{P}}^s,\hat{3}^-,\mathbf{5}^s)\nonumber\\
    =&\,i\,(-1)^{2s}\frac{ m^{4-2s}[24]^4\langle3|\mathbf{5}|2]^{4-2s}\left(\langle \boldsymbol{5}\boldsymbol{1} \rangle  \langle 3|\boldsymbol{5}|2]+\langle 3 \boldsymbol{5}\rangle  \langle 3\boldsymbol{1} \rangle  [3 2]\right)^{2s}}{[23]^2\tau_{14}\tau_{53}\left(\langle3|\mathbf{5}|2]s_{24}+\langle3|4|2]\tau_{53}\right)\left(\langle3|\mathbf{5}|2]\tau_{12}+\langle3|\mathbf{1}|2]\tau_{53}\right)}\nonumber\\
    =& M_{35}(\mathbf{1}^0,\hat{2}^+,\hat{3}^-,4^+,\mathbf{5}^0)\: Z_{35}^{2s} \ ,\label{eq:graviton-five-point-single-minus-bcfw-channel-35}
\end{align}
\begin{flalign*}
    &\text{where } \hspace{2mm} M_{35}(\mathbf{1}^0,\hat{2}^+,\hat{3}^-,4^+,\mathbf{5}^0)=\,-i\,\frac{
m^4[24]^3\langle3|\mathbf{5}|2]^4
}{
[23]^2\,\tau_{14}\,\tau_{53}
\langle3|\boldsymbol{5}(2+3)|4\rangle
\langle 3|\boldsymbol{5}(2+3)\boldsymbol{1}|2 ]
} \ .&&%\\
    %&\text{and } \hspace{2mm} %Z_{35}=\frac{\langle\mathbf{15}\rangle\langle3|\mathbf{5}|2]++\langle\mathbf{1}3\rangle\langle\mathbf{5}3\rangle[23]}{m\langle3|\mathbf{5}|2]} \quad .&&
\end{flalign*}

Now that we have calculated the contributions from all six channels in fig.~\ref{fig:graviton-five-point-single-minus-bcfw-channel-12}-- \ref{fig:graviton-five-point-single-minus-bcfw-channel-35}, we get the full amplitude following eq.~\eqref{eq:graviton-five-point-single-minus-bcfw-decomposition}.

\begin{equation}\label{eq:graviton-five-point-single-minus-bcfw-amplitude}
\boxed{M(\mathbf{1}^s,2^+,3^-,4^+,\mathbf{5}^s)
=
\sum_{\substack{
 u\in \{2,3\}\\
 v\in \{1,4,5\}
}}
M_{\mathcal{O}(u,v)} \, Z_{\mathcal{O}(u,v)}^{2s}}\ .
\end{equation}
where $\mathcal{O}(u,v)$ is a function which orders $u$ and $v$ in ascending order and $Z_{12}$, $Z_{25}$, $Z_{24}$, $Z_{34}$, $Z_{13}$ and $Z_{35}$ have the form mentioned in eq.~\eqref{eq:bcfw-channel-spin-structures}. The $M_{\mathcal{O}(u,v)}$ coefficients in eq.~\eqref{eq:graviton-five-point-single-minus-bcfw-amplitude} take the following forms: 
\begin{flalign*}
&\left\{
\begin{aligned}
M_{12}(\mathbf{1}^0,\hat{2}^+,\hat{3}^-,4^+,\mathbf{5}^0)
&= \,- i \,\frac{
\langle3|\mathbf{1}|2]^4
\langle3|\mathbf{5}|4]^4
}{
\langle23\rangle^2\,\langle34\rangle\,\tau_{12}\,\tau_{45}
[2|\boldsymbol{1}(2+3)|4]
\langle 3|\boldsymbol{5}(2+3)\boldsymbol{1}|2 ]
}\ , \\[6pt]
M_{25}(\mathbf{1}^0,\hat{2}^+,\hat{3}^-,4^+,\mathbf{5}^0)
&= \, -i \, \frac{
\langle3|\mathbf{5}|2]^4
\langle3|\mathbf{1}|4]^4
}{
\langle23\rangle^2\,\langle34\rangle\,\tau_{52}\,\tau_{14}
[2|\boldsymbol{5}(2+3)|4]
\langle 3|\boldsymbol{1}(2+3)\boldsymbol{5}|2 ]
}\ , \\[6pt]
M_{24}(\mathbf{1}^0,\hat{2}^+,\hat{3}^-,4^+,\mathbf{5}^0)
&= \, i \, \frac{
[24]^2
\langle3|\mathbf{15}|3\rangle^4
}{
\langle23\rangle^2\,s_{24}\,s_{15}
\langle3|\boldsymbol{5}(2+3)|4\rangle \langle3|\boldsymbol{1}(2+3)|4\rangle
}\ , \\[6pt]
M_{34}(\mathbf{1}^0,\hat{2}^+,\hat{3}^-,4^+,\mathbf{5}^0)
&= \, i \,\frac{
m^4[24]^6\langle3|4|2]^2
}{
[23]^2\,s_{15}\,s_{34}
[2|\boldsymbol{1}(2+3)|4]
[2|\boldsymbol{5}(2+3)|4]
}\ , \\[6pt]
M_{13}(\mathbf{1}^0,\hat{2}^+,\hat{3}^-,4^+,\mathbf{5}^0)
&= \, -i \, \frac{
m^4[24]^3\langle3|\mathbf{1}|2]^4
}{
[23]^2\,\tau_{45}\,\tau_{13}
\langle3|\boldsymbol{1}(2+3)|4\rangle
\langle 3|\boldsymbol{1}(2+3)\boldsymbol{5}|2 ]
}\ , \\[6pt]
M_{35}(\mathbf{1}^0,\hat{2}^+,\hat{3}^-,4^+,\mathbf{5}^0)
&= \, -i \, \frac{
m^4[24]^3\langle3|\mathbf{5}|2]^4
}{
[23]^2\,\tau_{14}\,\tau_{53}
\langle3|\boldsymbol{5}(2+3)|4\rangle
\langle 3|\boldsymbol{5}(2+3)\boldsymbol{1}|2 ]
}\ .
\end{aligned}
\right.
&&
\end{flalign*}

We have verified that eq.~\eqref{eq:graviton-five-point-single-minus-bcfw-amplitude} agrees with the amplitude in eq.~\eqref{single_minus_five_point_graviton_Compton} calculated using the KLT double copy.

\subsubsection{
Case II:
\texorpdfstring{$M(\boldsymbol{1}^s, 2^+, 3^+, 4^+, \boldsymbol{5}^s)$}
               {}
}\label{subsubsec:graviton-five-point-all-plus-bcfw}
In this case, again, we use the $[3,2\rangle$ BCFW shift. Even though a massless--massless shift usually leads to six channels in gravity, the $\hat{3}4$ channel drops out in this case, because of the anti-MHV kinematics in the three-point amplitude involved in that channel.
All the five remaining channels have the $(\langle \boldsymbol 1 \boldsymbol 5 \rangle / m)^{2s}$ spin dependence.
Upon summing their contributions, we get:
\begin{align}
    & \boxed{\begin{aligned}
        & M(\boldsymbol{1}^s,2^+,3^+,4^+,\boldsymbol{5}^s)\\
    =& i \, m^{4-2s}\left(\frac{s_{15}^3\: [2 4]^2}{s_{24} \:\langle 2 3\rangle ^2\: \langle 3|\boldsymbol{1}{(2+3)}|4\rangle \: \langle 3|\boldsymbol{5}{(2+3)}|4\rangle}\right.\\
    &\hspace{2cm}\left.- \Bigg\{\frac{1}{\langle 3|\boldsymbol{5}{(2+3)}\boldsymbol{1}|2]} \left(\frac{m^4 \:[2 3]^2 \:[2 4]^3}{\tau _{14}\: \tau _{53}\: \langle 3|\boldsymbol{5}{(2+3)}|4\rangle }+\frac{[2|\boldsymbol{1}{(2+3)}|4]^3}{\tau _{12} \:\tau _{54}\: \langle 2 3\rangle ^2 \:\langle 3 4\rangle }\right)\right.\\
    &\left.\hspace{10cm}+(1\leftrightarrow 5)\Bigg\}\right)\langle\boldsymbol{15}\rangle^{2s} \ . 
    \end{aligned}} \label{eq:graviton-five-point-all-plus-bcfw-amplitude}
\end{align}

We have verified that eq.~\eqref{eq:graviton-five-point-all-plus-bcfw-amplitude} agrees with the all-plus five-point graviton Compton amplitude in eq.~\eqref{all_plus_five_point_graviton_Compton} from KLT relations.

%----------------------------------------------------------------------------

\section{Conclusion}\label{sec:conclusion}
We have five-point Compton amplitudes for minimally coupled spinning matter in the massive spinor-helicity formalism using BCFW recursion relation in both gauge theory and gravity. We obtain results that are extremely compact relative to their covariant $D$-dimensional counterpart. The amplitudes are written in a form that is simultaneously valid for a range of massive spin $s$: $s \leq 1$ in gauge theory and $s \leq 2$ in gravity. These five-point Compton amplitudes are direct and full extensions of the AHH Compton amplitudes \cite{Arkani-Hamed:2017jhn} to the five-point case, whereas a few previous studies focused on specific massive spins \cite{Ochirov:2018uyq,Ballav:2020ese} or on classical limits of the amplitudes \cite{Bjerrum-Bohr:2023jau, Vazquez-Holm:2025ztz}. We also observe the universality of spin structures, shared between gauge theory and gravity, for a given helicity configuration under the same chosen BCFW shift.
%Here the spin-structures depend on the helicity configuration, chosen shift and the BCFW channels, but they are agnostic to the color-orderings in Yang-Mills theory.

We notice that BCFW recursion leads to unphysical spurious singularities in individual channels in gauge theory and gravity. But we have been able to eliminate those spurious singularities by analyzing the spurious singularities and the spin structures in case of the five-point Yang--Mills Compton amplitudes, through a cancellation mechanism that holds for all massive spins $s \leq 1$. Using the Yang--Mills Compton amplitudes free of spurious singularities, we have calculated five-point QED Compton amplitudes by performing permutation sum over the massless gluon legs and five-point graviton Compton amplitudes by using KLT relations, and the resulting amplitudes inherit the lack of spurious poles of the input Yang--Mills amplitudes, again for $s \leq 1$ in QED $s \leq 2$ in gravity. Alternatively, we have also calculated the QED and gravity amplitudes with BCFW recursion directly, which generally lead to more compact results, though with apparent spurious poles. Note that the lack of color ordering leads to a proliferation of BCFW channels in QED and gravity, which is another appeal of the former strategy of calculating Yang--Mills amplitudes first; this consideration will be even more relevant for future calculations at higher points, where the number of BCFW channels grows further. A calculation is already underway for the single-minus amplitude with an arbitrary number of massless gluons / gravitons \cite{Das:inprep}.

We have performed various cross-checks of these amplitudes against the $D$-dimensional amplitudes computed using dimensional reduction of massless amplitudes from the Mathematica package \texttt{IncreasingTrees} \cite{Edison:2020ehu}, and against various amplitude relations such as reflection, KK, BCJ and KLT double copy relations, using analytic parametrizations of spinor components in terms of a minimal number independent variables that eliminate ambiguities in spinor expressions.

Through generalized unitarity \cite{Bern:1994zx, Bern:1994cg, Bern:1997sc, Britto:2004nc}, the five-point tree-level graviton Compton amplitudes are important building blocks for various gravitational scattering processes in the post-Minkowskian expansion, e.g., one-loop / 2nd post-Minkowskian gravitational waveforms \cite{Brandhuber:2023hhy, Herderschee:2023fxh, Georgoudis:2023lgf, Bini:2023fiz, Bini:2024rsy, Alessio:2024wmz, Brunello:2025eso} , two-loop / 3rd post-Minkowskian black hole binary scattering \cite{Bern:2019nnu, Bern:2019crd, Cheung:2020gyp, Kalin:2020fhe, Bjerrum-Bohr:2021din, Brandhuber:2021eyq, DiVecchia:2021bdo, Herrmann:2021lqe, Herrmann:2021tct, Bern:2020buy, FebresCordero:2022jts, Akpinar:2024meg, Akpinar:2025bkt, Bjerrum-Bohr:2026fhx, Damour:2020tta, Mogull:2020sak, Jakobsen:2021smu, Jakobsen:2021lvp, Jakobsen:2022fcj, Jakobsen:2022psy}, as well as loop corrections to the five-point Compton amplitudes themselves. While $D$ dimensional tree amplitudes are most commonly used, it was shown in ref.~\cite{Bern:2019crd} that compact four-dimensional amplitudes are sufficient as building blocks for obtaining classical scattering results even at two loops. The spinning amplitudes of this paper can potentially enable elegant and efficient calculations of spin corrections in the aforementioned scattering processes.

One very natural extension of our calculation is to determine the Compton amplitudes for general massive higher spin, beyond $s=1$ in gauge theory and $s=2$ in gravity, by adding contact terms to the amplitudes to remove spurious poles. In principle, this can be done by constructing a general ansatz for the five-point Compton amplitude and then imposing appropriate conditions expected to be satisfied by the amplitudes. In the four-point case, this has been done through the formalism of chiral Lagrangians and massive higher-spin symmetry \cite{Ochirov:2022nqz, Cangemi:2023ysz, Cangemi:2022bew, Cangemi:2023bpe}. For gravity, it would be important to match the five-point amplitudes against the nonlinear source terms in second-order black-hole perturbation theory in gauge-invariant formalisms \cite{Garat:1999vr, Campanelli:1998jv, Spiers:2023mor, Spiers:2023cip}. Such a matching could also connect amplitude-based observables to nonlinear tidal response \cite{Riva:2023rcm, Iteanu:2024dvx, Combaluzier-Szteinsznaider:2024sgb, Parra-Martinez:2025bcu}.

%----------------------------------------------------------------------------

\acknowledgments
We are grateful to Alexander Ochirov, Rafael Aoude, Fabian Bautista and Dogan Akpinar for important discussion regarding this project. M.Z.'s work is supported in part by the U.K. Royal Society through Grant URF\textbackslash{}R\textbackslash{}251022. R.D.\ is supported by a PhD studentship funded by the same grant. For the purpose of open access, the authors have applied a Creative Commons Attribution (CC BY) license to any Author Accepted Manuscript version arising from this submission.

%----------------------------------------------------------------------------

\appendix

\section{Conventions}\label{app:conventions}

We work with mostly minus signature for the metric: $\eta_{\mu\nu}=diag\,(+,-,-,-)$. In the Weyl basis, the Dirac gamma matrices take the explicit form
\begin{equation}\label{eq:conventions-weyl-gamma-matrices}
    \gamma^{\mu}=\left(\begin{array}{cc}
0 & \left(\sigma^{\mu}\right)_{\alpha \dot{\alpha}}\\
\left(\bar{\sigma}^{\mu}\right)^{\dot{\alpha} \alpha} & 0
\end{array}\right) \ ,
\end{equation}
where $\sigma^{\mu}=\left(1, \sigma^{i}\right), \bar{\sigma}^{\mu}=\left(1,-\sigma^{i}\right)$, and $\sigma^{i}$ are the Pauli matrices. The gamma matrices obey the Clifford algebra $\left\{\gamma^{\mu}, \gamma^{\nu}\right\}=2 \eta^{\mu \nu}$.

The massless momenta can be expressed in terms of on-shell variables:
\begin{equation}\label{eq:conventions-massless-momentum-bispinor}
    \begin{aligned}
q_{\alpha \dot{\alpha}} \equiv q^{\mu}\left(\sigma_{\mu}\right)_{\alpha \dot{\alpha}} & =\lambda_{\alpha} \tilde{\lambda}_{\dot{\alpha}} \equiv|\lambda\rangle_{\alpha}\left[\left.\lambda\right|_{\dot{\alpha}}\right. \ ,\\
q^{\dot{\alpha} \alpha} \equiv q^{\mu}\left(\bar{\sigma}_{\mu}\right)^{\dot{\alpha} \alpha} & =\tilde{\lambda}^{\dot{\alpha}} \lambda^{\alpha} \equiv | \lambda]^{\dot{\alpha}}\left\langle\left.\lambda\right|^{\alpha}\right. \ ;
\end{aligned}
\end{equation}
where $\alpha, \dot{\alpha}$ are $SL(2, \mathbb{C})$ spinor indices.

On the other hand, massive momenta can be expressed as
\begin{equation}\label{eq:conventions-massive-momentum-bispinor}
    \begin{aligned}
p_{\alpha \dot{\alpha}} & =\lambda_{\alpha}{ }^{I} \tilde{\lambda}_{\dot{\alpha} I} \equiv|\lambda\rangle_{\alpha}^{I}\left[\left.\lambda\right|_{\dot{\alpha} I} \ ,\right.  \\
p^{\dot{\alpha} \alpha} & =\tilde{\lambda}_{I}^{\dot{\alpha}} \lambda^{\alpha I} \equiv |\lambda]_{I}^{\dot{\alpha}}\langle\left.\lambda\right|^{\alpha I} \ ,
\end{aligned}
\end{equation}
where $I$ is the $\operatorname{SU}(2)$ little group index. 

Spinor brackets can be contracted in the following manner:
\begin{equation}\label{eq:conventions-spinor-brackets}
    \begin{aligned}
\langle\lambda_{1} \lambda_{2}\rangle & \equiv\langle\lambda_{1}|^{\alpha} | \lambda_{2}\rangle_{\alpha}  \ ,\\
[\lambda_{1} \lambda_{2}] & \equiv[\lambda_{1}|_{\dot{\alpha}}| \lambda_{2}]^{\dot{\alpha}} \ .
\end{aligned}
\end{equation}

We raise and lower spinor and $\operatorname{SU}(2)$ little group indices using the Levi-Civita symbol, which is defined by
\begin{equation}\label{eq:conventions-levi-civita-symbol}
\epsilon^{12}=-\epsilon_{12}=1 \ .
\end{equation}

Spinors and $\operatorname{SU}(2)$ indices are raised and lowered by contracting with the Levi-Civita symbol in the following way,
\begin{equation}\label{eq:conventions-index-raising-lowering}
\lambda^{I}=\epsilon^{I J} \lambda_{J} \ , \quad \lambda_{I}=\epsilon_{I J} \lambda^{J}  \ .
\end{equation}

The on-shell conditions for the massive helicity variables are
\begin{equation}\label{eq:conventions-massive-spinor-onshell-relations}
    \begin{aligned}
\lambda^{\alpha I} \lambda_{\alpha J}=m \delta^{I}{ }_{J} \ , \quad \lambda^{\alpha I} \lambda_{\alpha}{ }^{J}=-m \epsilon^{I J} \ , \quad \lambda^{\alpha}{ }_{I} \lambda_{\alpha J}=m \epsilon_{I J} \ ,  \\
\tilde{\lambda}_{\dot{\alpha}}^{I} \tilde{\lambda}_{J}^{\dot{\alpha}}=-m \delta^{I}{ }_{J} \ , \quad \tilde{\lambda}_{\dot{\alpha}}^{I} \tilde{\lambda}^{\dot{\alpha} J}=m \epsilon^{I J} \ , \quad \tilde{\lambda}_{\dot{\alpha} I} \tilde{\lambda}_{J}^{\dot{\alpha}}=-m \epsilon_{I J} \ . 
\end{aligned}
\end{equation}

We will also choose the following convention instead of a more democratic convention with $i$-s.
\begin{equation}\label{eq:conventions-negated-momentum-spinors}
|-\mathbf{p}\rangle=-|\mathbf{p}\rangle \ , \quad \mid-\mathbf{p}]=\mid \mathbf{p}] \ .
\end{equation}

\textit{Note:} We will use the bold notation for massive spinors introduced in \cite{Arkani-Hamed:2017jhn} to suppress the symmetrization over $\operatorname{SU}(2)$ indices in amplitudes.

\paragraph{Conventions for Dirac Spinors} We will use the following conventions \cite{Ochirov:2018uyq} for the Dirac spinors. 

{\allowdisplaybreaks
\begin{align}
u_{p}^{I}=\left(\begin{aligned}
    {\lambda_{p \alpha}^{I}}\\{\tilde{\lambda}_{p}^{\dot{\alpha} I}}
\end{aligned}\right) \ , \quad \bar{u}_{p}^{I}=\left(\begin{aligned}
    {-\lambda_{p}^{\alpha I}} \quad{\tilde{\lambda}_{p \dot{\alpha}}^{I}}
\end{aligned}\right) \quad \Rightarrow \quad  \left\{\begin{aligned}
&(\slashed{p}-m) u_{p}^{I}=\bar{u}_{p}^{I}(\slashed{p}-m)=0 \ , \\
&\bar{u}_{p}^{I} u_{p}^{J}=2 m \epsilon^{I J} \ , \\
&\bar{u}_{p}^{I} \gamma^{\mu} u_{p}^{J}=2 p^{\mu} \epsilon^{I J} \ , \\
&u_{p}^{I} \bar{u}_{p I}=u_{p}^{I} \epsilon_{I J} \bar{u}_{p}^{J}=\slashed{p}+m \ , 
\end{aligned}\right. \label{eq:conventions-dirac-u-spinors}\\
\nonumber\\
v_{p}^{I}=\left(\begin{aligned}
    {-\lambda_{p \alpha}^{I}}\\{\tilde{\lambda}_{p}^{\dot{\alpha} I}}
\end{aligned}\right), \quad \bar{v}_{p}^{I}=\left(\begin{aligned}
    {\lambda_{p}^{\alpha I}} \quad {\tilde{\lambda}_{p}^{I}}
\end{aligned}\right) \quad \Rightarrow \quad\left\{\begin{aligned}
&(\slashed{p}+m) v_{p}^{I}=\bar{v}_{p}^{I}(\slashed{p}+m)=0 \ , \\
&\bar{v}_{p}^{I} v_{p}^{J}=2 m \epsilon^{I J} \ , \\
&\bar{v}_{p}^{I} \gamma^{\mu} v_{p}^{J}=-2 p^{\mu} \epsilon^{I J}  \ , \\
&v_{p}^{I} \bar{v}_{p I}=v_{p}^{I} \epsilon_{I J} \bar{v}_{p}^{J}=-\slashed{p}+m \ . 
\end{aligned}\right.\label{eq:conventions-dirac-v-spinors}
\end{align}
}

We can see that in our adopted convention in eq.~\eqref{eq:conventions-dirac-u-spinors} and eq.~\eqref{eq:conventions-dirac-v-spinors},
\begin{equation}
    v_{-p}^{I} = u_{p}^{I} \ ; \quad \bar{v}_{-p}^{I} = \bar{u}_{p}^{I} \ ,
\end{equation}
using the definition of spinors with negated momenta in eq.~\eqref{eq:conventions-negated-momentum-spinors}. When sewing across a cut spin-1/2 massive line, the usual fermion propagator is reproduced from little-group-contracted state sums via
\begin{equation}
  \frac {i (\slashed p + m)}{p^2-m^2} = \frac i {p^2-m^2} u_{p}^{I} \bar{u}_{p I}
  = \frac i {p^2-m^2} v_{-p}^{I} \bar{u}_{p I} \ .
\end{equation}

\paragraph{Conventions for Momentum and Polarization Vectors from Bi-spinors}
\label{app:bispinor-four-vector-construction}
For a four-vector $v^\mu$, one can construct bi-spinors $v_{\alpha\dot{\alpha}}=v_\mu (\sigma^\mu)_{\alpha\dot{\alpha}}$ and $v^{\alpha\dot{\alpha}}=v_\mu (\bar\sigma^\mu)^{\alpha\dot\alpha}$, where $\sigma^\mu=(1,\sigma^i)$ and $\bar\sigma^\mu=(1,-\sigma^i)$ for $i=1,2,3$. It is also straightforward to construct the four-vectors from the bi-spinors: $v_{\alpha\dot\alpha}$ and $v^{\alpha\dot\alpha}$. We will demonstrate how we implement this task with the bi-spinor $v_{\alpha\dot\alpha}$. We will then demonstrate how this method is applied to construct the four-vectors for massless and massive momenta, as well as massless and massive polarizations.

Let us take the bi-spinor 

\begin{equation}\label{eq:bispinor-four-vector-matrix}
    v_{\alpha\dot\alpha}= \begin{pmatrix}
v_{1\dot1} & v_{1\dot2}\\
v_{2\dot1} & v_{2\dot2}
\end{pmatrix} \ .
\end{equation}
%%%\begin{equation}
    %%%\text{where }\quad v_{1\dot1}=-v^0 +v^3\quad;\quad v_{2\dot2}=-v^0 -v^3\quad;\quad v_{1\dot2}=v^1-i v^2\quad;\quad v_{2\dot1}=v^1+i v^2\quad .
%%%\end{equation}
One can construct the four-vector $v^\mu=(v^0,v^1,v^2,v^3)$ corresponding to the bi-spinor $v_{\alpha\dot\alpha}$ in eq.~\eqref{eq:bispinor-four-vector-matrix}, where
\begin{equation}\label{eq:bispinor-to-four-vector-components}
    v^0=-\frac{v_{1\dot1}+v_{2\dot2}}{2}\quad;\quad v^1=\frac{v_{1\dot2}+v_{2\dot1}}{2}\quad;\quad v^2=\frac{v_{2\dot1}-v_{1\dot2}}{2i}\quad;\quad v^3=\frac{v_{1\dot1}-v_{2\dot2}}{2}\quad .
\end{equation}

Therefore, as long as we can construct the bi-spinor with the spinor building blocks, we can use eq.~\eqref{eq:bispinor-to-four-vector-components} to construct the four-vector. So, if we have all the spinors corresponding to a vector quantity, as long as we can build the bi-spinor, we, in turn, can construct the four-vector.

We will now briefly go through the bi-spinor corresponding to different quantities. First, we will discuss massless particles, then massive particles.

\textbf{Massless Case: }
Let us assume that $|\lambda\rangle_{\alpha}$ and $|\tilde{\lambda}]^{\dot{\alpha}}$ are the spinors corresponding to a massless particle.

In this case, the momentum bi-spinor $p_{\alpha\dot{\alpha}}$ is as follows:
\begin{equation}\label{eq:bispinor-massless-momentum}
    p_{\alpha\dot{\alpha}}=|\lambda\rangle_\alpha [\tilde\lambda|_{\dot{\alpha}} \ .
\end{equation}

The polarization bi-spinor for the particle is as follows:
\begin{align}\label{eq:bispinor-massless-polarization}
    \epsilon_{\alpha\dot{\alpha}}=\left\{\begin{aligned}
        & \sqrt{2} \ \frac{|\rho\rangle_\alpha [\tilde{\lambda}|_{\dot{\alpha}}}{\langle \rho \lambda \rangle} \quad \text{for helicity }+1\ ;\\
        & \\
        & \sqrt{2} \ \frac{|\lambda\rangle_\alpha [\tilde{\rho}|_{\dot{\alpha}}}{[\tilde{\lambda} \tilde{\rho} ]} \quad \text{for helicity }-1 \ .
    \end{aligned}\right.
\end{align}
In eq.~\eqref{eq:bispinor-massless-polarization}, $|\rho\rangle$ and $|\tilde{\rho}]$ are reference spinors.

\textbf{Massive Case: }
Let us assume that a massive spin-1 particle is represented by the spinors $|\lambda_I\rangle$ and $|\tilde{\lambda}_I]$, where $I=1,2$. Here, $I$ is the $SU(2)$ index, which can be raised and lowered using the Levi-Civita tensor.

The momentum bi-spinor of this particle can be expressed as
\begin{equation}\label{eq:bispinor-massive-momentum}
    p_{\alpha\dot{\alpha}}=|\lambda^I\rangle_\alpha [\tilde{\lambda}_I|_{\dot{\alpha}} \ .
\end{equation}

The polarization bi-spinor of this particle is as follows
\begin{equation}\label{eq:bispinor-massive-polarization}
    (\varepsilon_{IJ})_{\alpha\dot{\alpha}}=\frac{|\lambda_{(I}\rangle_\alpha [\tilde{\lambda}_{J)}|_{\dot{\alpha}}}{ m} \ .
\end{equation}
In eq.~\eqref{eq:bispinor-massive-polarization}, the $\operatorname{SU}(2)$ indices: $I$ and $J$ are symmetrized.

\section{Three-point massless Amplitudes}\label{app:three-point-ym-gravity-amplitudes}
We write down the forms of three-point MHV and anti-MHV sub-amplitudes involved in our BCFW recursion calculation. The color-ordered Yang--Mills amplitudes are denoted by $A[\dots]$ and the gravity amplitudes are denoted by $M(\dots)$.

\paragraph{MHV Amplitude}
In the case of fig.~\ref{fig_MHV}, $[12]=[23]=[34]=0$. 
\vspace{0.8cm}
\begin{figure}[H]
    \centering
    \begin{fmffile}{fig_B_01}
 \begin{fmfgraph*}(60,60)% units are now in cm
   \fmfleft{i2,o3}
   \fmfright{v1}
   \fmfblob{.30w}{g1}
   \fmf{gluon}{g1,i2}
   \fmfv{label=$1^-$}{i2}
   \fmf{gluon}{g1,v1}
   \fmfv{label=$3^+$}{v1}
   \fmf{gluon}{g1,o3}
   \fmfv{label=$2^-$}{o3}
\end{fmfgraph*}
\end{fmffile}
\vspace{0.8cm}
    \caption{{MHV Amplitude}}
    \label{fig_MHV}
\end{figure}
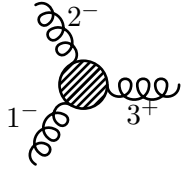
\begin{equation}
    A_{\rm MHV}[1^-,2^-,3^+]=i\,\frac{\langle12\rangle^4}{\langle12\rangle\langle23\rangle\langle31\rangle} \ . \label{eq:MHV_YM_Amp}
\end{equation}
\begin{equation}
    M_{\rm MHV}(1^-,2^-,3^+)=-i\,\left( \frac{\langle12\rangle^4}{\langle12\rangle\langle23\rangle\langle31\rangle}\right)^2 \ . \label{eq:MHV_GR_Amp}
\end{equation}

\paragraph{Anti-MHV Amplitude}
In the case of fig.~\ref{fig_Anti-MHV}, $\langle12\rangle=\langle23\rangle=\langle31\rangle=0$.
\vspace{0.8cm}
\begin{figure}[H]
    \centering
    \begin{fmffile}{fig_B_02}
 \begin{fmfgraph*}(60,60)% units are now in cm
   \fmfleft{i2,o3}
   \fmfright{v1}
   \fmfblob{.30w}{g1}
   \fmf{gluon}{g1,i2}
   \fmfv{label=$1^-$}{i2}
   \fmf{gluon}{g1,v1}
   \fmfv{label=$3^+$}{v1}
   \fmf{gluon}{g1,o3}
   \fmfv{label=$2^-$}{o3}
\end{fmfgraph*}
\end{fmffile}
\vspace{0.8cm}
    \caption{{Anti-MHV Amplitude}}
    \label{fig_Anti-MHV}
\end{figure}
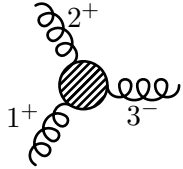
\begin{equation}
    A_{\rm anti-MHV}[1^+,2^+,3^-]=-i\,\frac{[12]^4}{[12][23][31]} \ .\label{eq:anti_MHV_YM_Amp}
\end{equation}
\begin{equation}
    M_{\rm anti-MHV}(1^+,2^+,3^-)=-i\,\left(\frac{[12]^4}{[12][23][31]}\right)^2 \ .\label{eq:anti_MHV_GR_Amp}
\end{equation}

%\section{4-vector Construction from Bi-spinors}

\section{Analytic Parametrization of Spinors}\label{app:analytic-parametrization}
Analytic parametrization of different quantities that are the building blocks of an amplitude is extremely important in the literature, as it serves the purpose in several checks of the amplitudes. With the numerical parametrization, it is tough to pinpoint discrepancies in calculations. In this regard, analytic parametrization provides better insights into the analytic structure and discrepancies in the calculation of amplitudes. We will present the analytic parametrization we used for various checks of our amplitudes in both the four-point and five-point cases. The spinor parametrizations we work with are certainly not unique. We start with the spinor parametrization for amplitudes with massless external legs, available in \cite{Abreu:2019rpt} with momentum twistor variables, and then turn it into the spinor parametrization for Compton scattering.

We present the spinor parametrization for massless scattering that we have chosen below:
\begin{align}
\textit{\textbullet \hspace{2mm}\underline{Four-point Case: }} & \:\begin{cases}
    & \lambda_{(4)}=\begin{pmatrix}
        x_2 & x_2 & 0 & x_2\left(1-1/x_1^2\right)\\
        & & & \\
        x_2 & 0 & x_2 & x_2
    \end{pmatrix} \ ,\\
    & \\
    & \tilde{\lambda}_{(4)}=\begin{pmatrix}
        -x_2(1-x_1^2) & 0 & x_2 & -x_2x_1^2\\
        & & & \\
        -x_2 & x_2 & x_2 & 0
    \end{pmatrix} \ ;
\end{cases}\\
\nonumber\\
\nonumber\\
\textit{\textbullet \hspace{2mm}\underline{Five-point Case: }} & \:\begin{cases}
       & \lambda_{(5)}=\begin{pmatrix}
           1 & -1/x_1^4 & -1/x_1^4 + 1/x_2 & -1/x_1^4 + 1/x_2 + 1/x_3 & 0 \\
           & & & & \\
           0 & 1 & 1 & 1 & 1
       \end{pmatrix} \ ,\\
       & \\
       & \tilde{\lambda}_{(5)}=\begin{pmatrix}
           -1 & -x_2 x_4 & x_3 (x_4 - 1) + x_2 x_4 & x_3 (1-x_4) & 0\\
           & & & & \\
           -x_5/x_4 & -x_1^4 & x_3 (1 - x_5/x_4) & x_3 (x_5/x_4 - 1) & x_1^4
       \end{pmatrix} \ .
   \end{cases}
\end{align}\\

The above parametrizations are provided for spinors with lower
(undotted or dotted) indices. For example,
\begin{equation}
  \lambda_{(4)} = \begin{pmatrix}
    \lambda_{1 \alpha} &
    \lambda_{2 \alpha} &
    \lambda_{3 \alpha} &
    \lambda_{4 \alpha}
  \end{pmatrix} \ .
\end{equation}

We then make the $1$st and $n$-th particle massive by doing the
following parametrization, for both $n=4$ and $n=5$:

\begin{align}
    \nonumber& 1\text{st massive leg: }\quad \begin{cases}
        & |\boldsymbol{1}^{I=1}\rangle=-x_1\frac{2x_0}{1-x_0^2}|1\rangle, \qquad |\boldsymbol{1}^{I=2}\rangle=-x_1\frac{1+x_0^2}{1-x_0^2}|n\rangle\\
        & |\boldsymbol{1}^{I=1}]=\frac{1}{x_1}\frac{1+x_0^2}{1-x_0^2}|n], \qquad |\boldsymbol{1}^{I=2}]=\frac{1}{x_1}\frac{2x_0}{1-x_0^2}|1]
    \end{cases}\\
    &\\
    \nonumber& n\text{-th massive leg: }\quad \begin{cases}
        & |\boldsymbol{n}^{I=1}\rangle=x_1\frac{1+x_0^2}{1-x_0^2}|1\rangle, \qquad |\boldsymbol{n}^{I=2}\rangle=x_1\frac{2x_0}{1-x_0^2}|n\rangle\\
        & |\boldsymbol{n}^{I=1}]=\frac{1}{x_1}\frac{2x_0}{1-x_0^2}|n], \qquad |\boldsymbol{n}^{I=2}]=\frac{1}{x_1}\frac{1+x_0^2}{1-x_0^2}|1]
    \end{cases}
\end{align}

Now that we have all the spinors associated with all the external legs, we need to define the contractions of angular and square spinors, which are:
\begin{align}
  \langle \eta \rho\rangle &= \det (\rho, \eta) = \eta_2 \rho_1 - \eta_1 \rho_2 = \eta^2 \rho^1 - \eta^1 \rho^2 \ ,\\
  [\tilde{\eta} \tilde{\rho}] &= \det (\tilde{\eta}, \tilde{\rho}) = \tilde{\eta}_1 \tilde{\rho}_2 - \tilde{\eta}_2 \tilde{\rho}_1  = \tilde{\eta}^1 \tilde{\rho}^2 - \tilde{\eta}^2 \tilde{\rho}^1 \ .
\end{align}
Note that the explicit component forms are not sensitive to whether you use lower or upper spinor indices. This is because the spinor lowering/uppering matrix $\epsilon$ has unit determinant, and $\det (\rho, \eta) = \det (\epsilon \, (\rho, \eta))$. The momentum of the $i$-th massive leg ($i=1$ or $n$), in the bi-spinor form, is
\begin{equation}
  p_{\alpha \dot \alpha} = \lambda^I_\alpha \tilde \lambda_{\dot \alpha I} =
  | \boldsymbol{i}^{I=2} \rangle_\alpha \langle \boldsymbol{i}^{I=1} |_{\dot \alpha} -
  | \boldsymbol{i}^{I=1} \rangle_\alpha \langle \boldsymbol{i}^{I=2} |_{\dot \alpha} \ .
\end{equation}
Our parametrization ensures that
\begin{equation}
  \det(\lambda^I_\alpha) = \det(\tilde \lambda^I_{\dot \alpha}) \ ,
\end{equation}
and that these determinant values are the same for particle 1 and $n$, so that the two massive legs have an equal mass.

With all this, we can calculate all the bi-spinors and four-vectors necessary in app.~\ref{app:bispinor-four-vector-construction}.

\section{Amplitude Relations Inherited via Dimensional Reduction}\label{app:dimensional-reduction}

The method of dimensional reduction can be used to reduce the Yang--Mills amplitudes in $(D+2)$-dimensions to gluon Compton amplitudes in $D$-dimensions with minimally coupled massive matter particles. This method shows how the gluon Compton amplitudes inherit the properties of Yang--Mills amplitudes, e.g. KK relations, reflection relations, cyclic symmetry, BCJ relations, KLT double copy, etc.

Tree-level color-ordered Yang--Mills amplitudes satisfy
\begin{enumerate}
    \item \textit{Kleiss--Kuijf (KK) Relations} $\quad A[1,\alpha,n,\beta]=(-1)^{|\beta|}\sum_{\sigma \in \alpha \shuffle \beta^T} A[1,\sigma,n] \ ,$
    \item \textit{Reflection Relations} $\quad A[1,2,\dots,n]=(-1)^n\,A[n,\dots,2,1] \ ,$
    \item \textit{Cyclic Symmetry} $\quad A[1,2,3,\dots,n]=A[2,3,\dots,n,1] \ ,$
    \item \textit{BCJ Relations} $\quad \sum_{j=1}^{n-1} \left(\sum_{k=1}^{j} p_k\cdot p_{n+1}\right) A[1,2,\dots,j,n+1,j+1,\dots,n] = 0 \ .$
\end{enumerate}

All of these relations for the Yang--Mills amplitude in $(D+2)$-dimensions can be translated to the case of gluon Compton amplitudes in $D$-dimensions by taking the projections of any of the two external legs in such a way that those legs have the same non-zero mass along with massive polarizations if they represent a massive vector bosonic field.

We can compute the KLT relations in $D$-dimensions for Compton amplitudes starting from the Yang--Mills amplitude in $(D+2)$-dimensions and projecting two external massless legs appropriately. We present the KLT relations for Compton amplitudes in app.~\ref{app:klt-double-copy} explicitly.

\section{The BCJ Basis for Yang--Mills Compton Amplitudes}\label{app:bcj-relations}

Solving the BCJ relations in the previous section for the five-point case gives us all Yang--Mills partial amplitudes in terms of $A[\boldsymbol{1}^s,2,3,4,\boldsymbol{5}^s]$ and $A[\boldsymbol{1}^s,4,3,2,\boldsymbol{5}^s]$.
\begin{empheq}[box=\fbox]{align}
    & A[\boldsymbol{1}^s, 3, 2, 4, \boldsymbol{5}^s]=-\frac{\tau _{12} \left(\tau _{14}+s_{34}\right) A[\boldsymbol{1}^s,2,3,4,\boldsymbol{5}^s]+\tau _{14} \tau _{52} A[\boldsymbol{1}^s,4,3,2,\boldsymbol{5}^s]}{\tau _{13} s_{24}} \ , \label{eq:bcj-relation-ordering-13245}\\
    &\nonumber\\
    & A[\boldsymbol{1}^s, 2, 4, 3, \boldsymbol{5}^s]=\frac{\left(\tau _{12}+s_{24}\right) \tau _{54} A[\boldsymbol{1}^s,2,3,4,\boldsymbol{5}^s]-\tau _{14} \tau _{52} A[\boldsymbol{1}^s,4,3,2,\boldsymbol{5}^s]}{s_{24} \tau _{53}} \ , \label{eq:bcj-relation-ordering-12435}\\
    &\nonumber\\
    &\nonumber A[3,\boldsymbol{1}^s,4,2,\boldsymbol{5}^s]=\frac{\left(\tau _{13} s_{23} s_{24}-\tau _{14} \left(s_{24}+\tau _{52}\right) \left(\tau _{13}+\tau _{53}\right)\right) A[\boldsymbol{1}^s,4,3,2,\boldsymbol{5}^s]}{\tau _{13} s_{24} \tau _{53}} \label{eq:bcj-relation-ordering-31425}\\
    &\qquad
    \qquad\qquad
    \qquad\qquad
    \qquad
    \qquad\qquad+\frac{\tau _{12} \left(\tau _{13}+\tau _{53}\right) \tau _{54} A[\boldsymbol{1}^s,2,3,4,\boldsymbol{5}^s]}{\tau _{13} s_{24} \tau _{53}} \ , \\
    &\nonumber\\
    & A[2, \boldsymbol{1}^s, 4, 3, \boldsymbol{5}^s]=\frac{\left(s_{23}-\tau _{14}\right) A[\boldsymbol{1}^s,4,3,2,\boldsymbol{5}^s]}{\tau _{53}}-\frac{\tau _{54} A[\boldsymbol{1}^s,2,3,4,\boldsymbol{5}^s]}{\tau _{53}} \ . \label{eq:bcj-relation-ordering-21435}
\end{empheq}
All other amplitudes in the $(2^+, 3^\pm, 4^+)$ case can be obtained from a $2 \leftrightarrow 4$ exchange.

\section{KLT Double Copy}\label{app:klt-double-copy}

We present the KLT double copy relations \cite{Kawai:1985xq, Bjerrum-Bohr:2004vlu, Bjerrum-Bohr:2010mia, Bjerrum-Bohr:2016axv, Cao:2021dcd} at tree level for three-, four-, and five-point Compton amplitudes \cite{Bern:2019crd}. One of the advantages of the Compton amplitudes calculated in the MSH formalism is that double-copying the YM-Compton amplitudes directly leads to the graviton Compton amplitudes without any by-products of axions and dilatons. 

\begin{empheq}[box=\fbox]{align}
    &\text{3-point: }\quad M(\boldsymbol{1}^s,\, 2,\, \boldsymbol{3}^s) = i\, A[\boldsymbol{1}^{s_1},\, 2,\, \boldsymbol{3}^{s_1}]\: A[\boldsymbol{1}^{s_2},\, 2,\, \boldsymbol{3}^{s_2}] \ ;\label{eq:klt-three-point-relation}\\
    &\text{4-point: }\quad   M(\boldsymbol{1}^s,\, 2,\, 3,\, \boldsymbol{4}^s)=  -i\,\tau_{12} \: A[\boldsymbol{1}^{s_1},\, 2,\, 3,\, \boldsymbol{4}^{s_1}]\: A[\boldsymbol{1}^{s_2},\, 2,\, \boldsymbol{4}^{s_2},\, 3] \ ;\label{eq:klt-four-point-relation}\\
    &\nonumber\text{five-point: }\quad M(\boldsymbol{1}^s,2,3,4,\boldsymbol{5}^s)= i \,\tau_{12} \,s_{34} \,A[\boldsymbol{1}^{s_1},2,3,4,\boldsymbol{5}^{s_1}]\: A[2,\boldsymbol{1}^{s_2},4,3,\boldsymbol{5}^{s_2}] \\
    &\hspace{5.5cm}+\: i\,\tau_{13}\, s_{24}\, A[\boldsymbol{1}^{s_1},3,2,4,\boldsymbol{5}^{s_1}]\: A[3,\boldsymbol{1}^{s_2},4,2,\boldsymbol{5}^{s_2}] \ ;\label{eq:klt-five-point-relation}
\end{empheq}
where $s=s_1 + s_2$.

We verify that the graviton Compton amplitudes we have calculated using BCFW recursion match those obtained from the double copy of the gluon Compton amplitudes.

\section{Three- and Four-Point Photon Amplitudes from Feynman Rules}\label{app:qed-feynman-rules}

We will calculate the spinor-helicity form of three-point amplitude starting from the QED Lagrangians. All external momenta are outgoing, as in the rest of the paper.

\subsection{Scalar three-point Amplitude}
First, we will start with scalar QED Lagrangian.
{
\allowdisplaybreaks
\begin{align}
    \mathcal{L} = \mathcal{D}\varphi \, (\mathcal{D}\varphi)^* -\frac{1}{4} F^{\mu\nu}F_{\mu\nu}  \ , \label{eq:scalar-qed-lagrangian}
\end{align}
}
where $\mathcal{D}^\mu = \partial^\mu - i \, e \, A^\mu$ and $F_{\mu\nu}$ is the gauge field strength.

Therefore, the interacting Lagrangian in this case is as follows.
{
\allowdisplaybreaks
\begin{align}
    \mathcal{L}_{int} = i \, e \, (\partial_\mu \varphi \, A^\mu \, \varphi^* - A_\mu \, \partial^\mu \varphi^* \, \varphi) \, + e^2 \, A_\mu A^\mu \varphi\varphi^*  \  . \label{eq:scalar-qed-interaction-lagrangian}
\end{align}
}

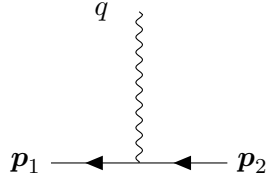
\begin{figure}[h]
    \centering
    \begin{tikzpicture}
\begin{feynman}
    % Define vertices
    \vertex (a) at (0,0) {\(\boldsymbol{p}_1\)};
    \vertex (b) at (3,0) {\(\boldsymbol{p}_2\)};
    \vertex (c) at (1,2) {\(q\)};
    
    % Internal vertices
    \vertex (v1) at (1.5,0);
    \vertex (v3) at (1.5,2);

    % Draw diagram
    \diagram*{
        (b) --  [fermion] (v1) -- [fermion] (a),
        (v1) -- [photon, out=90, in=-90] (v3),
    };
\end{feynman}
\end{tikzpicture} 
    \caption{Three-point Amplitude: Scalar QED}
    \label{fig:scalar-qed-three-point-amplitude}
\end{figure}

Now, following eq.~\eqref{eq:scalar-qed-interaction-lagrangian} we express the three-point Compton amplitude in fig.~\ref{fig:scalar-qed-three-point-amplitude}:
{
\allowdisplaybreaks
\begin{align}
    \mathcal{A}_{(3)} \, = \, i\,  (- i\, e) \, (i \boldsymbol{p}_1 \cdot \varepsilon (k) - i \boldsymbol{p}_2 \cdot \varepsilon (k)) = i \, e \, (\boldsymbol{p}_1 - \boldsymbol{p}_2)\cdot \varepsilon (k)  \  ,\label{eq:scalar-qed-three-point-amplitude}
\end{align}
}
where $\varepsilon$ is the polarization vector of the photon. The standard spinor-helicity forms of the polarization vectors for positive and negative helicities are also given in the bi-spinor component form in eq.~\eqref{eq:bispinor-massless-polarization}.
\[
  \varepsilon (k)_+^\mu \, = \, \frac{\langle \eta | \gamma^\mu | k ]}{\sqrt{2}\,\langle \eta k \rangle}, \qquad \qquad \varepsilon (k)_-^\mu \, = \, \frac{\langle k | \gamma^\mu | \eta ]}{\sqrt{2} \, [ k \eta]}  \  .
  \label{eq:masslessVectPol}
\]
Here, $|\eta\rangle$ and $|\eta ]$ are reference spinors.

Therefore, the three-point amplitude in eq.~\eqref{eq:scalar-qed-three-point-amplitude} takes the following spinor-helicity form:
\begin{equation}
    \mathcal{A}_{(3)}^{(+)} \, = \, i \, \sqrt{2} \, e \, \frac{\langle \eta | \boldsymbol{p}_1 | k ]}{\langle \eta k \rangle}, \qquad \mathcal{A}_{(3)}^{(-)} \, = \, i \, \sqrt{2} \, e \, \frac{\langle k | \boldsymbol{p}_1 | \eta ]}{[ k \eta]}  \  .
\end{equation}

\subsection{Spin-$1/2$ three-point Amplitude}
First, we will start with spin-$1/2$ QED Lagrangian.
\begin{align}
    \mathcal{L}\,=\,\bar{\psi} \, (i \slashed{\mathcal{D}} \, -\, m)\, \psi -\frac{1}{4} F^{\mu\nu}F_{\mu\nu}  \ , \label{eq:spinor-qed-lagrangian}
\end{align}
where $\mathcal{D}^\mu = \partial^\mu - i \, e \, A^\mu$ and $F_{\mu\nu}$ is the gauge field strength.

Therefore, the interacting Lagrangian in this case is as follows.
\begin{align}
    \mathcal{L}_{int} =  e \, \bar{\psi} \, \slashed{A} \, \psi  \  . \label{eq:spinor-qed-interaction-lagrangian}
\end{align}

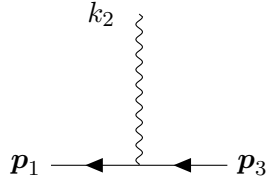
\begin{figure}[h]
    \centering
    \begin{tikzpicture}
\begin{feynman}
    % Define vertices
    \vertex (a) at (0,0) {\(\boldsymbol{p}_1\)};
    \vertex (b) at (3,0) {\(\boldsymbol{p}_3\)};
    \vertex (c) at (1,2) {\(k_2\)};
    
    % Internal vertices
    \vertex (v1) at (1.5,0);
    \vertex (v3) at (1.5,2);

    % Draw diagram
    \diagram*{
        (b) --  [fermion] (v1) -- [fermion] (a),
        (v1) -- [photon, out=90, in=-90] (v3),
    };
\end{feynman}
\end{tikzpicture} 
    \caption{Three-point Amplitude: Spin-$1/2$ QED}
    \label{fig:spinor-qed-three-point-amplitude}
\end{figure}

Now, following eq.~\eqref{eq:spinor-qed-interaction-lagrangian} we express the three-point Compton amplitude in fig.~\ref{fig:spinor-qed-three-point-amplitude}:
\begin{align}
    \mathcal{A}_{(3)} \, = \, i\,  \bar{u}(\boldsymbol{p}_1) \slashed{\varepsilon}(k_2)  v(\boldsymbol{p}_3) \ ,\label{eq:spinor-qed-three-point-amplitude}
\end{align}
where $\varepsilon$ is the polarization vector of the photon. We again use the photon polarizations in eq.~\eqref{eq:masslessVectPol}.
Using the explicit forms of the Dirac spinors $\bar u$ and $v$ in eqs.~\eqref{eq:conventions-dirac-u-spinors} and \eqref{eq:conventions-dirac-v-spinors}, the three-point amplitudes in eq.~\eqref{eq:spinor-qed-three-point-amplitude} take the following spinor-helicity forms:
{
\allowdisplaybreaks
\begin{align}
    \mathcal{A}_{(3)}^{(+)} \, \nonumber&= \, i \, e \, \left(\begin{aligned}
    {-\lambda_{1}^{\alpha}} \quad{\tilde{\lambda}_{1 \dot{\alpha}}}
\end{aligned}\right) \, \left(\begin{aligned}
    0 \quad \varepsilon_{\alpha \dot{\alpha}}^+\\
    \varepsilon^{\dot{\alpha}\alpha}_+ \quad 0
\end{aligned}\right) \, \left(\begin{aligned}
    {-\lambda_{3 \alpha}}\\{\tilde{\lambda}_{3}^{\dot{\alpha}}}
\end{aligned}\right) = \, i \, e\, (- \tilde{\lambda}_{1 \dot{\alpha}} \, \varepsilon^{\dot{\alpha}\alpha}_+ \, \lambda_{3 \alpha} \, - \, \lambda_{1}^{\alpha} \, \varepsilon_{\alpha \dot{\alpha}}^+ \, \tilde{\lambda}_{3}^{\dot{\alpha}})\\
\nonumber& = \, i \, \sqrt{2} \, e \, \frac{(-[\boldsymbol{1}  2] \langle \eta  \boldsymbol{3}\rangle \, - \, \langle\boldsymbol{1} \eta \rangle\, [2 \boldsymbol{3}])}{\langle\eta 2\rangle} \\
\implies\, \Aboxed{\mathcal{A}_{(3)}^{(+)}\, &= \, i \, \sqrt{2} \, e \, \frac{\langle\eta | \boldsymbol{1} | 2 ]}{m\, \langle\eta 2 \rangle} \langle\boldsymbol{13} \rangle}  \  .
\end{align}
}
To derive the boxed equation, we used three-point kinematics, $p_1+p_3+k_2=0$, which gives $p_1\!\cdot k_2=p_3\!\cdot k_2=0$. Hence $p_1|2]$ and $p_3|2]$ are proportional to $|2\rangle$, leading to $[\boldsymbol{1}2]=-x\langle\boldsymbol{1}2\rangle$ and $[2\boldsymbol{3}]=-x\langle\boldsymbol{3}2\rangle$, with
\begin{equation}
    x \equiv \frac{\langle\eta|\boldsymbol{1}|2]}{m\langle\eta 2\rangle} \ .
\end{equation}
The remaining numerator then reduces by Schouten to
$x\langle\boldsymbol{13}\rangle\langle\eta 2\rangle$. We can also derive the boxed equation using Schouten identities \footnote{ We use eq.~\eqref{eq:conventions-dirac-u-spinors}--\eqref{eq:conventions-dirac-v-spinors} in the first step. And then we use momentum conservation: $p_3 = - p_1 - k_2$, following the Schouten identity.
\begin{align*}
    & \frac{(-[\boldsymbol{1}  2] \langle \eta  \boldsymbol{3}\rangle \, - \, \langle\boldsymbol{1} \eta \rangle\, [2 \boldsymbol{3}])}{\langle\eta 2\rangle} = \frac{(\langle \boldsymbol{1} |\boldsymbol{1}|  2] \langle \eta  \boldsymbol{3}\rangle \, - \, \langle\boldsymbol{1} \eta \rangle\, [2 |\boldsymbol{3}| \boldsymbol{3}\rangle)}{m \, \langle\eta 2\rangle} = \frac{(\langle \boldsymbol{1} |\boldsymbol{1}|  2] \langle \eta  \boldsymbol{3}\rangle \, + \, \langle\boldsymbol{1} \eta \rangle\, [2 |\boldsymbol{1}| \boldsymbol{3}\rangle)}{m \, \langle\eta 2\rangle} \\
    =& \frac{ (-\langle \boldsymbol{3} |\boldsymbol{1}|  2] \langle \boldsymbol{1} \eta\rangle - \langle \eta |\boldsymbol{1}|  2] \langle \boldsymbol{3 1} \rangle) \, + \, \langle\boldsymbol{1} \eta \rangle\, [2 |\boldsymbol{1}| \boldsymbol{3}\rangle}{m \, \langle\eta 2\rangle} = \frac{ \langle \eta |\boldsymbol{1}|  2] \langle \boldsymbol{1 3} \rangle)}{m \, \langle\eta 2\rangle}  \  .
\end{align*}
}.

Similarly,
{
\allowdisplaybreaks
\begin{align}
    \mathcal{A}_{(3)}^{(-)} \, \nonumber&= \, i \, e \, \left(\begin{aligned}
    {-\lambda_{1}^{\alpha}} \quad{\tilde{\lambda}_{1 \dot{\alpha}}}
\end{aligned}\right) \, \left(\begin{aligned}
    0 \quad \varepsilon_{\alpha \dot{\alpha}}^-\\
    \varepsilon^{\dot{\alpha}\alpha}_- \quad 0
\end{aligned}\right) \, \left(\begin{aligned}
    {-\lambda_{3 \alpha}}\\{\tilde{\lambda}_{3}^{\dot{\alpha}}}
\end{aligned}\right) \, = \, i \, e \, (-\tilde{\lambda}_{1 \dot{\alpha}} \, \varepsilon^{\dot{\alpha}\alpha}_- \, \lambda_{3 \alpha} \, - \, \lambda_{1}^{\alpha} \, \varepsilon_{\alpha \dot{\alpha}}^- \, \tilde{\lambda}_{3}^{\dot{\alpha}}) \\
&\nonumber = \, i \, \sqrt{2} \, e  \; \frac{(-[\boldsymbol{1} \eta ]\, \langle 2 \boldsymbol{3}\rangle \, - \, \langle \boldsymbol{1} 2\rangle \, [\eta \boldsymbol{3}])}{[2 \eta]}\\
\implies\, \Aboxed{\mathcal{A}_{(3)}^{(-)}\, &= \, i \, \sqrt{2} \, e \, \frac{\langle 2 | \boldsymbol{1} | \eta ]}{m\, [ 2 \eta ]} [\boldsymbol{13}] }  \  .
\end{align}
}

\subsection{Spin-$1/2$ Compton Amplitude}\label{subsec:spinor-qed-compton}
\ 
% \[\text{QED Lagrangian: }
% \mathcal{L} = \bar{\psi}(i \slashed{D} - m) \psi -\frac{1}{4}F^{\mu\nu}F_{\mu\nu}\: , \] where $F^{\mu\nu}$ is the field strength and $D_\mu=\partial_\mu - i e A_\mu$ is the covariant derivative.

% Therefore the interaction Lagrangian is \[\mathcal{L}_{int} = e \bar{\psi} \gamma^\mu A_{\mu} \psi\]

\begin{figure}[h]
    \centering
    \begin{tikzpicture}
\begin{feynman}
    % Define vertices
    \vertex (a) at (0,0) {\(\boldsymbol{1}^I\)};
    \vertex (b) at (3,0) {\(\boldsymbol{3}^J\)};
    \vertex (c) at (1,2) {\(2\)};
    
    % Internal vertices
    \vertex (v1) at (1.5,0);
    \vertex (v3) at (1.5,2);

    % Draw diagram
    \diagram*{
        (b) --  [fermion] (v1) -- [fermion] (a),
        (v1) -- [photon, out=90, in=-90] (v3),
    };
\end{feynman}
\end{tikzpicture} 
    \caption{Three-point Amplitude}
    \label{fig:spinor-qed-compton-vertex}
\end{figure}
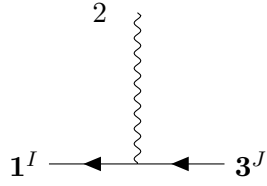

For an incoming fermion with momentum $-\boldsymbol{p}_3$, an outgoing fermion with momentum $\boldsymbol{p}_1$, and an outgoing photon with momentum $k_2$, the three-point amplitude in fig.~\ref{fig:spinor-qed-compton-vertex} is
\[\mathcal{A}_{(3)}(\boldsymbol{p}_1, k_2, \boldsymbol{p}_3) =  i e \bar{u}(\boldsymbol{p}_1) \gamma^\mu \varepsilon_\mu (k_2) v(\boldsymbol{p}_3) \ . \]

We now calculate the four-point Compton amplitude by summing Feynman diagrams in the $12$-channel, i.e.\ the $s$-channel, and the $13$-channel, i.e.\ the $u$-channel.

In the $12$-channel, the diagram in fig.~\ref{fig:12-Channel} evaluates to
\begin{equation}
    \mathcal{A}_{12}(\boldsymbol{1}^I, 2, 3, \boldsymbol{4}^J)
    =
    -ie^2 \bar{u}(\boldsymbol{p}_1)^I \slashed{\varepsilon}_2\ \frac{(\slashed{\boldsymbol{p}}_{12} + m)}{s_{12}-m^2} \slashed{\varepsilon}_3\ {v}(\boldsymbol{p}_{4})^J  \  .\label{eq:spinor-qed-compton-12-channel}
\end{equation}
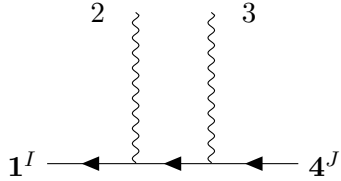
\begin{figure}[H]
    \centering
    \begin{tikzpicture}
\begin{feynman}
    % Define vertices
    \vertex (a) at (0,0) {\(\boldsymbol{1}^I\)};
    \vertex (b) at (4,0) {\(\boldsymbol{4}^J\)};
    \vertex (c) at (1,2) {\(2\)};
    \vertex (d) at (3,2) {\(3\)};
    
    % Internal vertices
    \vertex (v1) at (1.5,0);
    \vertex (v2) at (2.5,0);
    \vertex (v3) at (1.5,2);
    \vertex (v4) at (2.5,2);

    % Draw diagram
    \diagram*{
        (b) -- [fermion] (v2) -- [fermion] (v1) -- [fermion] (a),
        (v1) -- [photon, out=90, in=-90] (v3),
        (v2) -- [photon, out=90, in=-90] (v4),
    };
\end{feynman}
\end{tikzpicture} 
    \caption{12-Channel}
    \label{fig:12-Channel}
\end{figure}
To obtain the contribution from the $13$-channel in fig.~\ref{fig:13-Channel}, we will just swap legs $2$ and $3$.
\begin{figure}[H]
    \centering
    \begin{tikzpicture}
\begin{feynman}
    % Define vertices
    \vertex (a) at (0,0) {\(\boldsymbol{1}^I\)};
    \vertex (b) at (4,0) {\(\boldsymbol{4}^J\)};
    \vertex (c) at (1,2) {\(3\)};
    \vertex (d) at (3,2) {\(2\)};
    
    % Internal vertices
    \vertex (v1) at (1.5,0);
    \vertex (v2) at (2.5,0);
    \vertex (v3) at (1.5,2);
    \vertex (v4) at (2.5,2);

    % Draw diagram
    \diagram*{
        (b) -- [fermion] (v2) -- [fermion] (v1) -- [fermion] (a),
        (v1) -- [photon, out=90, in=-90] (v3),
        (v2) -- [photon, out=90, in=-90] (v4),
    };
\end{feynman}
\end{tikzpicture} 
    \caption{13-Channel}
    \label{fig:13-Channel}
\end{figure}
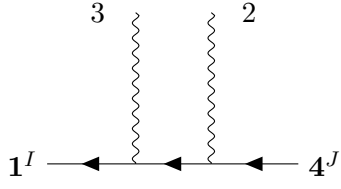
\begin{align}
    &\mathcal{A}_{13}(\boldsymbol{1}^I, 2, 3, \boldsymbol{4}^J)
    = -ie^2 \bar{u}(\boldsymbol{p}_1)^I \slashed{\varepsilon}_3\ \frac{(\slashed{\boldsymbol{p}}_{13} + m)}{s_{13}-m^2} \slashed{\varepsilon}_2\ {v}(\boldsymbol{p}_{4})^J  \  .\label{eq:spinor-qed-compton-13-channel}
\end{align}

Adding eq.~\eqref{eq:spinor-qed-compton-12-channel} and \eqref{eq:spinor-qed-compton-13-channel}, we get the total contribution for the four-point Compton amplitude: 
{
\allowdisplaybreaks
\begin{align}
    \nonumber&\mathcal{A}_{(4)}(\boldsymbol{1}^I, 2, 3, \boldsymbol{4}^J)\\
    =\nonumber& \mathcal{A}_{12}(\boldsymbol{1}^I, 2, 3, \boldsymbol{4}^J) + \mathcal{A}_{13}(\boldsymbol{1}^I, 2, 3, \boldsymbol{4}^J)\\
    =& -ie^2 \bar{u}(\boldsymbol{p}_1)^I \slashed{\varepsilon}_2\ \frac{(\slashed{\boldsymbol{p}}_{12} + m)}{s_{12}-m^2} \slashed{\varepsilon}_3\ {v}(\boldsymbol{p}_{4})^J + (2\leftrightarrow3)  \  . \label{eq:spinor-qed-compton-amplitude}
\end{align}
}

This is what we expect to get from the textbook Feynman rules in QED. This calculation also verifies the rule in eq.~\eqref{eq:conventions-negated-momentum-spinors} for gluing the massive legs in massive spinor-helicity language.

Now, we will reproduce the four-point different-helicity photon Compton amplitude from eq.~\eqref{eq:spinor-qed-compton-amplitude}.
{\allowdisplaybreaks
\begin{align}
    \nonumber&\mathcal{A}_{(3)}(\boldsymbol{1}^I, 2^+, 3^-, \boldsymbol{4}^J)\\
    =\nonumber& -ie^2 \bar{u}(\boldsymbol{p}_1)^I \slashed{\varepsilon}_2^+\ \frac{(\slashed{\boldsymbol{p}}_{12} + m)}{s_{12}-m^2} \slashed{\varepsilon}_3^-\ {v}(\boldsymbol{p}_{4})^J + (2\leftrightarrow3)\\
    =\nonumber& \frac{-i\,e^2}{s_{12}-m^2} \, \left(\begin{aligned}
    {-\lambda_{1}^{\alpha I}} \quad{\tilde{\lambda}_{1 \dot{\alpha}}^I}
\end{aligned}\right) \, \left(\begin{aligned}
    & 0 & \varepsilon_{2\alpha \dot{\beta}}^+\\
    &\varepsilon_{2^+}^{\dot{\alpha}\beta} & 0
\end{aligned}\right) \, \left(\begin{aligned}
    &0 & p_{12\beta \dot{\rho}}\\
    &p_{12}^{\dot{\beta}\rho} & 0
\end{aligned}\right) \, \left(\begin{aligned}
    &0 & \varepsilon_{3\sigma \dot{\rho}}^-\\
    &\varepsilon_{3^-}^{\dot{\rho}\sigma} & 0
\end{aligned}\right) \, \left(\begin{aligned}
    {-\lambda_{4 \sigma}^J}\\{\tilde{\lambda}_{4}^{\dot{\sigma J}}} 
\end{aligned}\right) + (2\leftrightarrow3)\\
    =\nonumber& \frac{-i\,e^2}{s_{12}-m^2} \, (-\tilde{\lambda}_{1 \dot{\alpha}}^I \; \varepsilon_{2^+}^{\dot{\alpha}\beta} \; p_{12\beta \dot{\rho}} \; \varepsilon_{3^-}^{\dot{\rho}\sigma} \; \lambda_{4 \sigma}^J \; - \; \lambda_{1}^{\alpha I} \; \varepsilon_{2\alpha \dot{\beta}}^+ \; p_{12}^{\dot{\beta}\rho} \; \varepsilon_{3\sigma \dot{\rho}}^- \; \tilde{\lambda}_{4}^{\dot{\sigma} J}) + (2\leftrightarrow3)\\
    =\nonumber& \frac{i\,e^2\,(\sqrt{2})^2}{s_{12}-m^2} \, \frac{[\boldsymbol{1}^I 2]\langle 3|\slashed{p}_{12}|2] \langle 3 \boldsymbol{4}^J \rangle + \langle \boldsymbol{1}^I 3 \rangle [2 | \slashed{p}_{12} | 3 \rangle [2 \boldsymbol{4}^J]}{ \langle 3 2 \rangle [ 3 2 ]} + (2\leftrightarrow3) \\
    =& -2 \, i \, e^2 \, \frac{\langle3 | \boldsymbol{1} | 2 ] }{(s_{12}-m^2) (s_{13}-m^2)} ([\boldsymbol{1}^I 2] \langle \boldsymbol{4}^J 3 \rangle + \langle \boldsymbol{1}^I 3 \rangle [ \boldsymbol{4}^J 2 ])  \  . \label{eq:spinor-qed-compton-spinor-helicity-amplitude}
\end{align}
}
In the above calculation the polarization for leg 2 and 3 are chosen as \[ \varepsilon_2^{+\mu} = \frac{\langle 3| \gamma^\mu | 2]}{\sqrt{2}\langle 3 2\rangle}, \qquad \varepsilon_3^{ - \mu} = \frac{\langle 3 | \gamma^\mu | 2 ]}{\sqrt{2} [3 2]} \ ,\]
The result in eq.~\eqref{eq:spinor-qed-compton-spinor-helicity-amplitude} agrees with the result in eq.~\eqref{eq:QED_4pt_Perm_Sum_Opp_Hel} derived from permutation sums of Yang--Mills amplitudes and the result in eq.~\eqref{eq:qed-four-point-bcfw-positive-negative-helicity} derived from BCFW recursion relations.

%----------------------------------------------------------------------------

\bibliographystyle{JHEP}
\bibliography{FivePointCompton1}

\end{document}